\documentclass[10pt,british,superscript,utf8,nomove,x11names,shortlabels]{article}

\usepackage[T1]{fontenc}
\usepackage[latin9]{inputenc}

\usepackage[a4paper]{geometry}
\usepackage{xcolor}
\usepackage{babel}
\usepackage{array}
\usepackage{longtable}
\usepackage{rotating}
\usepackage{float}
\usepackage{mathtools}
\usepackage{url}
\usepackage{enumitem}
\usepackage{multirow}
\usepackage{amsmath}
\usepackage{amsthm}
\usepackage{amssymb}
\usepackage[unicode=true,pdfusetitle,
 bookmarks=true,bookmarksnumbered=false,bookmarksopen=false,
 breaklinks=true,pdfborder={0 0 0},pdfborderstyle={},backref=false,colorlinks=true]
 {hyperref}
\hypersetup{
 colorlinks,citecolor=SpringGreen4,filecolor=black,linkcolor=blueTOC,urlcolor=blueTOC}

\makeatletter

\providecommand{\tabularnewline}{\\}

\usepackage{textcomp}
\usepackage[autolinebreaks,useliterate]{incl/mcode} 

\usepackage[space]{cite}
\setlist[enumerate,2]{leftmargin=0.2cm}

\usepackage{linegoal}
\usepackage{listings}
\usepackage{nameref}
\usepackage{titlesec}
\titlespacing\subsubsection{0pt}{12pt plus 4pt minus 2pt}{-2pt plus 2pt minus 2pt}

\usepackage{etoolbox}
\apptocmd{\thebibliography}{\normalfont}{}{}%

\newsavebox{\mylisting}

\definecolor{mygreen}{RGB}{28,172,0} 
\definecolor{mylilas}{RGB}{170,55,241}
\definecolor{mygrey}{RGB}{100,100,100}
\definecolor{myblue}{RGB}{91,146,178}
\definecolor{green_optForce}{RGB}{0, 158, 59}

\definecolor{NatPTiming}{HTML}{F46323}
\definecolor{NatPCritical}{HTML}{8A2293}
\definecolor{NatPCaution}{HTML}{C51425}
\definecolor{NatPTroubleshooting}{HTML}{2D52A9}
\definecolor{ForestGreen}{HTML}{228B22}

\definecolor{redContrib}{HTML}{CE0000}
\definecolor{orangeContrib}{HTML}{E07B17}
\definecolor{greenContrib}{HTML}{43A028}
\definecolor{blueContrib}{HTML}{0750EF}
\definecolor{pinkContrib}{HTML}{E12FF4}

\lstdefinelanguage{Julia}%
  {morekeywords={abstract,break,case,catch,const,continue,do,else,elseif,%
      end,export,false,for,function,immutable,import,importall,if,in,%
      macro,module,otherwise,quote,return,switch,true,try,type,typealias,%
      using,while},%
   sensitive=true,%
   alsoother={$},%
   morecomment=[l]\#,%
   morecomment=[n]{\#=}{=\#},%
   morestring=[s]{"}{"},%
   morestring=[m]{'}{'},%
}[keywords,comments,strings]%

\lstdefinestyle{juliaStyle}{
    language = Julia,
    basicstyle = \footnotesize\ttfamily,
    keywordstyle = \bfseries\color{blue},
    stringstyle = \color{magenta},
    commentstyle = \color{ForestGreen},
    showstringspaces = false,}

\lstdefinestyle{bashStyle} {
    language=bash,%
    basicstyle=\footnotesize\ttfamily,%
    breaklines=true,%
    identifierstyle=\color{black},%
    stringstyle=\color{red},%
    commentstyle=\color{mygrey},%
    showstringspaces=false,
    emph=[1]{\$},emphstyle=[1]\color{mylilas}, 
    emph=[2]{cd, git, clone, checkout, pull, fetch, commit, rebase}, emphstyle=[2]\color{myblue},
}
\usepackage{tikz}
\usepackage{xspace}
\usepackage[useregional]{datetime2}
\definecolor{blueTOC}{HTML}{1D7DDF}

\newcommand{\timingSymbol}{
\begin{tikzpicture}
\filldraw[fill=NatPTiming,draw=NatPTiming] circle (3pt);
\end{tikzpicture}
}

\newcommand{\criticalSymbol}{
\begin{tikzpicture}
\filldraw[fill=NatPCritical,draw=NatPCritical]  (0,0) --
(6pt,0) -- (3pt,6pt);
\end{tikzpicture}
}

\usepackage{relsize}

\newcommand{\timing}{\;\timingSymbol{\color{NatPTiming}\textbf{\hyperref[sec:timing]{\color{NatPTiming}TIMING}}}\xspace\normalsize}
\newcommand{\criticalStep}{\criticalSymbol{\color{NatPCritical}\textbf{CRITICAL STEP}}\xspace\normalsize}

\newcommand{\caution}{{\color{NatPCaution}\textbf{!\,CAUTION}}\xspace\normalsize}
\newcommand{\troubleshooting}{\;{\color{NatPTroubleshooting}\textbf{\hyperref[sec:troubleshooting]{?\,TROUBLESHOOTING}}}\xspace\normalsize}
\usepackage[font=small,labelfont=bf]{caption}

\definecolor{greenOptSys}{HTML}{75C047}
\newcounter{psteps}
\newcommand{\pStep}{\refstepcounter{psteps}{\vspace{.5\baselineskip}\noindent\large \bfseries {\color{greenOptSys}\arabic{psteps}$~|$}}~~\normalsize}
\newcommand{\pStepRef}[1]{\vspace{.5\baselineskip}\noindent\large \textbf{\color{greenOptSys}\ref*{#1}$~|$}~~\normalsize}

\makeatletter
\def\@seccntformat#1{\csname #1ignore\expandafter\endcsname\csname the#1\endcsname\quad}
\let\sectionignore\@gobbletwo
\def\@subseccntformat#1{\csname #1ignore\expandafter\endcsname\csname the#1\endcsname\quad}
\let\subsectionignore\@gobbletwo
\def\@subsubseccntformat#1{\csname #1ignore\expandafter\endcsname\csname the#1\endcsname\quad}
\let\subsubsectionignore\@gobbletwo
\let\latex@numberline\numberline
\def\numberline#1{\if\relax#1\relax\else\latex@numberline{#1}\fi}
\makeatother

\usepackage{array,booktabs}
\newcolumntype{M}[1]{>{\centering\arraybackslash}m{#1}}
\usepackage{hyphenat}

\makeatother

\begin{document}
\date{} 

\title{Creation and analysis of biochemical constraint-based models: the
COBRA Toolbox v3.0}
\maketitle
\begin{center}
\vspace*{-2cm}
\par\end{center}

\begin{center}
\today
\par\end{center}

Laurent Heirendt$^{1}$ \& Sylvain Arreckx$^{1}$, Thomas Pfau$^{2}$,
Sebasti\'{a}n N. Mendoza$^{3,18}$, Anne Richelle$^{4}$, Almut Heinken$^{1}$,
Hulda S. Haraldsd\'{o}ttir$^{1}$, Jacek Wachowiak$^{1}$, Sarah M. Keating$^{5}$,
Vanja Vlasov$^{1}$, Stefania Magnusd\'{o}ttir$^{1}$, Chiam Yu Ng$^{6}$,
German Preciat$^{1}$, Alise \v Zagare$^{1}$, Siu H.J. Chan$^{6}$,
Maike K. Aurich$^{1}$, Catherine M. Clancy$^{1}$, Jennifer Modamio$^{1}$,
John T. Sauls$^{7}$, Alberto Noronha$^{1}$, Aarash Bordbar$^{8}$,
Benjamin Cousins$^{9}$, Diana C. El Assal$^{1}$, Luis V. Valcarcel$^{10}$,
I\~{n}igo Apaolaza$^{10}$, Susan Ghaderi$^{1}$, Masoud Ahookhosh$^{1}$,
Marouen Ben Guebila$^{1}$, Andrejs Kostromins$^{11}$, Nicolas Sompairac$^{22}$,
Hoai M. Le$^{1}$, Ding Ma$^{12}$, Yuekai Sun$^{12}$, Lin Wang$^{6}$,
James T. Yurkovich$^{13}$, Miguel A.P. Oliveira$^{1}$, Phan T. Vuong$^{1}$,
Lemmer P. El Assal$^{1}$, Inna Kuperstein$^{22}$, Andrei Zinovyev$^{22}$,
H. Scott Hinton$^{14}$, William A. Bryant$^{15}$, Francisco J. Arag\'{o}n
Artacho$^{16}$, Francisco J. Planes$^{10}$, Egils Stalidzans$^{11}$,
Alejandro Maass$^{3,18}$, Santosh Vempala$^{9}$, Michael Hucka$^{17}$,
Michael A. Saunders$^{12}$, Costas D. Maranas$^{6}$, Nathan E. Lewis$^{4,19}$,
Thomas Sauter$^{2}$, Bernhard \O. Palsson$^{13,21}$, Ines Thiele$^{1}$,
Ronan M.T. Fleming$^{1}$

\vspace{2cm}
\begin{center}
\textbf{}\emph{This protocol is an update to Nature Protocols 2(3),
727\textendash 738, March (2007); doi:10.1038/nprot.2007.99; published
online 29 March 2007 and Nature Protocols 6 (9), 1290-1307 (2011);
doi:10.1038/protex.2011.234; published online 11 May 2011.}
\par\end{center}

\vspace{3cm}
\begin{center}
\textbf{Abstract}
\par\end{center}

COnstraint-Based Reconstruction and Analysis (COBRA) provides a molecular
mechanistic framework for integrative analysis of experimental data
and quantitative prediction of physicochemically and biochemically
feasible phenotypic states. The COBRA Toolbox is a comprehensive software
suite of interoperable COBRA methods. It has found widespread applications
in biology, biomedicine, and biotechnology because its functions can
be flexibly combined to implement tailored COBRA protocols for any
biochemical network. Version 3.0 includes new methods for quality
controlled reconstruction, modelling, topological analysis, strain
and experimental design, network visualisation as well as network
integration of chemoinformatic, metabolomic, transcriptomic, proteomic,
and thermochemical data. New multi-lingual code integration also enables
an expansion in COBRA application scope via high-precision, high-performance,
and nonlinear numerical optimisation solvers for multi-scale, multi-cellular
and reaction kinetic modelling, respectively. This protocol can be
adapted for the generation and analysis of a constraint-based model
in a wide variety of molecular systems biology scenarios. This protocol
is an update to the COBRA Toolbox 1.0 and 2.0. The COBRA Toolbox 3.0
provides an unparalleled depth of constraint-based reconstruction
and analysis methods.

\vspace{1cm}
\begin{center}
\textbf{Keywords}
\par\end{center}

Metabolic models, metabolic reconstruction, metabolic engineering,
gap filling, strain engineering, omics, data integration, metabolomics,
transcriptomics, constraint-based modelling, computational biology,
bioinformatics, biochemistry, human metabolism, and microbiome analysis.

\pagebreak{}

\vspace*{\fill}

\footnotesize

$^{1}$Luxembourg Centre for Systems Biomedicine, University of Luxembourg,
6 avenue du Swing, Belvaux, L-4367, Luxembourg.

\textbf{$^{2}$}Life Sciences Research Unit, University of Luxembourg,
6 avenue du Swing, Belvaux, L-4367, Luxembourg.

$^{3}$Center for Genome Regulation (Fondap 15090007), University
of Chile, Blanco Encalada 2085, Santiago, Chile.

\textbf{$^{4}$}Department of Pediatrics, University of California,
San Diego, School of Medicine, La Jolla, CA 92093, USA.

$^{5}$European Molecular Biology Laboratory, European Bioinformatics
Institute (EMBL-EBI), Hinxton, Cambridge, CB10 1SD, United Kingdom.

\textbf{$^{6}$}Department of Chemical Engineering, The Pennsylvania
State University, University, University Park, PA 16802, USA.

$^{7}$Department of Physics, University of California, San Diego,
9500 Gilman Dr., La Jolla, CA 92093, USA; Bioinformatics and Systems
Biology Program, University of California, San Diego, La Jolla, CA,
USA.

$^{8}$Sinopia Biosciences, San Diego, CA, USA.

\textbf{$^{9}$}School of Computer Science, Algorithms and Randomness
Center, Georgia Institute of Technology, Atlanta, GA, USA.

$^{10}$Biomedical Engineering and Sciences Department, TECNUN, University
of Navarra, Paseo de Manuel Lardizabal, 13, 20018, San Sebastian,
Spain.

$^{11}$Institute of Microbiology and Biotechnology, University of
Latvia, Jelgavas iela 1, Riga LV-1004, Latvia.

$^{12}$Department of Management Science and Engineering, Stanford
University, Stanford CA 94305-4026, USA.

$^{13}$Bioengineering Department, University of California, San Diego,
La Jolla, CA, USA.

$^{14}$Utah State University Research Foundation, 1695 North Research
Park Way, North Logan, Utah 84341, USA.

$^{15}$Centre for Integrative Systems Biology and Bioinformatics,
Department of Life Sciences, Imperial College London, London, United
Kingdom.

$^{16}$Department of Mathematics, University of Alicante, Spain.

$^{17}$California Institute of Technology, Computing and Mathematical
Sciences, MC 305-16, 1200 E. California Blvd., Pasadena, CA 91125,
USA.

$^{18}$Mathomics, Center for Mathematical Modeling, University of
Chile, Beauchef 851, 7th Floor, Santiago, Chile.

$^{19}$Novo Nordisk Foundation Center for Biosustainability at the
University of California, San Diego, La Jolla, CA 92093, United States.

$^{20}$Latvian Biomedical Research and Study Centre, Ratsupites iela
1, Riga, LV1067, Latvia.

$^{21}$Novo Nordisk Foundation Center for Biosustainability, Technical
University of Denmark, Kemitorvet, Building 220, 2800 Kgs. Lyngby,
Denmark.

$^{22}$Institut Curie, PSL Research University, Mines Paris Tech,
Inserm, U900, F-75005, Paris, France.\\
\\
Correspondence should be addressed to Ronan M.T. Fleming (ronan.mt.fleming@gmail.com).

\normalsize

\pagebreak{}

\section{INTRODUCTION}

\subsection{Development of the protocol}

In the past two decades, significant developments in molecular biology
have led to a deluge of data on the molecular composition of organisms
and their environment. Quantitative molecular measurements at genome
scale are now routine and include genomic, metabolomic, transcriptomic,
and proteomic data. This wealth of data has stimulated the development
of a wide variety of algorithms and software for data preprocessing
and analysis. Preprocessing converts the measured variables into estimates
of the quantity of each molecular species present in a sample. Analysis
takes this quantification of molecular species to derive new biological
knowledge. Despite many advances and novel discoveries, it remains
a major challenge for the biology community to be able to derive insight
from experiments in a way that leads to a deeper understanding of
a biological system. 

Analysis techniques may be split into those that do and those that
do not incorporate prior information on a system in question. All
else being equal, an integrative analysis technique that incorporates
prior information from complementary experiments will typically outperform
a technique that does not, because corroboration with data from complementary
experimental platforms can be used to increase the confidence that
an observation is real. Fundamentally, integrative analysis is successful
because it is an expression of a Bayesian statistical approach whereby
prior information from complementary experiments represents the knowledge
about a system before a new experiment. Thus, integrative analysis
is an inference procedure to update the knowledge about the system
in light of new data from an additional experiment. 

Integrative analysis can be either mechanistic or non-mechanistic.
Mechanistic techniques rely on using molecular mechanistic models
to represent prior information on the biochemical networks underlying
the data being analysed. Non-mechanistic integrative analysis is often
used to study 'omics' data and is essential when nothing is known
about the underlying molecular mechanisms. However, often a subset
of the underlying molecular mechanisms is known. Thus, a significant
limitation of non-mechanistic integrative analysis is that it ignores
the decades of prior information from molecular mechanistic experimental
studies in biochemistry, molecular biology, etc. As such, non-mechanistic
integrative analysis omits valuable prior information from its biochemical
network context and from the general physicochemical principles that
any biochemical network must obey. 

\begin{figure}[H]
\begin{centering}
\includegraphics[width=0.9\textwidth]{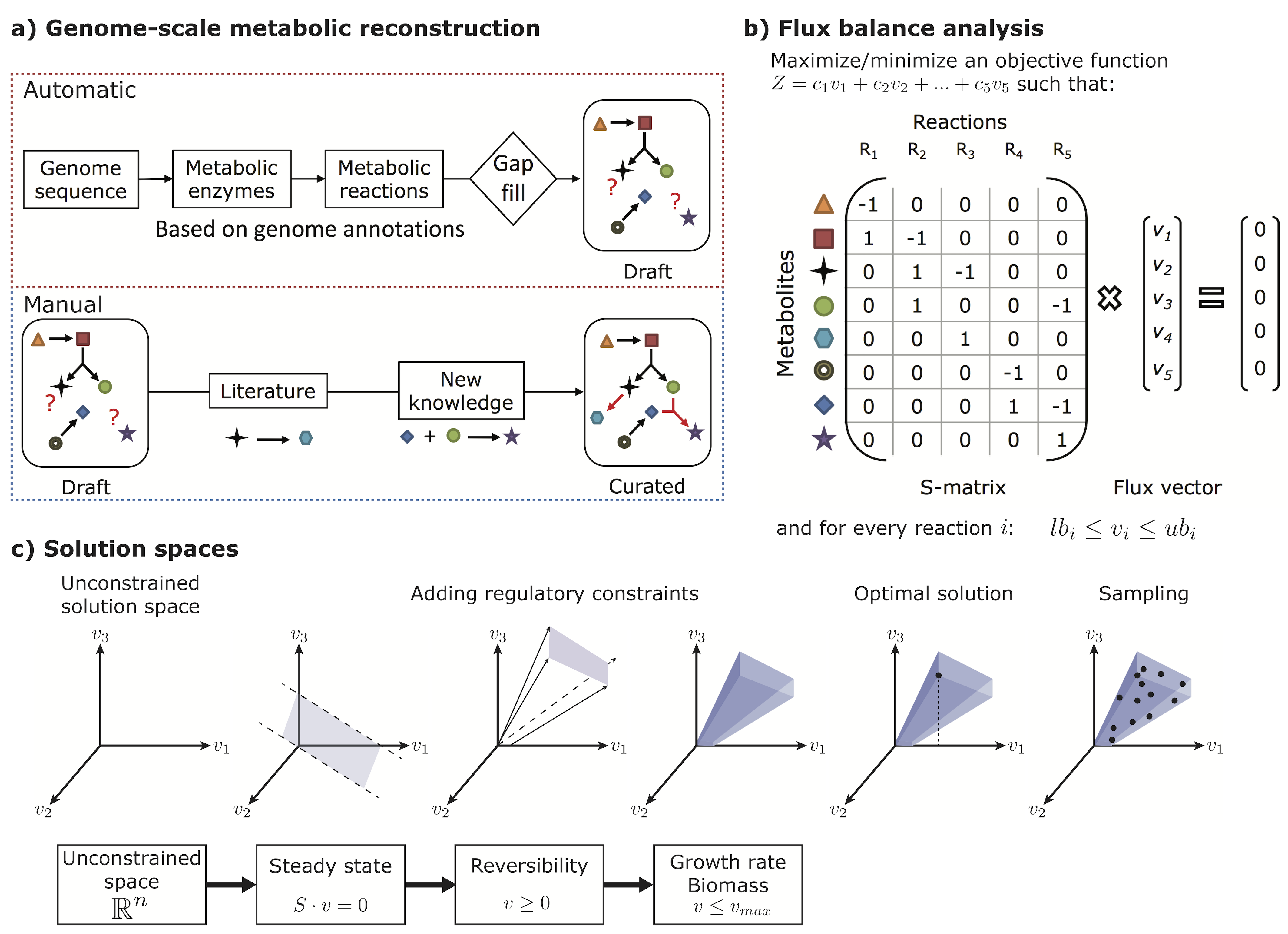}
\par\end{centering}
\caption{\label{fig:COBRA} Illustration of genome-scale metabolic reconstruction
and flux balance analysis. \textbf{a.} Genome-scale metabolic reconstructions
are built in two steps. First, several platforms exist for the generation
of a gap-filled draft metabolic reconstruction based on genome annotations.
Second, the draft reconstructions need to be refined based on known
experimental and biochemical data from literature\cite{thiele_protocol_2010}.
Novel experiments can be performed on the organism and the reconstruction
refined accordingly. \textbf{b.} In flux balance analysis (FBA), the
metabolic reconstruction containing $n$ reactions and $m$ metabolites
is converted to a stoichiometric matrix $S$. FBA solves an optimisation
problem where an objective function $Z$ is maximised or minimised.
$Z$ is formed by multiplying every reaction flux $v_{i}$ with a
predetermined constant, $c_{i}$, and adding the resulting values.
FBA solves a steady state, $Sv=0$. Every reaction $i$ is bound by
two values, an upper bound ($ub_{i}$) and lower bound ($lb_{i}$).
\textbf{c}. The solution space is defined by the mass balanced constraints
imposed for the upper and lower bounds of each reaction in the metabolic
reconstruction. Optimisation for a biological objective function (e.g.
biomass production, ATP consumption, heme production, etc.) identifies
an optimal flux within the solution space, whereas uniform sampling
provides an unbiased characterisation by sampling candidate fluxes
distributed in the solution space.}
\end{figure}

A genome-scale metabolic \emph{reconstruction} (see Figure \ref{fig:COBRA})
is a structured knowledge-base that abstracts pertinent information
on the biochemical transformations taking place within a chosen biochemical
system, e.g.,\emph{ }the human gut microbiome\cite{magnusdottir_generation_2017}.
Constraint-based reconstruction and analysis (COBRA\cite{palsson_systems_2015})
is a mechanistic integrative analysis framework that is applicable
to any biochemical system with prior mechanistic information, including
where mechanistic information is incomplete. The overall approach
is to mechanistically represent the relationship between genotype
and phenotype by mathematically and computationally modelling the
constraints that are imposed on the phenotype of a biochemical system
by physicochemical laws, genetics, and the environment\cite{obrien_using_2015}.
Emerging from multiple origins and catalysed by many original contributions
from a variety of fields, there now exists a wide variety of novel
methodological developments that can be organised and categorised
within the COBRA framework.

Early in the development of the COBRA framework, the need for ease
of reproducibility and demand for reuse of COBRA methods were recognised.
This necessity led to the COBRA Toolbox version 1.0\cite{becker_quantitative_2007},
an open source software package running in the MATLAB environment,
which facilitated quantitative prediction of metabolic phenotypes
using a selection of the COBRA methods available at the time. With
the expansion of the COBRA community and the growing phylogeny of
COBRA methods, the need was recognised for the amalgamation and transparent
dissemination of COBRA methods. This demand led to the COBRA Toolbox
version 2.0\cite{schellenberger_quantitative_2011}, with an enhanced
range of methods to simulate, analyse, and predict a variety of phenotypes
using genome-scale metabolic reconstructions. Since then, the increasing
functional scope and size of biochemical network reconstructions,
as well as the increasing breadth of physicochemical and biological
constraints that are represented within constraint-based models, naturally
result in the development of a broad arbour of new COBRA methods\cite{lewis_constraining_2012}.

\normalsize

The present protocol is an update to the COBRA Toolbox 1.0\cite{becker_quantitative_2007}
and 2.0\cite{schellenberger_quantitative_2011} illustrating the main
novel developments within version 3.0 of the COBRA Toolbox (see Table
\ref{tab:featureTable}), especially the expansion of functionality
to cover new modelling methods. In particular, this protocol includes
the input and output of new standards for sharing reconstructions
and models, an extended suite of supported general purpose optimisation
solvers, new optimisation solvers developed especially for constraint-based
modelling problems, enhanced functionality in the areas of computational
efficiency and high precision computing, numerical characterisation
of reconstructions, conversion of reconstructions into various forms
of constraint-based models, comprehensive support for flux balance
analysis and its variants, integration with omics data, uniform sampling
of high dimensional models, atomic resolution of metabolic reconstructions,
estimation and application of thermodynamic constraints, visualisation
of metabolic networks, and genome-scale kinetic modelling. 

With an increasing number of contributions from developers around
the world, the code base is evolving at a fast pace. The COBRA Toolbox
has evolved from monolingual MATLAB software to a multilingual software
suite via integration with C, FORTRAN, Julia, Perl and Python code,
as well as pre-compiled binaries, for specific purposes. For example,
the integration with quadruple precision numerical optimisation solvers,
implemented in FORTRAN, for robust and efficient modelling of multi-scale
biochemical networks, such as those obtained with integration\cite{thiele_multiscale_2012}
of metabolic\cite{feist_genome-scale_2007} and macromolecular synthesis\cite{thiele_genome-scale_2009}
reconstructions, which represent a new peak in terms of biochemical
comprehensiveness and predictive capacity\cite{yang_systems_2015}.
These developments warranted an industrial approach to software development
of the COBRA Toolbox. Therefore, we implemented a continuous integration
approach with the aim of guaranteeing a consistent, stable, and high-quality
software solution for a broad user community. All documentation and
code is released as part of the \emph{openCOBRA} project (\url{https://github.com/opencobra/cobratoolbox}).
Where reading the extensive documentation associated with the COBRA
Toolbox does not suffice, we describe the procedure for effectively
engaging with the community via a dedicated online forum (\url{https://groups.google.com/forum/#!forum/cobra-toolbox}).
Taken together, the COBRA Toolbox 3.0 provides an unparalleled depth
of interoperable COBRA methods and a proof-of-concept that knowledge
integration and collaboration by large numbers of scientists can lead
to cooperative advances impossible to achieve by a single scientist
or research group alone\cite{kitano_social_2011}.

\subsection{Applications of COBRA methods}

Constraint-based modelling of biochemical networks is broadly applicable
to a range of biological, biomedical, and biotechnological research
questions\cite{bordbar_constraint-based_2014} . Fundamentally, this
broad applicability arises from the common phylogenetic tree, shared
by all living organisms, that manifests in a set of shared mathematical
properties that are common to biochemical networks in normal, diseased,
wild-type, or mutant biochemical networks. Therefore, a COBRA method
developed primarily for use in one scenario can usually be quickly
adapted for use in a variety of related scenarios. Often, this adaptation
retains the mathematical properties of the optimisation problem underlying
the original constraint-based modelling method. By adapting the input
data and interpreting the output results in a different way, the same
method can be used to address a different research question.

Biotechnological applications of constraint-based modelling include
the development of sustainable approaches for chemical\cite{maia_silico_2016}
and biopharmaceutical production\cite{hefzi_consensus_2016,yusufi_mammalian_2017}.
Among these applications is the computational design of new microbial
strains for production of bioenergy feedstocks from non-food plants,
such as microbes capable of deconstructing biomass into their sugar
subunits and synthesising biofuels, either from cellulosic biomass
or through direct photosynthetic capture of carbon dioxide. 

\begin{table}
\begin{raggedright}
\fontsize{9}{11}\selectfont\hspace*{-0.3cm}%
\begin{tabular}{l|>{\raggedright}m{0.32\textwidth}|>{\raggedright}m{0.6\textwidth}}
\hline 
 & \textbf{\small{}Narrative} & \textbf{\small{}Novelty in the COBRA Toolbox 3.0 compared to 2.0}\tabularnewline
\hline 
\textbf{B} & {\small{}Initialise and verify the installation} & {\small{}Software dependency audit, e.g., solvers, binaries, git.}\tabularnewline
\hline 
\textbf{R} & {\small{}Input and output of reconstructions and models} & {\small{}Support for latest standards, e.g., SBML flux balance constraints\cite{bornstein_libsbml:_2008}.}\tabularnewline
\hline 
\textbf{R} & {\small{}Reconstruction: rBioNet} & {\small{}New software for quality controlled reconstruction\cite{thorleifsson_rbionet:_2011}.}\tabularnewline
\hline 
\textbf{R} & {\small{}Reconstruction: create a functional generic subnetwork} & {\small{}New methods for selecting different types of subnetworks.}\tabularnewline
\hline 
\textbf{R} & {\small{}Reconstruction exploration } & {\small{}New methods, e.g., find adjacent reactions.}\tabularnewline
\hline 
\textbf{R} & {\small{}Reconstruction refinement} & {\small{}Maintenance of internal model consistency, e.g., upon subnetwork
generation\cite{conde_constraint_2016}.}\tabularnewline
\hline 
\textbf{R} & {\small{}Numerical reconstruction properties} & {\small{}Flag a reconstruction requiring a multi-scale solver\cite{ma_reliable_2017}. }\tabularnewline
\hline 
\textbf{R} & {\small{}Convert a reconstruction into a flux balance analysis model} & {\small{}Identification of a maximal flux and stoichiometrically consistent
subset\cite{gevorgyan_detection_2008}.}\tabularnewline
\hline 
\textbf{I} & {\small{}Atomically resolve a metabolic reconstruction} & {\small{}New algorithms and methods for working with molecular structures,
atom mapping, identification of conserved moieties\cite{haraldsdottir_comparative_2014,haraldsdottir_identification_2016}.}\tabularnewline
\hline 
\textbf{I} & {\small{}Integration of metabolomic data} & {\small{}New methods for analysis of metabolomic data in a network
context\cite{aurich_metabotools:_2016,aurich_prediction_2014}.}\tabularnewline
\hline 
\textbf{I} & {\small{}Integration of transcriptomic and proteomic data} & {\small{}New algorithms for generation of context-specific models\cite{vlassis_fast_2014}.}\tabularnewline
\hline 
\textbf{A} & {\small{}Flux balance analysis and its variants} & {\small{}New flux balance methods, multi-scale model rescaling and
multi-scale solvers, additional solver interfaces, thermodynamically
feasible methods \cite{heirendt_distributedfba.jl:_2017,schellenberger_elimination_2011,guebila_model-based_2016,fleming_variational_2012,sun_robust_2013,thiele_multiscale_2012}.}\tabularnewline
\hline 
\textbf{A} & {\small{}Variation on reaction rate bounds in flux balance analysis} & {\small{}Increased computational efficiency.}\tabularnewline
\hline 
\textbf{A} & {\small{}Parsimonious flux balance analysis} & {\small{}New method for parsimonious flux balance analysis\cite{lewis_omic_2010}. }\tabularnewline
\hline 
\textbf{A} & {\small{}Sparse flux balance analysis} & {\small{}New method for sparse flux balance analysis.}\tabularnewline
\hline 
\textbf{A} & {\small{}Gap filling} & {\small{}Increased computational efficiency\cite{thiele_fastgapfill:_2014}.}\tabularnewline
\hline 
\textbf{A} & {\small{}Adding biological constraints to a flux balance model} & {\small{}New methods for coupling reaction rates\cite{magnusdottir_generation_2017,thiele_functional_2010}.}\tabularnewline
\hline 
\textbf{A} & {\small{}Testing biochemical fidelity} & {\small{}Human metabolic function test suite\cite{thiele_community-driven_2013}.}\tabularnewline
\hline 
\textbf{A} & {\small{}Testing basic properties of a metabolic model (sanity checks)} & {\small{}New methods to minimise occurrence of modelling artefacts}\cite{brunk_recon_2017}{\small{}.}\tabularnewline
\hline 
\textbf{A} & {\small{}Minimal spanning pathway vectors} & {\small{}New method for determining minimal spanning pathway vectors\cite{bordbar_minimal_2014}.}\tabularnewline
\hline 
\textbf{A} & {\small{}Elementary modes and pathway vectors} & {\small{}Extended functionality by integration with CellNetAnalyzer\cite{klamt_structural_2007}.}\tabularnewline
\hline 
\textbf{A} & {\small{}Minimal cut sets} & {\small{}Extended functionality by integration with CellNetAnalyzer\cite{ballerstein_minimal_2012,von_kamp_enumeration_2014},
and new algorithms for genetic MCSs \cite{apaolaza_-silico_2017}.}\tabularnewline
\hline 
\textbf{A} & {\small{}Flux variability analysis} & {\small{}Increased computational efficiency\cite{gudmundsson_computationally_2010}.}\tabularnewline
\hline 
\textbf{A} & {\small{}Uniform sampling of steady-state fluxes} & {\small{}New algorithm, guaranteed convergence to uniform distribution\cite{haraldsdottir_chrr:_2017}.}\tabularnewline
\hline 
\textbf{A} & {\small{}Thermodynamically constrain reaction directionality} & {\small{}New algorithms and methods for estimation of thermochemical
parameter estimation in multi-compartmental, genome-scale metabolic
models \cite{noor_consistent_2013,haraldsdottir_quantitative_2012}. }\tabularnewline
\hline 
\textbf{A} & {\small{}Variational kinetic modelling} & {\small{}New algorithms and methods for genome-scale kinetic modelling\cite{fleming_mass_2012,artacho_globally_2014,artacho_accelerating_2015,ahookhosh_local_2017}.}\tabularnewline
\hline 
\textbf{D} & {\small{}Metabolic engineering and strain design} & {\small{}New methods, e.g., OptForce, interpretation of new strain
designs. New modelling language interface to GAMS\cite{chowdhury_bilevel_2015}. }\tabularnewline
\hline 
\textbf{V} & {\small{}Human metabolic network visualisation: ReconMap} & {\small{}New method for genome-scale metabolic network visualisation\cite{noronha_reconmap:_2017,gawron_minervaplatform_2016,fujita_integrating_2014}.}\tabularnewline
\hline 
\textbf{V} & {\small{}Variable scope visualisation with automatic layout generation} & {\small{}New method for automatic visualisation of network parts\cite{kostromins_paint4net:_2012}.}\tabularnewline
\hline 
\multirow{2}{*}{} & {\small{}Contributing to the COBRA Toolbox with MATLAB.devTools} & {\small{}New software application enabling contributions by those
unfamiliar with version control software.}\tabularnewline
\cline{2-3} 
 & {\small{}Engaging with the COBRA Toolbox Forum} & {\small{}More than 800 posted questions with supportive replies connecting
problems and solutions. }\tabularnewline
\hline 
\end{tabular}
\par\end{raggedright}
\raggedright{}\caption{\label{tab:featureTable}Each method available in the COBRA Toolbox
3.0 is made accessible with a narrative tutorial that illustrates
how the corresponding function(s) are combined to implement each COBRA
method in the respective \emph{src} directories (\protect\url{https://github.com/opencobra/cobratoolbox/tree/master/src}):
\emph{analysis} (\textbf{A}), \emph{reconstruction} (\textbf{R}),
\emph{design} (\textbf{D}), \emph{visualisation} (\textbf{V}), \emph{dataIntegration}
(\textbf{I}), and \emph{base} (\textbf{B}).}
\end{table}

\newpage{}

Another prominent biotechnological application is the analysis of
interactions between organisms that form biological communities and
their surrounding environments, with a view toward utilisation of
such communities for bioremediation\cite{zhuang_genome-scale_2011}
or nutritional support of non-food plants for bioenergy feedstocks.
Biomedical applications of constraint-based modelling include the
prediction of the phenotypic consequences of single nucleotide polymorphisms\cite{jamshidi_systems_2006},
drug targets\cite{yizhak_model-based_2013}, enzyme deficiencies
\cite{shlomi_predicting_2009,sahoo_compendium_2012,thiele_community-driven_2013,pagliarini_genome-scale_2013},
as well as side and off-target effects of drugs\cite{shaked_metabolic_2016,chang_drug_2010,kell_systems_2006}.
COBRA has also been applied to generate and analyse normal and diseased
models of human metabolism\cite{duarte_global_2007,thiele_community-driven_2013,swainston_recon_2016,pornputtapong_human_2015,zielinski_systems_2017},
including organ-specific models\cite{mardinoglu_genome-scale_2014,karlstadt_cardionet:_2012,gille_hepatonet1:_2010},
multi-organ-models\cite{conde_constraint_2016,bordbar_multi-tissue_2011},
and personalised models\cite{yizhak_phenotype-based_2014,mardinoglu_integration_2013,bordbar_personalized_2015}.
Constraint-based modelling has also been applied to understanding
of the biochemical pathways that interlink diet, gut microbial composition,
and human health\cite{shoaie_quantifying_2015,nogiec_supplement_2013,heinken_systems-level_2013,heinken_functional_2014,magnusdottir_generation_2017}.

Today, genome-scale metabolic reconstructions can be automatically
created using many different platforms (see the Key features and comparisons
section). Automatic reconstruction tools have considerably sped up
the reconstruction process, which was previously very time-consuming\cite{thiele_protocol_2010}.
Various algorithms exist that facilitate gap filling of metabolic
networks, e.g., SMILEY\cite{reed_systems_2006}, GapFind\cite{satish_kumar_optimization_2007},
fastGapFill\cite{thiele_fastgapfill:_2014}, SONEC\cite{biggs_metabolic_2016},
and EnsemblFBA\cite{biggs_managing_2017}. Each algorithm applies
a different approach and thus the gap filling solutions proposed by
different algorithms may differ. However, several issues still need
to be manually resolved in automatically generated and gap-filled
reconstructions, e.g., ensuring stoichiometric consistency\cite{fleming_conditions_2016},
thermodynamically constraining reaction directionality\cite{fleming_quantitative_2009},
refining gene-protein-reaction associations, and confirming experimentally
determined biochemical functions of the reconstructed organism\cite{thiele_protocol_2010}.

\subsection{\label{subsec:Key-features-and}Key features and comparisons}

\begin{table}[h]
\begin{centering}
\begin{tabular}{l|l|l|l|l|>{\raggedright}p{1.2cm}}
\hline 
\textbf{Name} & \textbf{Implementation} & \textbf{Interface} & \textbf{Development} & \textbf{Dist.} & \textbf{OS}\tabularnewline
\hline 
The COBRA Toolbox & MATLAB (etc) & Script/Narrative & open source$^{\star}$ & git & all\tabularnewline
\hline 
RAVEN\cite{agren_raven_2013} & MATLAB  & Script & open source$^{\star}$ & git & all\tabularnewline
\hline 
CellNetAnalyzer\cite{klamt_structural_2007} & MATLAB (etc) & Script/GUI & closed source$^{\star}$ & zip & all\tabularnewline
\hline 
FBA-SimVis\cite{grafahrend-belau_fba-simvis:_2009} & Java + MATLAB & GUI & closed source$^{\dagger}$ & zip & Windows\tabularnewline
\hline 
OptFlux\cite{rocha_optflux:_2010} & Java & Script & open source$^{\star}$ & svn & all\tabularnewline
\hline 
COBRA.jl\cite{heirendt_distributedfba.jl:_2017} & Julia & Script/Narrative & open source$^{\star}$ & git & all\tabularnewline
\hline 
Sybil\cite{gelius-dietrich_sybil_2013} & R package & Script & open source$^{\star}$ & zip & all\tabularnewline
\hline 
COBRApy\cite{ebrahim_cobrapy:_2013} & Python packages & Script/Narrative & open source$^{\star}$ & git & all\tabularnewline
\hline 
CBMPy\cite{olivier_modelling_2005} & Python packages & Script & open source$^{\star}$ & zip & all\tabularnewline
\hline 
Scrumpy\cite{poolman_scrumpy:_2006} & Python packages & Script & open source$^{\star}$ & tar & all\tabularnewline
\hline 
SurreyFBA\cite{gevorgyan_surreyfba:_2011} & C++ & Script/GUI & open source$^{\star}$ & zip & all\tabularnewline
\hline 
FASIMU\cite{hoppe_fasimu:_2011} & C & Script & open source$^{\dagger}$ & zip & Linux\tabularnewline
\hline 
FAME\cite{boele_fame_2012} & Web-based & GUI & open source$^{\dagger}$ & zip & all\tabularnewline
\hline 
PathwayTools\cite{latendresse_construction_2012} & Web-based & GUI/Script & closed source$^{\star}$ & N/A & all\tabularnewline
\hline 
KBase\cite{arkin_doe_2016} & Web-based & Script/Narrative & open source$^{\star}$ & git & all\tabularnewline
\hline 
\end{tabular}
\par\end{centering}
\caption{\label{tab:cobraToolx}A selection of actively developed software
applications with constraint-based modelling capabilities. GUI, graphical
user interface. The COBRA Toolbox: \protect\url{https://opencobra.github.io/cobratoolbox},
RAVEN: \protect\url{https://github.com/SysBioChalmers/RAVEN}, CellNetAnalyzer:
\protect\url{https://www2.mpi-magdeburg.mpg.de/projects/cna/cna.html},
FBA-SimVis: \protect\url{https://immersive-analytics.infotech.monash.edu/fbasimvis},
OptFlux: \protect\url{ http://www.optflux.org}, COBRA.jl: \protect\url{https://opencobra.github.io/COBRA.jl},
Sybil: \protect\url{https://rdrr.io/cran/sybil}, COBRApy: \protect\url{http://opencobra.github.io/cobrapy},
CBMPy: \protect\url{http://cbmpy.sourceforge.net}, SurreyFBA: \protect\url{http://sysbio.sbs.surrey.ac.uk/sfba},
FASIMU: \protect\url{http://www.bioinformatics.org/fasimu}, FAME:
\protect\url{http://f-a-m-e.org}, Pathway Tools: \protect\url{http://bioinformatics.ai.sri.com/ptools},
KBase: \protect\url{https://kbase.us}. The symbol $^{\dagger}$ in
the development column refers to an inactive project and the $^{\star}$
to an active project. The label 'all' in the OS column means that
the applications is compatible with Windows, Linux and macOS operating
systems.}
\end{table}
Besides the COBRA Toolbox, constraint-based reconstruction and analysis
can be carried out with a variety of software tools. In 2012, Lakshmanan
et al.\cite{lakshmanan_software_2014} made a comprehensive, comparative
evaluation of the usability, functionality, graphical representation
and inter-operability of the tools available for flux balance analysis.
Each of these evaluation criteria is still valid when comparing the
current version of the COBRA Toolbox with other software with constraint-based
modelling capabilities. The rapid development of novel constraint-based
modelling algorithms requires continuity of software development.
Short term investment in new COBRA modelling software applications
has led to a plethora of COBRA modelling applications\cite{lakshmanan_software_2014}.
Each usually provides some unique capability initially, but many have
become antiquated due to lack of maintenance, failure to upgrade,
or failure to support new standards in model exchange formats (\url{http://sbml.org/Documents/Specifications}).
Therefore, we also restrict our comparison to software in active development
(see Table \ref{tab:cobraToolx}). 

Each software tool for constraint-based modelling has varying degrees
of dependency on other software. Web-based applications exist for
the implementation of a limited number of standard constraint-based
modelling methods. Their only local dependency is on a web browser.
The COBRA Toolbox depends on MATLAB (Mathworks Inc.), a commercially
distributed, general-purpose computational tool. MATLAB is a multi-paradigm
programming language and numerical computing environment that allows
matrix manipulations, plotting of functions and data, implementation
of algorithms, creation of user interfaces, and interfacing with programs
written in other languages, including C, C++, C\#, Java, Fortran,
and Python. All software tools for constraint-based modelling also
depend on at least one numerical optimisation solver. The most robust
and efficient numerical optimisation solvers for standard problems
are distributed commercially, but often with free licences available
for academic use, e.g., Gurobi Optimizer (\url{http://www.gurobi.com}).
Stand-alone constraint-based modelling software tools also exist and
their dependency on a numerical optimisation solver is typically satisfied
by GLPK (\url{https://gnu.org/software/glpk}), an open-source linear
optimisation solver. 

It is sometimes perceived that there is a commercial advantage to
depending only on open-source software. However, there are also commercial
costs associated with dependency on open-source software. That is,
in the form of increased computation times as well as increased time
required to install, maintain and upgrade open-source software dependencies.
This is an important consideration for any research group whose primary
focus is on biological, biomedical, or biotechnological applications,
rather than on software development. The COBRA Toolbox 3.0 strikes
a balance by depending on closed-source, general purpose, commercial
computational tools, yet all COBRA code is distributed and developed
in an open-source environment (\url{https://github.com/opencobra/cobratoolbox}).

The availability of comprehensive documentation is an important feature
in the usability of any modelling software. Therefore, a dedicated
effort has been made to ensure that all functions in the COBRA Toolbox
3.0 are comprehensively and consistently documented. Moreover, we
also provide a new suite of more than 35 tutorials (\url{https://opencobra.github.io/cobratoolbox/latest/tutorials})
to enable beginners, as well as intermediate and advanced users to
practise a wide variety of COBRA methods. Each tutorial is presented
in a variety of formats, including as a MATLAB live script, which
is an interactive document, or \emph{narrative}, (\url{https://mathworks.com/help/matlab/matlab_prog/what-is-a-live-script.html})
that combines MATLAB code with embedded output, formatted text, equations,
and images in a single environment viewable with the MATLAB Live Editor
(version R2016a or later). MATLAB live scripts are similar in functionality
to Mathematica Notebooks (Wolfram Inc.) and Jupyter Notebooks (\url{https://jupyter.org}).
The latter support interactive data science and scientific computing
for more than 40 programming languages. To date, only the COBRA Toolbox
3.0, COBRApy\cite{ebrahim_cobrapy:_2013}, KBase\cite{arkin_doe_2016},
and COBRA.jl\cite{heirendt_distributedfba.jl:_2017} offer access
to constraint-based modelling algorithms via narratives.

KBase is a collaborative, open environment for systems biology of
plants, microbes and their communities\cite{arkin_doe_2016}. It also
has a suite of analysis tools and data that support the reconstruction,
prediction, and design of metabolic models in microbes and plants.
These tools are tailored toward the optimisation of microbial biofuel
production, the identification of minimal media conditions under which
that fuel is generated, and predict soil amendments that improve the
productivity of plant bioenergy feedstocks. In our view, KBase is
currently the tool of choice for the automatic generation of draft
microbial metabolic networks, which can then be imported into the
COBRA Toolbox for further semi-automated refinement, which has recently
successfully been completed for a suite of gut microbial organisms\cite{magnusdottir_generation_2017}.
However, KBase\cite{arkin_doe_2016} currently offers a modest depth
of constraint-based modelling algorithms.

MetaFlux\cite{latendresse_construction_2012} is a web-based tool
for the generation of network reconstructions directly from pathway
and genome databases, proposing network refinements to generate functional
flux balance models from reconstructions, predict steady-state reaction
rates with flux balance analysis and interpret predictions in a graphical
network visualisation. MetaFlux is tightly integrated within the PathwayTools\cite{karp_pathway_2016}
environment, which provides a broad selection of genome, metabolic
and regulatory informatics tools. As such, PathwayTools provides breadth
in bioinformatics and computational biology, while the COBRA Toolbox
3.0 provides depth in constraint-based modelling, without providing,
for example, any genome informatics tools. Although an expert can
locally install a PathwayTools environment, the functionality is closed
source and only accessible via an application programming interface.
This approach does not permit the level of repurposing possible with
open-source software. As recognised in the computational biology community
\cite{sandve_ten_2013}, open-source development and distribution
is scientifically important for tractable reproducibility of results
as well as reuse and repurposing of code \cite{ince_case_2012}. 

Lakshmanan et al.\cite{lakshmanan_software_2014} consider the availability
of a graphical user interface to be an important feature in the usability
of modelling software. For example, SurreyFBA\cite{gevorgyan_surreyfba:_2011}
provides a command line tool and graphical user interface for constraint-based
modelling of genome-scale metabolic reaction networks. The time lag
between the development of a new modelling method and its availability
via a graphical user interface necessarily means that graphically
driven COBRA tools permit a limited depth of novel constraint-based
modelling methods. While MATLAB provides a generic graphical user
interface, the COBRA Toolbox is controlled either by scripts or narratives,
rather than graphically. Exceptions include the input of manually-curated
data during network reconstruction\cite{thorleifsson_rbionet:_2011},
the assimilation of genome-scale metabolic reconstructions\cite{sauls_assimilating_2014},
and the visualisation of simulation results in biochemical network
maps\cite{noronha_reconmap:_2017} via specialised network visualisation
software\cite{gawron_minervaplatform_2016}. 

Due to the relative simplicity of the MATLAB programming language,
new COBRA Toolbox users, including those without software development
experience, can rapidly become familiar with the basics of constraint-based
modelling. This initial learning effort is worth it for the flexibility
it opens up, especially considering the broad array of constraint-based
modelling methods now available within the COBRA Toolbox 3.0. Although
it should be technically possible to generate a computational specification
of the point-and-click analysis steps that are required to generate
results using a graphical user interface, to our knowledge, none of
the graphically-driven modelling tools in Table \ref{tab:cobraToolx}
offers this facility. Such a specification would be required for another
scientist to reproduce the same results using the same tool. This
weakness limits the ability to reproduce analytical results, as verbal
specification is not sufficient for reproducibility\cite{ince_case_2012}.

Each language-specific COBRA implementation has its benefits and drawbacks,
which are mainly associated with the programming language itself.
PySCeS-CBM\cite{olivier_modelling_2005} and COBRApy\cite{ebrahim_cobrapy:_2013}
both provide support for a set of COBRA methods implemented in the
Python programming language. Python is a multi-paradigm, interpreted
programming language for general-purpose programming. It has a broad
development community and a wide range of open-source libraries, especially
in bioinformatics. As such, it is well suited for the amalgamation
and management of heterogeneous experimental data. At present, the
COBRA software tools in Python provide access to standard COBRA methods.
In COBRApy\cite{ebrahim_cobrapy:_2013}, this functionality can be
extended by using Python to invoke MATLAB and use the COBRA Toolbox.
Achieving such interoperability between COBRA software implemented
in different programming languages and developed together by a united
open source community is the primary objective of the openCOBRA project
(\url{https://opencobra.github.io}). 

Sybil\cite{gelius-dietrich_sybil_2013} is an open-source, object-oriented
software library that implements a limited set of standard constraint-based
modelling algorithms in the programming language R, which is a free,
platform independent environment for statistical computing and graphics.
Sybil is available for download from the comprehensive R archive network
(CRAN), but does not follow an open-source development model. The
COBRA Toolbox is primarily implemented in MATLAB, a proprietary, multi-paradigm,
programming language which is interpreted for execution rather than
compiled prior to execution. As such, MATLAB code typically runs slower
than compiled code, but the main advantage is the ability to rapidly
and flexibly implement sophisticated numerical computations by leveraging
the extensive libraries for general-purpose numerical computing, supplied
commercially within MATLAB (Mathworks Inc.), and distributed freely
by the community (\url{https://mathworks.com/matlabcentral}).

For the application of computationally-demanding constraint-based
modelling methods to high-dimensional or high-precision constraint-based
models, the COBRA Toolbox 3.0 comes with an array of integrated, pre-compiled
extensions and interfaces that employ complementary programming languages
and tools. These include a quadruple precision Fortran 77 optimisation
solver implementation for constraint-based modelling of multi-scale
biochemical networks\cite{ma_reliable_2017}, and a high-level, high-performance,
open-source implementation of flux balance analysis in Julia \cite{heirendt_distributedfba.jl:_2017}.
The latter is tailored to solve multiple flux balance analyses on
a subset or all the reactions of large- and huge-scale networks, on
any number of threads or nodes. To enumerate elementary modes or minimal
cut-sets, we provide an interface to CellNetAnalyzer\cite{klamt_structural_2007,klamt_application_2011}
(\url{https://www2.mpi-magdeburg.mpg.de/projects/cna/cna.html}),
which excels at computationally-demanding, enumerative, discrete geometry
calculations of relevance to biochemical networks. In addition, we
included an updated implementation of the genetic minimal cut-sets
approach \cite{apaolaza_-silico_2017}, which extends the concept
of minimal cut-sets to gene knockout interventions.

In summary, the COBRA Toolbox 3.0 provides an unparalleled depth of
constraint-based reconstruction and analysis methods, has a highly
active and supportive open-source development community, is accompanied
by extensive documentation and narrative tutorials, it leverages the
most comprehensive library for numerical computing, and it is distributed
with extensive interoperability with a range of complementary programming
languages that exploit their particular strengths to realise specialised
constraint-based modelling methods. A list of the main COBRA methods
now available in the COBRA Toolbox is given in Table \ref{tab:featureTable}.
Moreover, all of this functionality is provided within one accessible
software environment.

\subsection{Experimental Design}

The COBRA Toolbox 3.0 is designed for flexible adaptation into customised
pipelines for constraint-based reconstruction and analysis in a wide
range of biological, biochemical, or biotechnological scenarios, from
single organisms to communities of organisms. To become proficient
in adapting the COBRA Toolbox to generate a protocol specific to one's
situation, it is wise to first familiarise oneself with the principles
of constraint-based modelling. This can best be achieved by studying
the educational material already available. The textbook \emph{Systems
Biology: Constraint-based Reconstruction and Analysis\cite{palsson_systems_2015}}
is an ideal place to start. It is accompanied by a set of lecture
videos that accompany each chapter \url{http://systemsbiology.ucsd.edu/Publications/Books/SB1-2LectureSlides}.
The textbook \emph{Optimization Methods in Metabolic Networks}\cite{maranas_optimization_2016}
provides the fundamentals of mathematical optimisation and its application
in the context of metabolic network analysis. A study of this educational
material will accelerate one's ability to utilise any software application
dedicated to COBRA. 

Once one is cognisant of the conceptual basis of COBRA, one can then
proceed with this protocol, which summarises a subset of the key methods
that are available within the COBRA Toolbox. To adapt this protocol
to one's situation, users can combine the COBRA methods implemented
within the COBRA Toolbox in numerous ways. The adaption of this protocol
to one's situation may require the development of new customised MATLAB
scripts that combine existing methods in a new way. Due to the aforementioned
benefits of narratives, the first choice should be to implement these
customised scripts in the form of MATLAB live scripts. To get started,
the existing tutorial narratives, described in Table \ref{tab:featureTable},
can be repurposed as templates for new analysis pipelines. Narrative
figures and tables can then be used within the main text of scientific
articles and converted into supplementary material to enable full
reproducibility of computational results. The narratives specific
to individual scientific articles can also be shared with peers within
\url{https://github.com/opencobra/cobratoolbox/tree/master/papers}.
New tutorials can be shared with the COBRA community by contributing
to the future development of the COBRA Toolbox (cf. the Software architecture
section). Depending on one's level of experience, or the novelty of
an analysis, the adaptation of this protocol to a particular situation
may require the adaption of existing COBRA methods, or development
of new COBRA methods, or both. 
\begin{table}
\begin{centering}
\begin{tabular}{>{\raggedright}p{3.7cm}|l|>{\raggedright}p{2cm}|>{\raggedright}p{7.4cm}}
\hline 
\textbf{Field name} & \textbf{Size} & \textbf{Data Type} & \textbf{Field description}\tabularnewline
\hline 
\mcodeENUM{.b}  & $m\times1$ & double & The coefficients of the constraints of the metabolites ($Sv=b$).\tabularnewline
\hline 
\mcodeENUM{.csense}  & $m\times1$ & char & The sense of the constraints represented by $b$, each row is either
\mcodeENUM{'E'} (equality), \mcodeENUM{'L'} (less than) or \mcodeENUM{'G'}
(greater than).\tabularnewline
\hline 
\mcodeENUM{.metCharges} & $m\times1$ & numeric & The charge of the respective metabolite (\mcodeENUM{NaN} if unknown).\tabularnewline
\hline 
\mcodeENUM{.metFormulas} & $m\times1$ & cell of char & Elemental formula for each metabolite.\tabularnewline
\hline 
\mcodeENUM{.metInChIString} & $m\times1$ & cell of char & Formula for each metabolite in the InCHI strings format.\tabularnewline
\hline 
\mcodeENUM{.metNames} & $m\times1$ & cell of char & Full name of each corresponding metabolite.\tabularnewline
\hline 
\mcodeENUM{.mets}  & $m\times1$ & cell of char & Identifiers of the metabolites.\tabularnewline
\hline 
\mcodeENUM{.metSmiles} & $m\times1$ & cell of char & Formula for each metabolite in SMILES Format.\tabularnewline
\hline 
\mcodeENUM{.c}  & $n\times1$ & double & The objective coefficient of the reactions.\tabularnewline
\hline 
\mcodeENUM{.grRules} & $n\times1$ & cell of char & A string representation of the GPR rules defined in a readable format.\tabularnewline
\hline 
\mcodeENUM{.lb}  & $n\times1$ & double & Lower bounds for fluxes through the reactions.\tabularnewline
\hline 
\mcodeENUM{.rxnConfidenceScores}  & $n\times1$ & numeric & Confidence scores for reaction presence (0-5, with 5 being the highest
confidence).\tabularnewline
\hline 
\mcodeENUM{.rxnECNumbers} & $n\times1$ & cell of char & E.C. number for each reaction.\tabularnewline
\hline 
\mcodeENUM{.rxnNames}  & $n\times1$ & cell of char & Full name of each corresponding reaction.\tabularnewline
\hline 
\mcodeENUM{.rxnNotes} & $n\times1$ & cell of char & Description of each corresponding reaction.\tabularnewline
\hline 
\mcodeENUM{.rxnReferences} & $n\times1$ & cell of char & Description of references for each corresponding reaction.\tabularnewline
\hline 
\mcodeENUM{.rxns} & $n\times1$ & cell & Identifiers of the reactions.\tabularnewline
\hline 
\mcodeENUM{.subSystems} & $n\times1$ & cell of cell of char & subSystem assignments for each reaction.\tabularnewline
\hline 
\mcodeENUM{.ub}  & $n\times1$ & double & Upper bounds for fluxes through the reactions.\tabularnewline
\hline 
\mcodeENUM{.S}  & $m\times n$ & numeric & The stoichiometric matrix containing the model structure (for large
models a sparse format is suggested).\tabularnewline
\hline 
\mcodeENUM{.geneNames} & $g\times1$ & cell of char & Full name of each corresponding gene.\tabularnewline
\hline 
\mcodeENUM{.genes} & $g\times1$ & cell of char & Identifiers of the genes in the model.\tabularnewline
\hline 
\mcodeENUM{.proteinNames} & $g\times1$ & cell of char & Full name for each protein.\tabularnewline
\hline 
\mcodeENUM{.proteins} & $g\times1$ & cell of char & Proteins associated with each gene (one protein per gene).\tabularnewline
\hline 
\mcodeENUM{.rxnGeneMat}  & $n\times g$ & numeric or logical & Matrix with rows corresponding to reactions and columns corresponding
to genes.\tabularnewline
\hline 
\mcodeENUM{.compNames} & $c\times1$ & cell of char & Descriptions of the Compartments (\mcodeENUM{compNames(m)} is associated
with \mcodeENUM{comps(m)}).\tabularnewline
\hline 
\mcodeENUM{.comps} & $c\times1$ & cell of char & Symbols for compartments.\tabularnewline
\hline 
\mcodeENUM{.osenseStr}  & $1\times3$ & char & The objective sense: either \mcodeENUM{'max'} (maximisation) or \mcodeENUM{'min'}
(minimisation).\tabularnewline
\hline 
\end{tabular}
\par\end{centering}
\caption{\label{tab:Fields,-dimensions,-data}A description of the main fields
of a standard model structure.}
\end{table}

\subsubsection{\label{subsec:Software-architecture-of}Software architecture of
the COBRA Toolbox 3.0}

\paragraph{}

The source code of the COBRA Toolbox (\url{https://github.com/opencobra/cobratoolbox/tree/master/src})
is divided into several top-level folders, which either mimic the
main classes of COBRA methods (\emph{reconstruction}, \emph{dataIntegration},
\emph{analysis}, \emph{visualisation}, \emph{design}) or contain the
basic functions (\emph{base}) available for use within many COBRA
methods. For example, the input or output of reconstructions and models
in various formats as well as all the interfaces to optimisation solvers
is contained within the \emph{base} folder. The \emph{reconstruction}
folder contains all of the methods associated with the reconstruction
and refinement of a biochemical network to match experimental data,
as well as the conversion of a reconstruction into various forms of
constraint-based models (see Table \ref{tab:Fields,-dimensions,-data}
for a description of the main fields of a COBRA model). The \emph{dataIntegration}
folder contains the methods for integration of metabolomic, transcriptomic,
proteomic, and thermodynamic data with a reconstruction or model.
The \emph{analysis} folder contains all of the methods for interrogation
of the properties of a reconstruction or model, and combinations thereof,
as well as the prediction of biochemical network states using constraint-based
models. The \emph{visualisation} folder contains all of the methods
for the visualisation of predictions within a biochemical network
context, using various biochemical cartography tools that interoperate
with the COBRA Toolbox. The \emph{design} folder contains {\small{}new
strain design methods and a new modelling language interface to GAMS.}

\subsubsection{\label{subsec:Open-source-software}Open-source software development
with the COBRA Toolbox}

Understanding how the COBRA Toolbox is developed is most important
for developers, so beginners may skip this section at first. The COBRA
Toolbox is version controlled using Git (\url{https://git-scm.com}),
a free and open-source distributed, version control system, which
tracks changes in computer files and is used for coordinating work
on those files by multiple people. The continuous integration environment
facilitates contributions from the fork of COBRA developers to a development
branch, whilst ensuring that robust, high-quality, well-tested code
is released to end users on the \emph{master} branch. To lower the
technological barrier to the use of the aforementioned software development
tools, we have developed MATLAB.devTools (\url{https://github.com/opencobra/MATLAB.devTools}),
a new user-friendly software extension that enables submission of
new COBRA software and tutorials. A server-side, semi-automated continuous
integration environment ensures that the code in each new submission
is first verified automatically, via a comprehensive test suite that
detects bugs and integration errors, and second, is reviewed manually
by at least one domain expert, before integration with the development
branch. Thirdly, each new contribution to the \textit{development}
branch is evaluated in practice by active COBRA researchers, before
it becomes part of the \textit{master} branch. 

\begin{figure}
\begin{centering}
\includegraphics[scale=0.15]{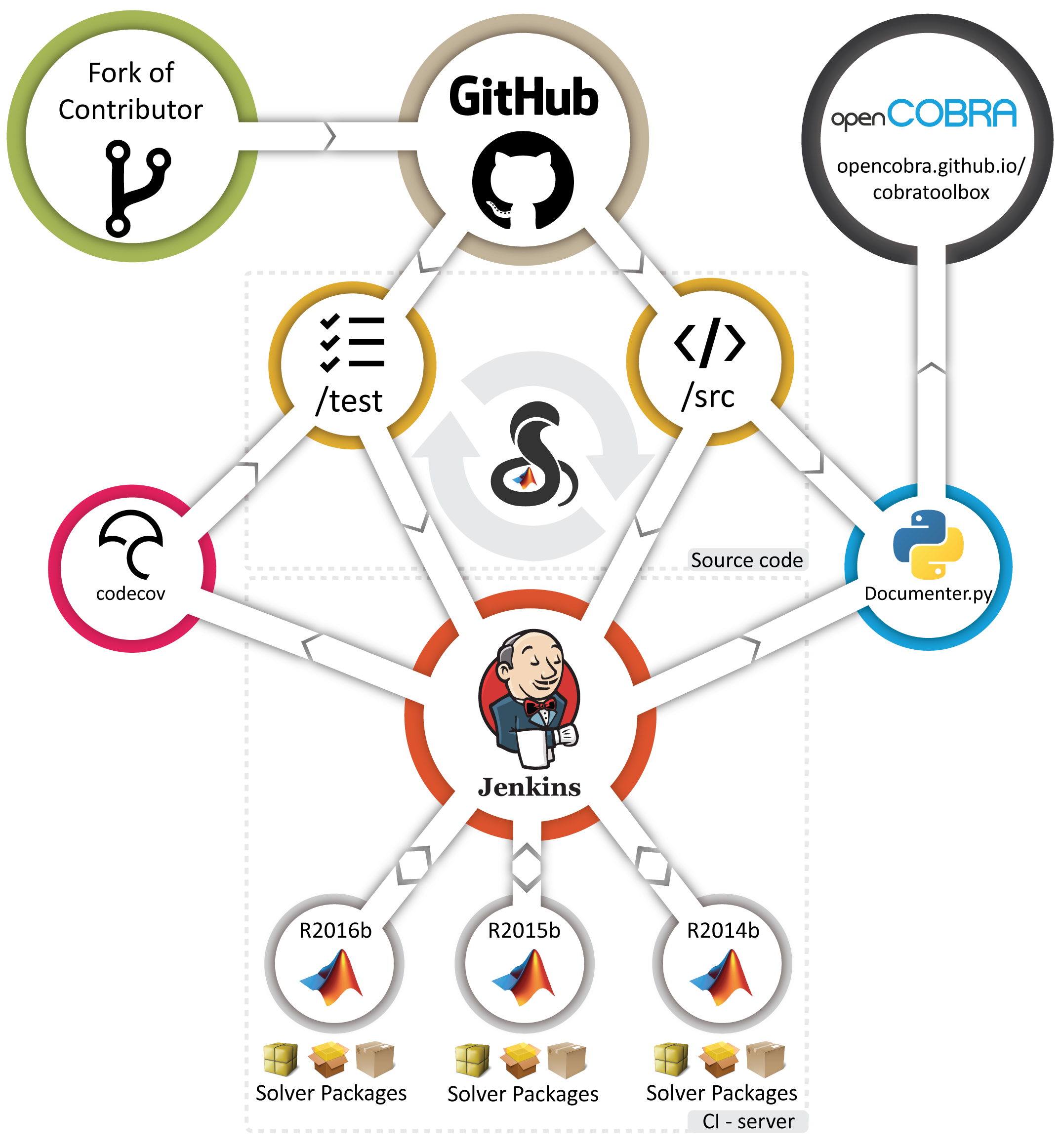}
\par\end{centering}
\caption{\label{fig:Continuous-integration-setup.}Continuous integration is
performed on a dedicated server running Jenkins (\protect\url{https://jenkins.io}).
The main code is located in the \emph{src} folder and tests functions
in the \emph{test} folder. A test not only runs a function (first
degree testing), but tests the output of that function (second degree
testing). The continuous integration setup relies on end-of-year releases
of MATLAB only. Soon after the latest stable version of MATLAB is
released, full support will be provided for the COBRA Toolbox. After
a successful run of tests on the three latest end-of-year releases
of MATLAB using various solver packages, the documentation based on
the headers of the functions (docstrings) is extracted, generated,
and automatically deployed. Immediate feedback through code coverage
reports (\protect\url{https://codecov.io/gh/opencobra/cobratoolbox})
and build statuses are reported on GitHub. With this setup, the impact
of local changes in the code base is promptly revealed. }
\end{figure}

Until recently, the code quality checks of the COBRA Toolbox have
been primarily \emph{static}: the code has been reviewed by experienced
users and developers while occasional code inspections led to discoveries
of bugs. The continuous integration setup defined in Figure \ref{fig:Continuous-integration-setup.}
aims at \emph{dynamic} testing with automated builds, code evaluation,
and documentation deployment. Often, a function runs properly independently
and yields the desired output(s), but when called within a different
part of the code, logical errors are thrown. The unique advantage
of continuous integration is that logical errors are mostly avoided.

Besides automatic testing, manual usability testing is performed regularly
by users and is key to provide a tested and usable code base to the
end user. These users provide feedback on the usability of the code
base, as well as the documentation, and report eventual issues online
(\url{https://github.com/opencobra/cobratoolbox/issues}). The documentation
is automatically deployed to \url{https://opencobra.github.io/cobratoolbox}
based on function headers. Moreover, each of the narrative tutorials
is presented in a format suitable for web browsers (\url{https://opencobra.github.io/cobratoolbox/latest/tutorials}). 

\subsubsection{Controls}

COBRA is part of an iterative systems biology cycle\cite{palsson_systems_2015}.
As such, it can be used as a framework for integrative analysis of
experimental data in the context of prior information on the biochemical
network underlying one or many complementary experimental datasets.
Moreover, it can be used to predict the outcome of new experiments,
or it can be used in both of these scenarios at once. Assuming all
of the computational steps are errorless, the appropriate control
for any prediction derived from a computational model is the comparison
with independent experimental data, that is, experimental data that
was not used for the model-generated predictions. It is also important
to introduce quality controls to check that the computational steps
are free from certain errors that may arise during adaptation of existing
COBRA protocols or development of new ones. 

There are various strategies for the implementation of computational
quality controls. Within the COBRA Toolbox 3.0, significant effort
has been devoted to automatically test the functionality of existing
COBRA methods. We have also embedded a large number of \emph{sanity
checks,} which evaluate whether the input data could possibly be appropriate
for use with a function. These sanity checks have been accumulated
over more than a decade of continuous development of the COBRA Toolbox.
Their objective is to rule out certain known classes of obviously
false predictions that might result from an inappropriate use of a
COBRA method, but they do not (and are not intended) to catch every
such error, as it is impossible to imagine all of the eventual erroneous
inputs that may be presented to a COBRA Toolbox function. It is advisable
to add own narratives with additional sanity checks, which will depend
heavily on the modelling scenario. 

\subsection{Required expertise}

Most of this protocol can be implemented by anyone with a basic familiarity
with the principles of constraint-based modelling. Some methods are
only for advanced users. If one is a beginner with respect to MATLAB,
Supplementary Manual 1 provides pointers to get started. MATLAB is
a relatively simple programming language to learn, but it is also
a powerful language for an expert due to the large number of software
libraries for numerical and symbolic computing that it provides access
to. Certain specialised methods within this protocol, such as thermodynamically
constraining reaction directionality, depend on the installation of
other programming languages and software, which may be too challenging
for a beginner with a non-standard operating system. 

If the documentation and tutorials provided within the COBRA Toolbox
are not sufficient, then Steps \ref{subsec:Search-for-existing} and
\ref{subsec:Suggest-new-solutions} guide the user toward sources
of COBRA community support. The computational demands associated with
the implementation of this protocol for one's reconstruction or model
of choice is dependent on the size of the network concerned. For a
genome-scale model of metabolism, usually a desktop computer is sufficient.
However, for certain models, such as a community of genome-scale metabolic
models, a multi-scale model of metabolism and macromolecular synthesis,
or a multi-tissue model, more powerful processing and extensive memory
capacity is required, ranging from a workstation to a dedicated computational
cluster. Embarrassingly parallel, high-performance computing is feasible
for most model analysis methods implemented in the COBRA Toolbox,
which will run in isolation with invocation from a distributed computing
engine. It is currently an ongoing topic of research, beyond the scope
of this protocol, to fully exploit high-performance computing environments
with software developed within the wider openCOBRA environment, though
some examples\cite{heirendt_distributedfba.jl:_2017} are already
available for interested researchers to consult. 

\subsection{Limitations}

A protocol for the generation of a high-quality, genome-scale reconstruction,
using various software applications, including the COBRA Toolbox,
has previously been disseminated\cite{thiele_protocol_2010}; therefore,
this protocol focuses more on modelling than reconstruction. The COBRA
Toolbox is not meant to be a general-purpose computational biology
tool as it is focussed on constraint-based reconstruction and analysis.
For example, although various forms of generic data analysis methods
are available within MATLAB, the input data for integration with reconstructions
and models within the COBRA Toolbox is envisaged to have already been
preprocessed by other tools. Within its scope, the COBRA Toolbox aims
for complete coverage of COBRA methods. The first comprehensive overview
of the COBRA methods available for microbial metabolic networks\cite{lewis_constraining_2012}
requires an update to encompass many additional methods that have
been reported to date, in addition to the COBRA methods targeted toward
other biochemical networks. The COBRA Toolbox 3.0 provides the most
extensive coverage of published COBRA methods. However, there are
certainly some methods that have yet to be incorporated directly as
MATLAB implementations, or indirectly via a MATLAB interface to a
software dependency. Although in principle any COBRA method could
be implemented entirely within MATLAB, it may be more efficient to
leverage the core strength of another programming language that could
provide intermediate results that can be incorporated into the COBRA
Toolbox via various forms of MATLAB interfaces. Such a setup would
enable one to overcome any current limitation in coverage of existing
methods.

\subsection{Overview}

This protocol consists of a set of methods that are introduced in
sequence but can be combined in a multitude of ways. The overall purpose
is to enable the user to generate a biologically relevant, high-quality
model that enables novel predictions and hypotheses generation. Therefore,
we implement and enforce standards in reconstruction and simulation
that have been developed by the COBRA community over the past two
decades. All explanations of a method are also accompanied by explicit
computational commands.

First, we explain how to initialise and verify the installation of
the COBRA Toolbox in MATLAB (Mathworks, Inc.). The main options to
import and explore the content of a biochemical network reconstruction
are introduced. For completeness, a brief summary of methods for manual
and algorithmic reconstruction refinement are provided, with reference
to the established reconstruction protocol \cite{thiele_protocol_2010}.
We also explain how to characterise the numerical properties of a
reconstruction, especially with respect to detection of a reconstruction
requiring a multi-scale numerical optimisation solver. We explain
how to semi-automatically convert a reconstruction into a constraint-based
model suitable for flux balance analysis. This is followed by an extensive
explanation of how to carry out flux balance analysis and its variants.
The procedure to fill gaps in a reconstruction, due to missing reactions,
is also explained. 

We provide an overview of the main methods to integrate metabolomic,
transcriptomic, proteomic, and thermochemical data to generate context-specific,
constraint-based models. Various methods are explained for the addition
of biological constraints to a constraint-based model. We then explain
how to test the chemical and biochemical fidelity of the model. Now
that a high-quality model is generated, we explain how to interrogate
the discrete geometry of its stoichiometric subspaces, how to efficiently
measure the variability associated with the prediction of steady state
reaction rate using flux variability analysis, and how to uniformly
sample steady-state fluxes. We introduce various approaches for prospective
uses of a constraint-based model, such as strain and experimental
design. 

We explain how to atomically resolve a metabolic reconstruction by
connecting it with molecular species structures and how to use cheminformatic
algorithms for atom mapping and identification of conserved moieties.
Using molecular structures for each metabolite, and established thermochemical
data, we estimate the transformed Gibbs energy of each subcellular
compartment specific reaction in a model of human metabolism in order
to thermodynamically constrain reaction directionality and constrain
the set of feasible kinetic parameters. Sampled kinetic parameters
are then used for variational kinetic modelling, in an illustration
of the utility of recently published algorithms for genome-scale kinetic
modelling. We also explain how to visualise predicted phenotypic states
using a recently developed approach for metabolic network visualisation.
We conclude with an explanation of how to engage with the community
of COBRA developers, as well as contribute code to the COBRA Toolbox
with MATLAB.devTools, a newly developed piece of software for community
contribution of COBRA methods to the COBRA Toolbox.

\section{\label{sec:MATERIALS}MATERIALS}

\subsection{\label{subsec:EQUIPMENT}Equipment}

\setlist[enumerate]{leftmargin=*}

\subsubsection{\label{subsec:Hardware}Required hardware}
\begin{itemize}
\item A computer with any 64-bit Intel or AMD processor and at least 8 GB
of RAM. \caution

Depending on the size of the reconstruction or model, more processing
power and more memory may be needed, especially if it is also desired
to store the results of analysis procedures within the MATLAB workspace. 
\item A hard drive with free storage of at least 10 GB. 
\item \caution A working and stable internet connection is required during
installation and while contributing to the COBRA Toolbox.
\end{itemize}

\subsubsection{\label{subsec:Required-Software}Required software}
\begin{itemize}
\item A \emph{Linux}, \emph{macOS} or \emph{Windows} operating system that
is MATLAB qualified (\url{https://mathworks.com/support/sysreq.html}).
\item MATLAB (MathWorks Inc. - \url{https://mathworks.com/products/matlab.html}),
version R2014b or above is required. Version R2016a or above is required
for running MATLAB live scripts (tutorials \emph{.mlx} files). Note
that the tutorials can be run on R2014b using the provided \emph{.m}
files. \caution\textbf{ }No support is provided for versions older
than R2014b. MATLAB is released on a twice-yearly schedule. After
the latest release (version \emph{b}), it may be a couple of months
before certain methods with dependencies on other software become
compatible. For example, the latest releases of MATLAB may not be
compatible with the existing solver interfaces, necessitating an update
of the MATLAB interface provided by the solver developers, or an update
of the COBRA Toolbox, or both.
\item The COBRA Toolbox (\url{https://github.com/opencobra/cobratoolbox})
version 3.0 or above. 
\item A working \emph{bash} terminal (or shell) with UNIX tools.\emph{ curl}
version 7.0 or above must be installed to ensure connectivity between
the COBRA Toolbox and the remote Github server. The version control
software \emph{git} 1.8 or above is required to be installed and accessible
through system commands. \caution\textbf{ }On \emph{Windows}, the
shell integration included with \emph{git Bash} (\url{https://git-for-windows.github.io})
utilities must be installed. On \emph{macOS}, a working installation
of Xcode (\url{https://developer.apple.com/xcode}) version 8.0 or
above and command line tools is mandatory. 
\end{itemize}

\subsubsection{Optional software\label{subsec:Optional-SoftwareEquip}}

The following third-party software and MATLAB toolboxes are only required
for one or more optional steps of the procedure. 
\begin{itemize}
\item Reading and writing models in SBML (Systems Biology Markup Language)
format requires the MATLAB interface from the libSBML application
programming interface, version 5.15.0 or above. The COBRA Toolbox
3.0 supports the latest SBML Level 3 Flux Balance Constraints Version
2 package (\url{http://sbml.org/Documents/Specifications/SBML_Level_3/Packages/fbc}).
The COBRA Toolbox developers work closely with the SBML Team to ensure
that the COBRA Toolbox supports the latest standards, and moreover
that standard development is also focused on meeting the evolving
requirements of the constraint-based modelling community. \caution\textbf{
}After the latest release of MATLAB, there may be a short time lag
before input and output become fully compatible. For example, the
input and output of \emph{.xml} files in the SBML standard formats
relies on platform dependent binaries that we maintain (\url{https://github.com/opencobra/COBRA.binary})
for each major platform, but the responsibility for maintenance of
the source code \cite{bornstein_libsbml:_2008} lies with the SBML
team (\url{http://sbml.org}), who have a specific forum for raising
interoperability issues (\url{https://groups.google.com/forum/#!forum/sbml-interoperability}). 
\item The MATLAB Image Processing Toolbox, the Parallel Computing Toolbox,
the Statistics and Machine Learning Toolbox, and the Optimization
Toolbox and Bioinformatics Toolbox (\url{https://mathworks.com/products})
must be licensed and installed to ensure certain model analysis functionality,
such as topology based algorithms, flux variability analysis, or sampling
algorithms.
\item The Chemaxon Calculator Plugins (\url{https://chemaxon.com/products/calculator-plugins}
- ChemAxon Ltd), version 16.9.5.0 or above, is a suite offering a
range of cheminformatics tools. Standardizer is ChemAxon\textquoteright s
solution to transform chemical structures into customised, canonical
representations to achieve best reliability with chemical databases.
A licence is freely available for academics.
\item Java (\url{https://java.com/en/download/help/download_options.xml}),
version 8 or above, is a programming language which enables platform
independent applications.
\item Python (\url{https://python.org/downloads}), version 2.7, is a high-level
programming language for general-purpose programming and is required
to run NumPy or generate the documentation locally (relevant when
contributing).
\item NumPy (\url{https://scipy.org/install.html}), version 1.11.1 or above,
is the fundamental package for scientific computing with Python.
\item OpenBabel (\url{https://openbabel.org}), version 2.3 or above, is
a chemical toolbox designed to speak the many languages of chemical
data.
\item Reaction Decoder Tool (RDT - \url{https://github.com/asad/ReactionDecoder/releases}),
version 1.5.0 or above, is a Java-based, open-source atom mapping
software tool. 
\end{itemize}

\subsubsection{Solvers}

Table \ref{tab:Optimisation-Solvers.} provides an overview of supported
optimisation solvers. For optimal performance, we recommend to install
an industrial-strength numerical optimisation solver. At least one
linear programming (LP) solver is required for basic constraint-based
modelling methods. By default, the COBRA Toolbox installs the free
LP and MILP solver GLPK (\url{https://gnu.org/software/glpk}) as
well as DQQ, MINOS, PDCO, and QPNG. On Windows, the OPTI solver suite
(\url{https://inverseproblem.co.nz/OPTI}) must be installed separately
in order to use the OPTI interface. \caution\textbf{ }Depending on
the type of optimisation problem underlying a COBRA method, an additional
numerical optimisation solver may be required.

\begin{table}
\begin{centering}
\begin{tabular}{c|ll|l|c|c|c|c|c}
\hline 
 & \textbf{Name} & \textbf{Version} & \textbf{Interface} & \textbf{LP } & \textbf{MILP } & \textbf{QP } & \textbf{MIQP } & \textbf{NLP}\tabularnewline
\hline 
\multirow{10}{*}{\begin{turn}{90}
Active Support
\end{turn}} & DQQ & - & dqqMinos & $\star$ &  &  &  & \tabularnewline
 & GLPK & 2.7+ & glpk & $\star$ & $\star$ &  &  & \tabularnewline
 & GUROBI & 7.0+ & gurobi & $\star$ & $\star$ & $\star$ & $\star$ & \tabularnewline
 & ILOG CPLEX & 12.7.1+ & ibm\_cplex & $\star$ & $\star$ & $\star$ &  & \tabularnewline
 & MATLAB & R2014b+ & matlab & $\star$ &  &  &  & $\star$\tabularnewline
 & MINOS & - & quadMinos & $\star$ &  &  &  & $\star$\tabularnewline
 & MOSEK & 8.0+ & mosek & $\star$ & $\star$ & $\star$ &  & \tabularnewline
 & PDCO & - & pdco & $\star$ &  & $\star$ &  & $\star$\tabularnewline
 & \multirow{2}{*}{Tomlab CPLEX} & \multirow{2}{*}{8.0+} & cplex\_direct & $\star$ & $\star$ & $\star$ & $\star$ & \tabularnewline
 &  &  & tomlab\_cplex & $\star$ & $\star$ & $\star$ & $\star$ & \tabularnewline
\hline 
\multirow{3}{*}{\begin{turn}{90}
Passive 
\end{turn}} & OPTI & 2.27+ & opti & $\star$ & $\star$ & $\star$ & $\star$ & $\star$\tabularnewline
 & QPNG & - & qpng &  &  & $\star$ &  & \tabularnewline
 & Tomlab SNOPT & 8.0+ & tomlab\_snopt &  &  &  &  & $\star$\tabularnewline
\hline 
\multirow{4}{*}{\begin{turn}{90}
Legacy
\end{turn}} & GUROBI & 7.0+ & gurobi\_mex & $\star$ & $\star$ & $\star$ & $\star$ & \tabularnewline
 & \multirow{2}{*}{LINDO} & \multirow{2}{*}{2.0+} & lindo\_old & $\star$ &  &  &  & \tabularnewline
 &  &  & lindo\_legacy & $\star$ &  &  &  & \tabularnewline
 & MATLAB & R2014b+ & lp\_solve & $\star$ &  &  &  & \tabularnewline
\hline 
\end{tabular}
\par\end{centering}
\caption{\textbf{\label{tab:Optimisation-Solvers.}} An overview of the types
of optimisation problems solved by each optimisation solver. The interface
to certain standard optimisation solvers is actively supported, whereas
the interface to other non-standard solvers requires testing by the
end user to ensure compatibility, while a legacy solver interface
might require refinement before it becomes compatible with newer solver
or MATLAB releases.}
\end{table}

\subsubsection{Application specific software}

Certain solvers have additional software requirements, and some binaries
provided in the COBRA.binary (\url{https://github.com/opencobra/COBRA.binary})
repository might not be compatible with your system. 
\begin{itemize}
\item The \emph{dqqMinos} and \emph{Minos} solvers may only be used on Unix.
The C-shell \emph{csh} (\url{http://bxr.su/NetBSD/bin/csh}) is required.
\item The GNU C-compiler \emph{gcc} 7.0 or above (\url{https://gcc.gnu.org}).
The library of the gcc compiler is required for generating new binaries
of \emph{fastFVA} with a different version of the CPLEX solver than
officially supplied. The GNU Fortran compiler \emph{gfortran} 4.1
or above (\url{https://gcc.gnu.org/fortran}). The library of the
\emph{gfortran} compiler is required for running \emph{dqqMinos.}
\end{itemize}

\subsubsection{Contributing software}
\begin{itemize}
\item MATLAB.devTools (\url{https://github.com/opencobra/MATLAB.devTools})
is highly recommended for contributing code to the COBRA Toolbox in
a user-friendly and convenient way, even for those without basic knowledge
of \emph{git}. 
\end{itemize}

\subsection{\label{subsec:EQUIPMENT-SETUP}Equipment setup}

\subsubsection{Required software }

Here we describe specific installation instructions for software described
in previous sections.
\begin{itemize}
\item \caution Make sure that the operating system is compatible with the
MATLAB version by checking the requirements on \url{https://mathworks.com/support/sysreq/previous_releases.html}.
Follow the upgrade and installation procedures on the supplier's website
or ask your system administrator for help if required.
\item Install MATLAB and its licence by following the official installation
instructions (\url{https://mathworks.com/help/install/ug/install-mathworks-software.html})
or ask your system administrator.
\item Install the COBRA Toolbox by following the procedures given on \url{https://github.com/opencobra/cobratoolbox}.
\caution Make sure that all system requirements outlined under \url{https://opencobra.github.io/cobratoolbox/docs/requirements.html}
are met. If an installation of the COBRA Toolbox is already present,
there is no need to re-clone the full repository. Instead, you can
update the repository from MATLAB or from the terminal.

\begin{enumerate}
\item Update from within MATLAB by running:

\begin{lstlisting}
>> updateCobraToolbox
\end{lstlisting}
\item Update from the terminal (or shell) by running from within the \emph{cobratoolbox}
directory

\begin{lstlisting}[style=bashStyle]
$ cd cobratoolbox  # change to the cobratoolbox directory
$ git checkout master  # switch to the master branch
$ git pull origin master  # retrieve changes
\end{lstlisting}
\end{enumerate}
\criticalStep

The official repository must be cloned as explained in the installation
instructions in Steps \ref{subsec:Installation-of-the}-\ref{subsec:Update-the-fork}.
The COBRA Toolbox can only be updated if no changes have been made
locally in the cloned repository. Steps \ref{subsec:Installation-of-the}-\ref{subsec:Update-the-fork}
provide explanations on how to contribute.

In case the update of the COBRA Toolbox fails or cannot be completed,
clone the repository again.
\end{itemize}
\begin{itemize}
\item On \emph{Linux} and \emph{macOS}, a \emph{bash} terminal with \emph{git}
and \emph{curl }is readily available. Supplementary Manual 2 provides
a brief guide to the basics of using a terminal. \caution On Windows,
the command line tools such as \emph{git} or \emph{curl} will be be
installed together with \emph{git Bash}. Make sure that you select
\emph{<Use git Bash and optional Unix tools from the Windows Command
prompt} \emph{during the installation process>} of git Bash. After
installing \emph{git Bash}, restart MATLAB. On \emph{macOS}, install
the \emph{Xcode} command line tools by following the instructions
on \url{https://railsapps.github.io/xcode-command-line-tools.html}.
\end{itemize}

\subsubsection{Optional software \label{subsec:Optional-Software}}

Some software is only required for one or more optional steps of the
procedure.
\begin{itemize}
\item The libSBML package, version 5.15.0 or above is already packaged with
the COBRA Toolbox via the COBRA.binary submodule for all common operating
systems. Alternatively, binaries can be downloaded separately and
installed by following the procedure on \url{http://sbml.org/Software/libSBML}.
\item The individual MATLAB toolboxes can be installed during the MATLAB
installation process. If MATLAB is already installed, the toolboxes
can be managed using the built-in MATLAB add-on manager as described
on \url{https://mathworks.com/help/matlab/matlab_env/manage-your-add-ons.html}.
\item The Chemaxon Calculator Plugins, version 16.9.5.0 or above can be
installed by following the installation procedures outlined in the
user guide on \url{https://chemaxon.com/products/calculator-plugins}.
\item Install Java, version 8 or above, by following the procedures given
on \url{https://java.com/en/download/help/index_installing.xml}.
\item Python, version 2.7 is already installed on \emph{Linux} and \emph{macOS}.
On Windows, follow the instructions on \url{https://wiki.python.org/moin/BeginnersGuide/Download}.
\item NumPy may be installed by following the procedures on \url{https://docs.scipy.org/doc/numpy-1.10.1/user/install.html}.
\item OpenBabel, version 2.3 or above, may be installed by following the
installation instructions on \url{http://openbabel.org/wiki/Category:Installation}.
\item Latest version of the Reaction Decoder Tool (RDT) can be installed
by following the procedures on \url{https://github.com/asad/ReactionDecoder#installation}.
\end{itemize}
Most steps of the solver installation require superuser or administrator
rights (\mcode{sudo}) and eventually setting environment variables.
Detailed instructions and links to the official installation guidelines
for installing Gurobi, Mosek, Tomlab and IBM Cplex can be found on
\url{https://opencobra.github.io/cobratoolbox/docs/solvers.html}.
\caution Make sure that environment variables are properly set in
order for the solvers to be properly recognised by the COBRA Toolbox.

\subsubsection{Application specific software}
\begin{itemize}
\item On \emph{Linux} or \emph{macOS}, the C-shell \emph{csh} can be installed
by following the instructions on \url{https://en.wikibooks.org/wiki/C_Shell_Scripting/Setup}.
\item The \emph{gcc} and \emph{gfortran} compilers can be installed by following
the links given on \url{https://opencobra.github.io/cobratoolbox/docs/compilers.html}.
\end{itemize}

\subsubsection{Contributing software}
\begin{itemize}
\item The MATLAB.devTools can be installed by following the instructions
given on \url{https://github.com/opencobra/MATLAB.devTools#installation}.
Alternatively, if the COBRA Toolbox is already installed, then the
MATLAB.devTools can be installed directly from within MATLAB by typing:

\begin{lstlisting}
>> installDevTools()
\end{lstlisting}
\end{itemize}

\section{\label{sec:PROCEDURE}PROCEDURE}

\subsubsection[Initialisation of the COBRA Toolbox]{Initialisation of the COBRA Toolbox \timing $5-30$ s}

\pStep \label{step:Init1}At the start of each MATLAB session, the
COBRA Toolbox must be initialised. The initialisation can be done
either manually (option A) or automatically (option B).
\begin{enumerate}[resume]
\item Automatically initialising the COBRA Toolbox

For a regular user who primarily uses the official openCOBRA repository,
automatic initialisation of the COBRA Toolbox is recommended. 
\begin{enumerate}[resume, label=(i)]
\item Edit the MATLAB \emph{startup.m} file and add a line with \mcode{initCobraToolbox}
so that the COBRA Toolbox is initialised each time that MATLAB is
started. \begin{lstlisting}
>> edit startup.m
\end{lstlisting}
\end{enumerate}
\end{enumerate}
\begin{enumerate}
\item Manually initialising the COBRA Toolbox

It is highly recommended to manually initialise when contributing
(see Steps \ref{subsec:Installation-of-the}-\ref{subsec:Update-the-fork}),
especially when the official version and a clone of the fork are present
locally. 
\begin{enumerate}[label=(i)]
\item Navigate to the directory where you installed the COBRA Toolbox and
initialise by running:\begin{lstlisting}
>> initCobraToolbox;
\end{lstlisting}
\end{enumerate}
\end{enumerate}
\criticalStep During initialisation, a check for software dependencies
is made and reported to the command window. It is not necessary that
all possible dependencies are satisfied before beginning to use the
toolbox, e.g., satisfaction of a dependency on a multi-scale linear
optimisation solver is not necessary for modelling with a mono-scale
metabolic model. However, other software dependencies are essential
to be satisfied, e.g., dependency on a linear optimisation solver
must be satisfied for any method that uses flux balance analysis.
\troubleshooting 

\pStep \label{subsec:checkSolvers}At initialisation, one from a
set of available optimisation solvers will be selected as the default
solver. If Gurobi is installed, it is used as the default solver for
LP, QP, and MILP problems. Otherwise, the GLPK solver is selected
for LP and MILP problems. It is important to check if the solvers
installed are satisfactory. A table stating the solver compatibility
and availability is printed to the user during initialisation. Check
the currently selected solvers with 

\begin{lstlisting}   
>> changeCobraSolver;
\end{lstlisting}

\criticalStep A dependency on at least one linear optimisation solver
must be satisfied for flux balance analysis.

\subsubsection[Verify and test the COBRA Toolbox]{Verify and test the COBRA Toolbox \timing $\sim10^{3}$ s}

\pStep \label{subsec:Verify-and-test} Optionally test the functionality
of the COBRA Toolbox locally, especially if one encounters an error
running a function. The test suite runs tailored tests that verify
the output and proper execution of core functions on the locally configured
system. The full test suite can be invoked by typing:

\begin{lstlisting}  
>> testAll  
\end{lstlisting}

\troubleshooting

\subsubsection[Importing a reconstruction or model]{Importing a reconstruction or a model \timing $10-10^{2}$ s}

\pStep\label{step:ImportAReconstruction} The COBRA Toolbox offers
support for several commonly used model data formats, including models
in Systems Biology Markup Language (\emph{SBML}), Excel Sheets (\emph{.xls})
and different Simpheny(\emph{c}) formats. The COBRA Toolbox fully
supports the standard format documented in the SBML Level 3 Version
1 with the Flux Balance Constraints (\emph{fbc}) package version 2
specifications (\url{www.sbml.org/specifications/sbml-level-3/version-1/fbc/sbml-fbc-version-2-release-1.pdf}).
In order to load a model with a \mcode{fileName} into the MATLAB
workspace as a COBRAv3 model structure, run: 

\begin{lstlisting}
>> model = readCbModel(fileName);
\end{lstlisting}

When \mcode{filename} is left blank, a file selection dialogue window
is opened. If no file extension is provided, the code will automatically
determine the appropriate format from the given filename. The \mcode{readCbModel}
function also supports reading normal MATLAB files for convenience,
and checks whether those files contain valid COBRA models. Legacy
model structures saved in a \emph{.mat }file are loaded and converted.
The fields are also checked for consistency with the current definitions.

\criticalStep It is advisable that \mcode{readCbModel()} is used
to load new models. This is also valid for models provided in \mcode{.mat}
files, as \mcode{readCbModel} checks the model for consistency with
the COBRA Toolbox 3.0 field definitions and automatically performs
necessary conversions for models with legacy field definitions or
field names. In order to develop future-proof code, it is good practice
to use \mcode{readCbModel()} instead of the built-in function \mcode{load}.

\troubleshooting

Old MATLAB models saved as \mcode{.mat} files sometimes contain deprecated
fields or fields which have invalid values. Some of these instances
are checked and corrected during \mcode{readCbModel} but there might
be instances, where \mcode{readCbModel} fails. If this happens, it
is advisable, to \mcode{load} the mat file, run the \mcode{verifyModel}
function on the loaded model, and manually adjust all indicated inconsistent
fields. After this procedure, we suggest to \mcode{save} the model
again and use \mcode{readCbModel} to load the model. 

If an SBML file produces an error during IO, please check that the
file is valid SBML using the SBML\-Validator (\url{http://sbml.org/Facilities/Validator}). 

\textbf{ANTICIPATED RESULTS}

After reading a model, a new MATLAB struct should be present in the
MATLAB workspace. This struct should contain at least the following
fields: \mcode{S, lb, ub, c, osense, b, csense, rxns, mets, genes}
and \mcode{rules}. Additional fields from Table \ref{tab:Fields,-dimensions,-data}
can also be present, if the source file contains the corresponding
information.

\subsubsection[Exporting a reconstruction or model]{Exporting a reconstruction or a model \timing $10-10^{2}$ s}

\pStep \label{step:ExportingAReconstruction}The COBRA Toolbox offers
a set of different output methods. The most commonly used formats
are SBML \emph{.xml} and MATLAB \emph{.mat} files. SBML is the preferred
output format, as it can be read by most applications in the field
of computational systems biology. However, some information cannot
be encoded in standard SBML, so a \emph{.mat} file might contain information
not present in the corresponding SBML output. In order to output a
COBRA model structure in either format, use: 

\begin{lstlisting}
>> writeCbModel(model, fileName);
\end{lstlisting}

The extension of the \mcode{fileName} provided is used to identify
the type of output requested. The model will consequently be converted
and saved in the respective format. When exporting a reconstruction
or model, it is necessary that model adheres to the \mcode{model}
structure in Table \ref{tab:Fields,-dimensions,-data}, and that fields
contain valid data. For example, all cells of the the \mcode{rxnNames}
field should only contain data of type \mcode{char} and not data
of type \mcode{double}. 

\troubleshooting

If \mcode{writeCbModel} throws an error, please check the model using
the \mcode{verifyModel}, and make sure, that all fields adhere to
the field specifications detailed above.

\textbf{ANTICIPATED RESULTS}

After writing a file with the specified format containing the model
information, the file is present either in the indicated folder or
the current directory.

\subsubsection[Use of rBioNet to add reactions to a reconstruction]{Use of rBioNet to add reactions to a reconstruction\timing $1-10^{3}$
s}

\pStep \label{step:rBioNet}We highly recommend using \emph{rBioNet}\cite{thorleifsson_rbionet:_2011}
(a graphical user interface-based reconstruction tool) for the addition
or removal of reactions and of gene-reaction associations. 

\criticalStep If you do not have existing \emph{rBioNet} metabolite,
reaction, and compartment databases, the first step is to create these
files. Please refer to the \emph{rBioNet} tutorial provided in the
COBRA Toolbox for instructions on how to add new metabolites and reactions
to an \emph{rBioNet} database. Make sure that all the relevant metabolites
and reactions that you wish to add to your reconstruction are present
in your \emph{rBioNet} databases.

There are two options for using \emph{rBioNet} functionality to add
reactions to a reconstruction:
\begin{enumerate}
\item Adding reactions from an \emph{rBioNet} database to a reconstruction
using the \emph{rBioNet} graphical user interface.
\begin{enumerate}
\item Verify your \emph{rBioNet} settings

First, make sure the paths to your \emph{rBioNet} reaction, metabolite,
and compartment databases are set correctly

\begin{lstlisting} 
>> rBioNetSettings;
\end{lstlisting}
\item Load the \emph{.mat} files that hold your reaction, metabolite, and
compartment databases.
\item To add reactions from an rBioNet database to a reconstruction, invoke
the \emph{rBioNet} graphical user interface with:

\begin{lstlisting} 
>> ReconstructionTool;
\end{lstlisting}

Select File > Open Model Creator.
\item Load your reconstruction by selecting File > Open Model > Complete
Reconstruction.
\item Add reactions from the \emph{rBioNet} database by selecting 'Add Reaction'
and selecting a reaction. Repeat for all reactions that should be
added to the reconstruction.
\item Save your updated reconstruction by selecting File > Save > As Reconstruction
Model. As \emph{rBioNet} was created using the old COBRA model structure,
use the following command to convert your model to the new model structure:

\begin{lstlisting} 
>> model = convertOldStyleModel(model);
\end{lstlisting}
\end{enumerate}
\item Adding reactions from the \emph{rBioNet} database without using the
\emph{rBioNet} interface.

If you wish to add the reactions only to the \emph{rBioNet} database,
hence benefiting from the included quality control and assurance measures,
but then afterwards use the COBRA Toolbox commands to add reactions
to the reconstruction, use the following commands.
\begin{enumerate}
\item Load (or create) a list of reaction abbreviations \mcode{ReactionList}
to be added from the \emph{rBioNet} reaction database:

\begin{lstlisting} 
>> load('Reactions.mat');
\end{lstlisting}
\item Load the \emph{rBioNet} reaction database \mcode{'rxnDB'}:

\begin{lstlisting} 
>> load('rxnDB.mat');
\end{lstlisting}
\item Then, add new reactions:

\begin{lstlisting} 
>> for i = 1:length(ReactionList)                 
    model = addReaction(model, ReactionList{i}, 'reactionFormula', rxnDB(find(ismember(rxn(:, 1), ReactionList{i})), 3));    
   end
\end{lstlisting}
\end{enumerate}
\end{enumerate}
A stoichiometric representation of a reconstructed biochemical network
is contained within the \mcode{model.S} matrix. This is a stoichiometric
matrix with \mcode{m} rows and \mcode{n} columns. The entry \mcode{model.S(i,j)}
corresponds to the stoichiometric coefficient of the $i^{th}$ molecular
species in the $j^{th}$ reaction. The coefficient is negative when
the molecular species is consumed in the reaction and positive if
it is produced in the reaction. If \mcode{model.S(i,j) == 0}, then
the molecular species does not participate in the reaction.In order
to manipulate an existing reconstruction in the COBRA Toolbox, one
can use \emph{rBioNet}, use a spreadsheet, or generate scripts with
reconstruction functions. Each approach has its advantages and disadvantages.
When adding a new reaction or gene-protein-reaction association \emph{rBioNet}
ensures that reconstruction standards are satisfied, but it may make
the changes less tractable when many reactions are added. A spreadsheet-based
approach is tractable, but only allows for the addition, and not the
removal, of reactions. In contrast, using reconstruction functions
provides an exact specification for all of the refinements made to
a reconstruction. One can also combine these approaches by first formulating
the reactions and gene-protein-reaction associations with \emph{rBioNet}
and then adding sets of reactions using reconstruction functions.

\subsubsection[Use of a spreadsheet to add reactions to a reconstruction]{Use of a spreadsheet to add reactions to a reconstruction\timing
$1-10^{3}$ s}

\pStep \label{subsec:Load-reactions}Load reactions from a spreadsheet
with a pre-specified format\cite{thiele_community-driven_2013} into
a new model structure \mcode{modelNewR}:\begin{lstlisting} 
>> modelNewR = xls2model('NewReactions.xlsx'); 
\end{lstlisting}

\pStep \label{step:Merge-recon}Merge the existing reconstruction
\mcode{model} with the new model structure \mcode{modelNewR} to
obtain a reconstruction with expanded content \mcode{modelNew}: \begin{lstlisting} 
>> modelNew = mergeTwoModels(model, modelNewR, 1); 
\end{lstlisting}

\subsubsection[Use of scripts with reconstruction functions]{Use of scripts with reconstruction functions\timing $1-10^{2}$
s}

\pStep\label{step:reconstrFunctions} In order to ensure traceability
of all manipulations to a reconstruction, generate, execute and save
a script that calls reconstruction functions rather than using the
command line. The function \mcode{addReaction} can be used to add
a reaction to a reconstruction:\begin{lstlisting} 
>> model = addReaction(model, 'GAPDH', 'metaboliteList', {'g3p[c]', 'nad[c]', 'pi[c]', '13bpg[c]', 'nadh[c]', 'h[c]'}, 'stoichCoeffList', [-1; -1; -2; 1; 1; 1]);
\end{lstlisting}

The use of \mcode{metaboliteList} provides a cell array of compartment
specific molecular species abbreviations, while \mcode{stoichCoeffList}
is used to provide a numeric array of stoichiometric coefficients.
If particular metabolites do not exist in \mcode{model.mets}, then
this function will add them to the list of metabolites. In the function
\mcode{addReaction()}, duplicate reactions are recognized even when
the order of metabolites or the abbreviation of the reaction are different.
Certain types of reactions, such as exchange, sink, and demand reactions\cite{aurich_metabotools:_2016},
may also be added by using the functions \mcode{addExchangeRxn},
\mcode{addSinkReactions}, or \mcode{addDemandReaction}, respectively.

After adding one or multiple reactions to a reconstruction, it is
important to verify that these reactions can carry flux; that is,
that they are functionally connected to the remainder of the network. 

\pStep Check whether the added reaction(s) can carry flux. For each
newly added reaction \mcode{NewRxn} change it to be the objective
function using:

\begin{lstlisting} 
>> model = changeObjective(model, 'NewRxn');
\end{lstlisting}

then maximise \mcode{'max'} and minimise \mcode{'min'} the flux
through this reaction.

\begin{lstlisting} 
>> FBA = optimizeCbModel(model, 'max');
>> FBA = optimizeCbModel(model, 'min');
\end{lstlisting}

If the reaction should have a negative flux value (e.g., a reversible
metabolic reaction or an uptake exchange reaction), then the minimisation
should result in a negative objective value \mcode{FBA.f < 0}. If
both maximisation and minimisation return an optimal flux value of
zero (i.e., \mcode{FBA.f == 0}), then this newly added reaction cannot
carry a non-zero flux value under the given simulation condition and
the cause for this must be identified. 

If the reaction(s) can carry non-zero fluxes, please repeat Steps
\ref{step:inconsflux} and \ref{Step:subsec:-It-is}\textbf{ }to ensure
stoichiometric consistency, as well as the chemical and biochemical
fidelity.

\pStep Remove reactions. In order to remove reactions from a reconstruction,
use:\begin{lstlisting}
>> modelOut = removeRxns(model, rxnRemoveList);
\end{lstlisting} \pStep Remove metabolites. In order to remove metabolites only,
run:\begin{lstlisting}
>> model = removeMetabolites(model, metaboliteList, removeRxnFlag);
\end{lstlisting}

Note that the removal of one or more metabolites makes sense only
if they do not appear in any reactions or if one wishes to remove
all reactions associated with one or more metabolites.

\pStep\label{Step.-Remove-trivial} Remove trivial stoichiometry.
If metabolites with zero rows, or reactions with zero columns are
present in a stoichiometric matrix, they can be removed with:\begin{lstlisting}
>> modelOut = removeTrivialStoichiometry(model);
\end{lstlisting}

After removing one or more reactions (or metabolites) from the reconstruction,
please repeat Steps \ref{step:reconstrFunctions} to \ref{Step.-Remove-trivial}
in order to check that these modifications did not alter existing
metabolic functions of the reconstruction-derived models.

\subsubsection[Check the scaling of a reconstruction ]{Check the scaling of a reconstruction \timing $1-10^{2}$ s}

\pStep \label{step:checkScaling}Most optimisation solvers are designed
to work with data (e.g., stoichiometric coefficients, bounds, and
objective coefficients in linear optimisation problems) that is well
scaled. Standard solvers are based on 16-digit double-precision floating-point
arithmetics, so the input data should not require a solution with
more than 8 significant digits in order to ensure that solutions are
accurate to the remaining 8 digits of precision. Such a solution approach
is sufficient for most metabolic models, except, for instance, if
micro and macronutrients are simultaneously being considered. Multi-scale
models of metabolism and macromolecular synthesis require higher precision
solvers, but they only need to be used when necessary, so it is useful
to check the scaling of a new reconstruction or model.

The scaling of a stoichiometric matrix can be checked with:

\begin{lstlisting}   
>> [precisionEstimate, solverRecommendation, scalingProperties] = checkScaling(model);
\end{lstlisting}

\subsubsection[Select a double- or quad-precision optimisation solver]{Select a double- or quad-precision optimisation solver\timing $1-5$
s}

\pStep\label{subsec:selectSolver} The COBRA Toolbox is integrated
with a wide variety of different optimisation solvers (cf. Table$~$\ref{tab:Optimisation-Solvers.}).
Quad MINOS\cite{ma_reliable_2017} is a quadruple-precision version
of the general-purpose, industrial-strength linear and nonlinear optimisation
solver MINOS. This solver operates with 34 digits of precision, and
was developed with multi-scale constraint-based modelling problems
in mind. Higher precision solvers are more precise but less computationally
efficient than standard solvers. They must be used when necessary,
i.e., with multi-scale reconstructions and models. To solve multi-scale
linear optimisation problems, the COBRA Toolbox offers a Double-Quad-Quad
MINOS method (DQQ) that combines the use of Double and Quad solvers
in order to improve efficiency while maintaining high accuracy in
the solution. One can set the optimisation solver used by the COBRA
Toolbox as \mcode{solverStatus = changeCobraSolver(solverName,} \mcode{solverType)}
where \mcode{solverName} specifies the solver to be used, while \mcode{solverType}
specifies the type of problems to solve with the solver specified
by \mcode{solverName} (\mcode{'LP'} for linear optimisation problem,
\mcode{'MILP'} for mixed integer linear problems, \mcode{'QP'} for
quadratic problems , \mcode{'MIQP'} for mixed integer quadratic problems,
\mcode{'NLP'} for non-linear problems, or \mcode{'ALL'} to change
the solver for all the previously mentioned problem types. Depending
on the \mcode{precisionEstimate}, there are two options: choose a
double precision solver (option A) or a quad precision solver (option
B).
\begin{enumerate}
\item The \mcode{solverRecommendation} is \mcode{double}
\begin{enumerate}
\item The recommendation shows that a double precision solver is probably
sufficient. For example, the Gurobi solver can be set to solve linear
programming problems with:

\begin{lstlisting}    
>> solverStatus = changeCobraSolver('gurobi', 'LP');
\end{lstlisting}A positive \mcode{solverStatus} also indicates that the COBRA Toolbox
will use Gurobi as the default linear optimisation solver. 
\end{enumerate}
\end{enumerate}
\begin{enumerate}[resume]
\item The \mcode{solverRecommendation} is \mcode{quad}
\begin{enumerate}
\item The recommendation shows that a higher precision solver is required.
For example, the quad-precision optimisation solver \mcode{dqqMinos}
may be selected for solving linear optimisation problems with:

\begin{lstlisting}   
>> solverStatus = changeCobraSolver('dqqMinos', 'LP');
\end{lstlisting}
\end{enumerate}
\end{enumerate}
\criticalStep A dependency on at least one linear optimisation solver
must be satisfied for flux balance analysis. If any numerical issues
arise while using a double precision solver, then a higher precision
solver should be tested. For instance, a double precision solver may
incorrectly report that an ill-scaled optimisation problem is infeasible
although it actually might be feasible for a higher precision solver.
The \mcode{checkScaling} function may be used on all operating systems,
but the \mcode{dqqMinos} or \mcode{quadMinos} interfaces are only
available on UNIX operating systems. \troubleshooting

\subsubsection[Identify stoichiometrically consistent and inconsistent reactions]{Identify stoichiometrically consistent and inconsistent reactions\timing
$1-10^{5}s$}

\pStep \label{step:stoichConsistency}All biochemical reactions conserve
mass; therefore, it is essential that each biochemical reaction in
a model does actually conserve mass. Reactions that do not conserve
mass\cite{fleming_mass_2012} are, however, often added to a reconstruction
in order to represent the flow of mass into and out of a system, e.g.,
during flux balance analysis. Every reaction that does not conserve
mass, but is added to a model in order to represent the exchange of
mass across the boundary of a biochemical system, is henceforth referred
to as an \emph{external reaction,} e.g., $D\rightleftharpoons\emptyset$,
where $\emptyset$ represents null. Every reaction that is supposed
to conserve mass is referred to as an \emph{internal reaction}. Besides
exchange reactions, a reconstruction may contain mass imbalanced internal
reactions due to incorrect or incompletely specified stoichiometry.
This situation results in one or more sets of \emph{stoichiometrically
inconsistent} reactions\cite{gevorgyan_detection_2008}. For instance,
the reactions $A+B\rightleftharpoons C$ and $C\rightleftharpoons A$
are stoichiometrically inconsistent because it is impossible to assign
a positive molecular mass to all species whilst ensuring that each
reaction conserves mass. By combining flux through both of the former
reactions in the forward direction, the net effect is $B\rightarrow\emptyset$,
that is, inadvertent exchange of $B$ across the boundary of the model. 

In order to distinguish between the reactions in a model that are
stoichiometrically consistent and stoichiometrically inconsistent,
there are three options:
\begin{enumerate}
\item Use stoichiometric matrix or reaction names
\begin{enumerate}
\item A heuristic approach to pinpoint external reactions is to identify
reactions with only one stoichiometric coefficient, or reactions with
the \mcode{model.rxns} abbreviation prefixes \mcode{EX_}, \mcode{DM_}
and \mcode{sink_}, for exchange, demand and sink reactions, respectively:
\begin{lstlisting}   
>> model = findSExRxnInd(model);
\end{lstlisting}In the result, \mcode{model.SIntRxnBool} gives a boolean vector of
reactions that are heuristically thought to be internal. \troubleshooting 
\end{enumerate}
\end{enumerate}
\begin{enumerate}[resume]
\item Use the \mcode{checkMassChargeBalance} function
\begin{enumerate}
\item When \mcode{model.metFormulas} is populated with the chemical formulae
of molecular species, it is possible to check which reactions are
elementally imbalanced with: \begin{lstlisting}   
>> [massImbalance] = checkMassChargeBalance(model);
\end{lstlisting}The output \mcode{massImbalance} is a $n\times t$ matrix with a
non-zero entry for any elemental imbalance in a reaction. The other
outputs from this function can also be used to analyse imbalanced
reactions to suggest modifications to the stoichiometric specification
that can resolve the imbalance. A resolution of mass imbalance should
ensure that the reaction stoichiometry is consistent with the known
biochemical mechanism of the reaction. \troubleshooting 
\end{enumerate}
\item Use the \mcode{findStoichConsistentSubset} function
\begin{enumerate}
\item Given stoichiometry alone, a non-convex optimisation problem can be
used to approximately identify the largest set of reactions in a reconstruction
that are stoichiometrically consistent.\begin{lstlisting}   
>>[~, SConsistentRxnBool, ~, SInConsistentRxnBool, ~, unknownSConsistencyRxnBool, model] = findStoichConsistentSubset(model, massBalanceCheck);
\end{lstlisting}When checking for stoichiometric inconsistency, external reactions
identified via Option B can be used to warm start the algorithm for
Option C if \mcode{massBalanceCheck == 1}. The non-zero entries of
\mcode{unknownSConsistencyRxnBool} and \mcode{unknownSConsistencyMetBool}
denote reactions and uniquely involved molecular species where consistency
could not be established. 
\end{enumerate}
\end{enumerate}
\criticalStep Any supposedly internal reaction that is actually stoichiometrically
inconsistent with the remainder of a reconstruction should be omitted
from a model that is intended to be subjected to flux balance analysis,
otherwise erroneous predictions may result due to inadvertent violation
of the steady-state mass conservation constraint. \troubleshooting 

\subsubsection[Identify stoichiometrically consistent and inconsistent molecular
species]{Identify stoichiometrically consistent and inconsistent molecular
species\timing $1-10^{3}~$s}

\pStep \label{step:ConsistentSpecies}The molecular species that
only participate in reactions that are stoichiometrically inconsistent
can be identified using:

\begin{lstlisting}   
>>[SConsistentMetBool, ~, SInConsistentMetBool, ~, unknownSConsistencyMetBool, ~, model] = findStoichConsistentSubset(model, massBalanceCheck);
\end{lstlisting}

\subsubsection[Set simulation constraints]{Set simulation constraints\timing $1-10^{3}$ s}

\pStep\label{step:set-simulation} In order to set the constraints
on a model:

\begin{lstlisting}   
>> model = changeRxnBounds(model, rxnNameList, value, boundType);
\end{lstlisting}

The list of reactions for which the bounds should be changed is given
by \mcode{rxnNameList}, while the vector \mcode{value} contains
the new boundary reaction rate values. This type of bound can be set
to a lower (\mcode{'l'}) or upper bound (\mcode{'u'}). Alternatively,
both bounds can be changed simultaneously (\mcode{'b'}).

\criticalStep The more biochemically realistic the applied constraints
are with respect to a particular context, the more likely network
states that are specific to that context are to be predicted, as opposed
to those predicted from a generic model. All else being equal, a model
derived from a comprehensive yet generic reconstruction will be less
constrained than a model derived from a less comprehensive yet generic
reconstruction. That is, in general, the more comprehensive a reconstruction
is, the greater attention must be paid to setting simulation constraints.

\subsubsection[Identify molecular species that leak, or siphon, across the boundary
of the model]{Identify molecular species that leak, or siphon, across the boundary
of the model\timing $1-10^{3}$ s}

\pStep\label{step:identify-molecular-species} Identification of
internal and external reactions using \mcode{findSExRxnInd} in Step
A is the fastest option, but may not always be accurate. It is therefore
wise to check whether there exist molecular species that can be produced
from nothing (leak) or consumed givin nothing (siphon) in a reconstruction,
with all external reactions blocked. If \mcode{modelBoundsFlag == 1},
then the leak testing uses the model bounds on internal reactions,
and if \mcode{modelBoundsFlag == 0}, then all internal reactions
are assumed reversible.

\begin{lstlisting}
>> modelBoundsFlag = 1;  
>> [leakMetBool, leakRxnBool, siphonMetBool, siphonRxnBool] = findMassLeaksAndSiphons(model, model.SIntMetBool, model.SIntRxnBool, modelBoundsFlag);
\end{lstlisting}

\criticalStep Non-zero entries in \mcode{leakMetBool}, or \mcode{siphonMetBool},
indicate that the corresponding molecular species can be produced
from nothing, or consumed giving nothing, and may invalidate any flux
balance analysis prediction.

\subsubsection[Identify flux inconsistent reactions]{Identify flux inconsistent reactions\timing $1-10^{3}$ s}

\pStep\label{step:inconsflux} In flux balance analysis, the objective
is to predict reaction fluxes subject to a steady state assumption
on internal molecular species and a mass balance assumption for molecular
species exchanged across the boundary of the model. It is therefore
useful to know, before making any flux balance analysis prediction,
which reactions do not admit a non-zero steady state flux, i.e., the
reactions that are flux inconsistent, also known as blocked reactions.
In order to identify these reactions that do not admit a non-zero
flux, use:

\begin{lstlisting}   
>>[fluxConsistentMetBool, fluxConsistentRxnBool, fluxInConsistentMetBool, fluxInConsistentRxnBool] =  findFluxConsistentSubset(model);
\end{lstlisting}

\subsubsection[Flux balance analysis]{Flux balance analysis\timing $1-10^{2}$ s}

\pStep\label{step:FBA} In standard notation, flux balance analysis\cite{orth_what_2010}
is the linear optimisation problem

\begin{equation}
\begin{array}{ll}
\underset{v\in\mathbb{R}^{n}}{\textrm{max}} & \rho(v)\coloneqq c^{T}v\\
\text{s.t.} & Sv=0,\\
 & l\leq v\leq u,
\end{array}\label{eq:FBA}
\end{equation}
where $c\in\mathbb{R}^{n}$ is a parameter vector that linearly combines
one or more reaction fluxes to form the objective function, denoted
$\rho(v)$. In the COBRA Toolbox, \mcode{model.c} contains the objective
coefficients. $S\in\mathbb{R}^{m\times n}$ is the stoichiometric
matrix stored in \mcode{model.S}, and the lower and upper bounds
on reaction rates, $l,u\in\mathbb{R}^{n}$ are stored in \mcode{model.lb}
and \mcode{model.ub}, respectively. The equality constraint represents
a steady state constraint (production = consumption) on internal metabolites
and a mass balance constraint on external metabolites (production
+ input = consumption + output). The solution to Problem (\ref{eq:FBA})
can be obtained using a variety of linear programming (LP) solvers
that have been interfaced with the COBRA Toolbox. Table \ref{tab:Optimisation-Solvers.}
gives the various options. A typical application of flux balance analysis
is to predict an optimal steady-state flux vector that optimises a
microbial biomass production rate\cite{feist_biomass_2010}, subject
to literature derived bounds on certain reaction rates. Deciphering
the most appropriate objective function for a particular context is
an important open research question. The objective function in Problem
(\ref{eq:FBA}) can be modified by changing \mcode{model.c} directly,
or using the convenient function:\begin{lstlisting}
>> model = changeObjective(model, rxnNameList, objectiveCoeff);
\end{lstlisting}

A cell array \mcode{rxnNameList} and numeric array \mcode{objectiveCoeff}
are used to give the reaction abbreviation and corresponding linear
objective coefficient for one or more reactions to be optimised. By
default, \mcode{objectiveCoeff(p) > 0} and \mcode{objectiveCoeff(q) < 0}
correspond to maximisation and minimisation of the $p^{th}$ and $q^{th}$
reaction abbreviation in \mcode{rxnNameList} .

\pStep\label{step:optimizeCbModel} Flux balance analysis, and many
of its variants, can be computed using the versatile function\textbf{
}\mcode{optimizeCbModel}. That is, the default method implemented
by \mcode{optimizeCbModel} is flux balance analysis, as defined in
Problem (\ref{eq:FBA}), but depending on the optional arguments provided
to \mcode{optimizeCbModel}, many methods that are variations on flux
balance analysis are also implemented and accessible with slight changes
to the input arguments.
\begin{enumerate}
\item Computing a flux balance analysis solution
\begin{enumerate}
\item A solution to the flux balance analysis Problem (\ref{eq:FBA}) can
be computed using:

\begin{lstlisting}
>>  FBAsolution = optimizeCbModel(model);
\end{lstlisting}

\criticalStep Assuming the constraints are feasible, the optimal
objective value \mcode{FBAsolution.f} is unique; however, the optimal
flux vector \mcode{FBAsolution.v} is most likely not unique. It is
unwise to base any biological interpretation on a single optimal flux
vector if it is one of many alternative optima, because the optimal
vector returned can vary depending on the solver chosen to solve the
problem. Therefore, when a flux vector is interpreted, it should be
a unique solution to some optimisation problem. 
\end{enumerate}
\end{enumerate}
\begin{enumerate}[resume]
\item Computing the unique flux balance analysis solution
\begin{enumerate}
\item In order to predict a unique optimal flux vector, it is necessary
to regularise the objective by subtracting a strictly concave function
from it. That is $\rho(v)=c^{T}v-\theta(v),$ where $\theta(v)$ is
a strictly convex function. This can be achieved with:

\begin{lstlisting}
>> osenseStr = 'max';
>> minNorm = 1e-6;
>> solution = optimizeCbModel(model, osenseStr, minNorm);
\end{lstlisting}Assuming the constraints are feasible, the optimal objective value
\mcode{solution.f} and the optimal flux vector \mcode{solution.v}
are unique. Setting \mcode{minNorm} to $10^{-6}$ is equivalent to
maximising the function $\rho(v)\coloneqq c^{T}v-\frac{\sigma}{2}v^{T}v$
with $\sigma=10^{-6}$ and $\theta(v)=\frac{\sigma}{2}v^{T}v$ is
a regularisation function. With high-dimensional models, it is wise
to ensure that the optimal value of the regularisation function is
smaller than the optimal value of the original linear objective in
Problem (\ref{eq:FBA}), that is $\rho(v^{\star})\gg\theta(v^{\star})$.
A pragmatic approach is to select \mcode{minNorm = 1e-6;}, then reduce
it if necessary. 
\end{enumerate}
\end{enumerate}
The solution structure \mcode{FBAsolution} from \mcode{optimizeCbModel}
always has the same form, even if the meaning of the fields changes
depending on the optional input arguments to the function. The field
\mcode{.stat} contains a standardised solver status. If \mcode{FBAsolution.stat == 1},
then an optimal solution has been found and will be returned. The
field \mcode{.v} is a flux vector such that the optimal value of
the objective function is attained, \mcode{.y} yields the vector
of dual variables for the equality constraints, and \mcode{.w} contains
the vector of optimal dual variables for the inequality constraints.
The field \mcode{.stat} is translated from the solver specific status
\mcode{.origStat}. The latter is idiosyncratic to each numerical
optimisation solver, and this is translated to the standardised solver
status in order to enable other functions within the COBRA Toolbox
to operate in a manner invariant with respect to the underlying solver,
to the maximum extent possible. If \mcode{FBAsolution.stat == 2},
then the lower and upper bounds are insufficient to limit the value
of the objective function and the problem is unbounded, so no optimal
solution is returned. If \mcode{FBAsolution.stat == 0}, then the
constraints in Problem (\ref{eq:FBA}) do not admit any feasible steady
state flux vector and therefore no optimal solution exists. If \mcode{FBAsolution.stat == -1},
then no solution is reported, due to a time limit or numerical issues.
\troubleshooting 

\subsubsection[Relaxed flux balance analysis]{Relaxed flux balance analysis\timing $1-10^{3}$ s}

\pStep\label{step:relaxedFBA} Every solution to Problem (\ref{eq:FBA})
must satisfy $Sv=0$ and $l\le v\le u$, independent of any objective
chosen to optimise over the set of constraints. It may occur that
these constraints are not all simultaneously feasible, i.e., the system
of inequalities is infeasible. This situation might be caused by an
incorrectly specified reaction bound. In order to resolve the infeasibility,
one can use \emph{relaxed flux balance analysis,} which is an optimisation
problem that minimises the number of bounds to relax in order to render
a flux balance analysis problem feasible. The optimisation problem
is

\begin{equation}
\begin{array}{ll}
\min\limits _{v,p,q} & \alpha\Vert p\Vert_{0}+\alpha\Vert q\Vert_{0}\\
\text{s.t.} & Sv=0\\
 & l-p\leq v\leq u+q\\
 & p,q\geq0,
\end{array}\label{eq:relaxFBA}
\end{equation}
where $p,q\in\mathbb{R}^{n}$ denote the relaxations on the lower
and upper bounds of the reaction rates vector $v$, and where $r\in\mathbb{R}^{m}$
denotes the relaxations of the mass balance constraint. A non-negative
vector parameter $\alpha\in\mathbb{R}_{+}^{n}$ may be used to prioritise
relaxation of bounds on some reactions rather than others, e.g., relaxation
of bounds on exchange reactions rather than internal reactions. The
optimal choice of parameters depends heavily on the biochemical context.
A relaxation of the minimum number of constraints is desirable because,
ideally, one should be able to justify the relaxation of each bound
with reference to the literature. The scale of this task is proportional
to the number of bounds proposed to be relaxed, motivating the sparse
optimisation problem to minimise the number of relaxed bounds. Relaxed
flux balance analysis can be implemented with:

\begin{lstlisting}
>> solution = relaxFBA(model, relaxOption);
\end{lstlisting}

The structure \mcode{relaxOption} can be used to prioritise the relaxation
of one type of bound over another. For example, in order to disallow
relaxation of bounds on all internal reactions, set the field \mcode{.internalRelax}
to \mcode{0} and to allow the relaxation of bounds on all exchange
reactions set the field \mcode{.exchangeRelax} to \mcode{2}. If
there are certain reaction bounds that should not be relaxed, then
this can be specified using the boolean vector field \mcode{.excludedReactions}.
The first application of \mcode{relaxFBA} to a model may predict
bounds to relax that are not supported by literature or other experimental
evidence. In this case the field \mcode{.excludedReactions} can be
used to disallow the relaxation of bounds on certain reactions.

\subsubsection[Sparse flux balance analysis]{Sparse flux balance analysis\timing $1-10^{3}$ s}

\pStep\label{step:sparseFBA} The prediction of the minimal number
of active reactions required to carry out a particular set of biochemical
transformations\cite{melendez-hevia_game_1985}, consistent with an
optimal objective derived from flux balance analysis, is based on
a cardinality minimisation problem termed \emph{sparse flux balance
analysis}

\begin{equation}
\begin{array}{ll}
\min\limits _{v} & \Vert v\Vert_{0}\\
\text{s.t.} & Sv=b\\
 & l\leq v\leq u\\
 & c^{T}v=\rho^{\star},
\end{array}\label{eq:sparseFBA}
\end{equation}
where the last constraint is optional and represents the requirement
to satisfy an optimal objective value $\rho^{\star}$ derived from
any solution to Problem (\ref{eq:FBA}). The optimal flux vector can
be considered as a steady-state biochemical pathway with minimal support,
subject to the bounds on reaction rates and satisfaction of the optimal
objective of Problem (\ref{eq:FBA}). There are many possible applications
of such an approach; here, we consider one example.

Sparse flux balance analysis is used to find the smallest active stoichiometrically
balanced cycle that can produce ATP at a maximal rate using the ATP
synthase reaction (\url{https://vmh.uni.lu/#reaction/ATPS4m}). We
use the \mcode{Recon3Dmodel.mat}\cite{brunk_recon_2017}(naming subject
to change), which does not have such a cycle active due to bound constraints,
but does contain such an active cycle with all internal reactions
set to be irreversible. First the model is loaded, then the internal
reactions are identified and blocked and finally the objective is
set to maximise the ATP synthase reaction rate. Thereafter, the sparse
flux balance analysis solution is computed.

\begin{lstlisting}
>> model = readCbModel('Recon3Dmodel.mat');
>> model = findSExRxnInd(model);
>> modelClosed = model;
>> modelClosed.lb(model.SIntRxnBool) = 0; 
>> modelClosed.ub(model.SIntRxnBool) = 0;
>> modelClosed_ATPS4mi = changeObjective(modelClosed, 'ATPS4mi', 1);
>> osenseStr = 'max';
>> minNorm = 'zero';
>> sparseFBAsolution = optimizeCbModel(modelClosed_ATPS4mi, osenseStr, minNorm);
\end{lstlisting}

\subsubsection[Identify dead-end metabolites and blocked reactions]{Identify dead-end metabolites and blocked reactions\timing $\sim$$10^{2}$
s}

\pStep\label{Step:subsec:-Given-a} Manually curated as well as automatically
created genome-scale metabolic reconstructions contain \emph{dead-end
metabolites}, which can either only be produced or only be consumed
in the metabolic network (including transport to/from the system boundary).
Given a \mcode{model}, the function \mcode{detectDeadEnds} identifies
all \emph{dead-end metabolites} in a model:\begin{lstlisting}   
>> deadEndMetabolites = detectDeadEnds(model);
\end{lstlisting}

\pStep The \mcode{deadEndMetabolites} may be split into \mcode{downstreamGaps}
and \mcode{rootGaps}. Metabolites that cannot be produced or consumed
by any of the reactions in the network are referred to as \mcode{rootGaps}.
\begin{lstlisting}   
>> [deadEndMetabolites, rootGaps, downstreamGaps] = gapFind(model, 'true');
\end{lstlisting}

Dead-end metabolites listed in \mcode{deadEndMetabolites} are metabolites
that are both produced and consumed based on network topology alone
but are still dead-end metabolites because there are not any two reactions
that can actively produce and consume the metabolite in any steady
state.

\pStep\label{step:blockedReaction} Both the root and the downstream
metabolites are part of reactions that cannot carry any flux (i.e.,
blocked reactions) given the network topology subject to the current
bounds on reaction rates. In order to identify blocked reactions,
use: \begin{lstlisting}   
>> blockedReactions = findBlockedReaction(model);
\end{lstlisting}

\subsubsection[Gap fill a metabolic network]{Gap fill a metabolic network\timing $10^{2}-10^{5}$ s}

\pStep \label{step:DeadEndMetabolites}Dead-end metabolites show
that there are missing reactions in the network that must enable their
consumption/production. Thus, they define the boundaries of network
gaps that must be filled with one or more reactions to complete our
representation of the full metabolic network. These gaps are due to
incompleteness of our current knowledge, even in well-studied model
organisms\cite{orth_systematizing_2010}. This is partially due to
\emph{orphan enzymes}, whose biochemical functions have been described
but no corresponding gene sequences have yet been found in any organism\cite{yamada_prediction_2012}.
Such biochemical functions cannot be added to reconstructions by automatic
(sequence-based) inference, but must be added manually or by some
non-sequence related computational approach. Moreover, gene annotations
have been experimentally validated in only a limited number of organisms,
which may lead to annotation errors when annotations are propagated
across a large number of genes using sequence based methods only\cite{liberal_simple_2013}.
Genome-scale metabolic reconstructions can assist in identifying missing
knowledge by detecting and filling network gaps, as has been demonstrated
for various organisms, including \emph{E. coli}\cite{reed_systems_2006,orth_gap-filling_2012},
\emph{Chlamydomonas reinhardtii}\cite{chang_metabolic_2011}, and
\emph{Homo sapiens}\cite{rolfsson_human_2011,rolfsson_inferring_2013}.

The COBRA Toolbox facilitates the identification and filling of gaps
using \mcode{gapFind}\cite{satish_kumar_optimization_2007} and \mcode{fastGapFill}\cite{thiele_fastgapfill:_2014}.
\mcode{fastGapFill} uses a reference database (\textbf{\emph{U}},
e.g. KEGG REACTION) and a transport and exchange reaction database
\textbf{\emph{X}} that consists of transport and exchange reactions
for each metabolite in both the reference database and the reconstruction.
Reactions and pathways are proposed for addition to the metabolic
reconstruction during gap filling from the combined \textbf{\emph{UX
}}database. \mcode{fastGapFill} works for both compartmentalised
and decompartmentalised reconstructions. It relies on \emph{fastcc.m},
which was developed within \mcode{fastCORE} in order to approximate
the most compact (i.e., least) number of reactions to be added to
fill the highest possible number of gaps. 

Prioritise reaction types in the reference database to use for filling
gaps using a \mcode{weights} parameter structure. The parameters
\mcode{weights.MetabolicRxns}, \mcode{weights.ExchangeRxns}, and
\mcode{weights.TransportRxns} allow different priorities to be set
for internal metabolic reactions, exchange reactions, and transport
reactions, respectively. Transport reactions include intracellular
and extracellular transport reactions. The lower the weight for a
reaction type, the higher is its priority. Generally, a metabolic
reaction should be prioritised in a solution over transport and exchange
reactions, with for example:\begin{lstlisting}   
>> weights.MetabolicRxns = 0.1;
>> weights.ExchangeRxns = 0.5;
>> weights.TransportRxns = 10;
\end{lstlisting}\pStep\label{Step:subsec:-Prioritise-reaction} Use the function
\mcode{prepareFastGapFill} to prepare a gap filling problem. A reconstruction
is given as a \mcode{model} structure along with the optional inputs:
list of compartments (\mcode{listCompartments}), a parameter \mcode{epsilon}
that is needed for the fastCORE algorithm, the \mcode{fileName}
for the universal database (e.g., KEGG; default: \emph{\mcode{'reaction.lst'}}),
\mcode{dictionaryFile}, which lists the universal database IDs and
their counterpart in the reconstruction as defined in \mcode{model.mets}
(default: \mcode{'KEGG_dictionary.xls'}), and \mcode{blackList},
which permits the exclusion of certain reactions from the universal
database (default: no blacklist).\begin{lstlisting}   
>> [consistModel, consistMatricesSUX, blockedRxns] = ...
   prepareFastGapFill(model, listCompartments, epsilon, fileName, dictionaryFile, blackList);
\end{lstlisting}The first output variable is \mcode{consistModel}, which contains
a flux consistent subnetwork of the input model. \\
\mcode{consistMatricesSUX} represents the flux consistent \textbf{SUX}
matrix, which contains the flux consistent S matrix (model), the universal
database placed in all cellular compartments along with transport
reactions for each metabolite from cytosol to compartment and exchange
reactions for all extracellular metabolites. Finally, \mcode{blockedRxns}
lists again the blocked reactions in the input model.

\pStep\label{Step:-The-main} The main aim of the \mcode{fastGapFill}
function is to find a compact set of reactions from the \textbf{UX}
matrix to be added to the input model to close the gaps in the model.
Gap filling may be carried out using one of two options, depending
on the amount of metadata required to aid the interpretation of proposed
reactions to be added to the model to fill gaps. The two options are:
\begin{enumerate}
\item Without returning additional metadata
\begin{enumerate}
\item In order to fill gaps without returning additional metadata, run:\begin{lstlisting}   
>> epsilon = 1e-4; 
>> addedRxns = fastGapFill(consistMatricesSUX, epsilon, weights);
\end{lstlisting}The parameter \mcode{epsilon} defines the minimum non-zero flux requested
in a blocked reaction when filling gaps. In a multi-scale model, the
value of epsilon may need to be decreased, when using a quadruple
precision solver (see Step \ref{subsec:selectSolver}). The output
\mcode{addedRxns} contains the reactions from the \textbf{UX} matrix
added to fill the gap(s).
\end{enumerate}
\end{enumerate}
\begin{enumerate}[resume]
\item With returning additional metadata
\begin{enumerate}[resume]
\item In order to return additional metadata for assistance with the manual
evaluation of proposed reactions, use:\begin{lstlisting}  
>> addedRxnsExtended = postProcessGapFillSolutions(addedRxns, model, blockedRxns);
\end{lstlisting}The output structure \mcode{addedRxnsExtended} contains the information
present in \mcode{addedRxns} as well as the statistics and whether
desired pathways contain the flux vectors.
\end{enumerate}
\end{enumerate}
The main result is a list of candidate reactions to be added to the
metabolic reconstructions. These reactions need to be evaluated for
their biological and physiological plausibility in the organism, or
cell-type, under consideration. 

\criticalStep Algorithmic approaches help to identify new candidate
reactions, but these candidates must be manually curated before being
added to a reconstruction. This step is critical for obtaining a high-quality
metabolic reconstruction. Adding the least number of reactions to
fill gaps may not be the most appropriate assumption from a biological
viewpoint. Consequently, the reactions proposed to be added to reconstruction
require further manual assessment. Proposed gap filling solutions
must be rejected if they are biologically incorrect. 

The mapping between the metabolite abbreviations in the universal
database (e.g., KEGG) and the reconstruction metabolite abbreviations
in \mcode{model.mets}, will ultimately limit how many blocked reactions
might be resolved with \mcode{fastGapFill}. The larger the number
of metabolites that map between these different namespaces, the larger
the pool of metabolic reactions from the universal database that can
be proposed to fill gaps. The mapping between the reconstruction and
universal metabolite database can be customised using the \mcode{dictionaryFile},
which lists the universal database identifiers and their counterparts
in the reconstruction.

\subsubsection[Extracellular metabolomic data]{Extracellular metabolomic data\timing $10^{3}-10^{5}$ s}

\pStep \emph{\label{step:extraMetabolomic}Metabolomics} is an indispensable
analytical method in many biological disciplines including microbiology,
plant sciences, biotechnology, and biomedicine. In particular, extracellular
metabolomic data are often generated from cell lines in order to characterise
and phenotype them under different experimental conditions (e.g.,
drug treatment or hypoxia). However, the analysis and interpretation
of metabolomic data is still in its infancy, limiting the interpretation
to potential metabolic pathways rather than providing a comprehensive
understanding of the underlying mechanistic basis of the observed
data.

\emph{MetaboTools} is a COBRA Toolbox extension\cite{aurich_metabotools:_2016},
that integrates semi-quantitative and quantitative extracellular metabolomic
data with metabolic models. The resulting models allow for the interpretation
and generation of experimentally testable mechanistic hypotheses.
With \emph{MetaboTools}, extracellular metabolomic data are integrated
with a COBRA model structure, e.g., the generic human metabolic model\cite{brunk_recon_2017},
in a way that ensures the integration of a maximal number of measured
metabolites, while adding a minimal number of additional uptake and
secretion metabolites such that the specified constraints on the metabolic
network can be sustained.  

It is assumed that the extracellular metabolomic experiments are carried
out with a defined fresh medium and that the corresponding model can
only take up the components of the medium (plus dissolved gases).
To apply constraints that are representative of the chemical composition
of the fresh medium used in an experiment, use the \mcode{setMediumConstraints}
function:\begin{lstlisting}   
>> modelMedium = setMediumConstraints(starting_model, set_inf, current_inf, medium_composition, met_Conc_mM, cellConc, t, cellWeight, mediumCompounds, mediumCompounds_lb);
\end{lstlisting}The \mcode{starting_model} is the model before addition of fresh
medium constraints. The \mcode{current_inf} input argument allows
one to specify a value for the large magnitude finite number that
is currently used to represent an effectively infinite reaction rate
bound, then harmonise them to a new value specified by \mcode{set_inf}.
When no information on the bounds of a reaction is known, the ideal
way to set reaction bounds is \mcode{model.lb(j) = -inf;} and \mcode{model.ub(j) = inf;}.
However, depending on the optimisation solver, an infinite lower or
upper bound may or may not be accepted. Therefore, when no information
on the bounds of a reaction are known, except perhaps the directionality
of the reaction, then the upper or lower bound may be a large magnitude
finite number, e.g., \mcode{model.ub(j) = 1000;}. 

The fresh medium composition must be specified with a vector of exchange
reaction abbreviations for metabolites in the cell medium \mcode{medium_composition}
and the corresponding millimolar concentration of each medium component
\mcode{met_Conc_mM}. The density of the culture (\mcode{cellConc},
cells per mL), the time between the beginning and the end of the experiment
(\mcode{t}, hours), and the measured cellular dry weight (\mcode{cellWeight},
gDW) must also be specified. Basic medium components (\mcode{mediumCompound}),
such as protons, water and bicarbonate, and the corresponding lower
bounds on exchange reactions (\mcode{mediumCompounds_lb}), must also
be specified. Even though they are present, they are not usually listed
in the specification of a commercially defined medium, but they are
needed for cells and the generic human metabolic model in order to
support the synthesis of biomass. The \mcode{modelMedium} is a new
model with external reaction bounds set according to the defined fresh
medium.

\pStep Next, prepare the quantitative exometabolomic data using the
\mcode{prepIntegrationQuant} function:\begin{lstlisting}   
>> prepIntegrationQuant(modelMedium, metData, exchanges, sampleNames, test_max, test_min, outputPath);
\end{lstlisting}The fluxes for each metabolite are given as uptake (negative) and
secretion (positive) flux values in a metabolomic data matrix \mcode{metData},
in which each column represents a sample in \mcode{sampleNames} and
each row in \mcode{exchanges} represents an exchanged metabolite.
The units used for fluxes must be consistent within a model. For the
input model in \mcode{modelMedium}, the \mcode{prepIntegrationQuant}
function tests whether the qualitative uptake (\mcode{test_max},
e.g., +/-500) and secretion (\mcode{test_min}, e.g., $10^{-5}$)
of the metabolites is possible for each sample defined in the metabolomic
data matrix \mcode{metData}. If a metabolite cannot be secreted or
taken up, it will be removed from the data matrix for that particular
sample. Possible reasons for this could be missing production or degradation
pathways, or blocked reactions. For each sample, the uptake and secretion
profile compatible with the input model in \mcode{modelMedium} is
saved to the location specified in \mcode{outputPath} using the unique
sample name.

\pStep \label{step:quantConstraint}The model constrained by the
defined fresh medium composition \mcode{modelMedium} and the output
of the \mcode{prepIntegrationQuant} function can now be used to generate
a set of functional, contextualised, condition-specific models using:\begin{lstlisting}   
>> [ResultsAllCellLines, OverViewResults] = setQuantConstraints(modelMedium, samples, tol, minGrowth, obj, no_secretion, no_uptake, {}, {}, 0, outputPath);
\end{lstlisting}A subset of samples can be specified with \mcode{samples}. All fluxes
smaller than \mcode{tol} will be treated as zero. A lower bound (\mcode{minGrowth},
e.g., 0.008 per hour) on a specified objective function, e.g., \mcode{obj = biomass_reaction2;}
needs to be defined, along with metabolites that should not be secreted,
e.g., \mcode{no_secretion = 'EX_o2[e]'}, or taken up (\mcode{no_uptake = 'EX_o2s'}).
The function returns a \mcode{ResultsAllCellLines} structure containing
the context-specific models as well as an overview of model statistics
in \mcode{OverViewResults}. For each sample, a condition-specific
model is created, in which the constraints have been set in accordance
with the medium specification and the measured extracellular metabolomic
data. This set of condition-specific models can then be phenotypically
analysed using the various additional functions present in the COBRA
Toolbox as detailed in the \emph{MetaboTools} protocol\cite{aurich_metabotools:_2016}.

\subsubsection[Intracellular metabolomic data]{Intracellular metabolomic data\timing $10^{2}-10^{4}$ s}

\textsf{\textcolor{black}{}}
\begin{figure}
\centering{}\textsf{\textcolor{black}{\caption{\label{fig:uFBAsteps}Conceptual overview of the main steps involved
in the unsteady-state flux balance analysis (uFBA) method. }
}}
\end{figure}

\pStep \label{step:intraMetabolomic}COBRA methods have also been
developed for integration with intracellular metabolomic measurements\cite{willemsen_metdfba:_2014,kleessen_integration_2015,bordbar_elucidating_2017},
further improving the ability of the COBRA Toolbox to be used for
the integration and interpretation of metabolomic data. In particular,
unsteady-state flux balance analysis (uFBA\cite{bordbar_elucidating_2017})
enables the integration of absolutely quantified time-course metabolomic
data into a metabolic model, generating constraints on intracellular
fluxes even when intracellular metabolite levels are not at steady-state.
The main steps in the \mcode{uFBA} method are illustrated in Figure
\ref{fig:uFBAsteps}.

\textsf{\textcolor{black}{}}
\begin{figure}
\centering{}\textsf{\textcolor{black}{\includegraphics[scale=0.8]{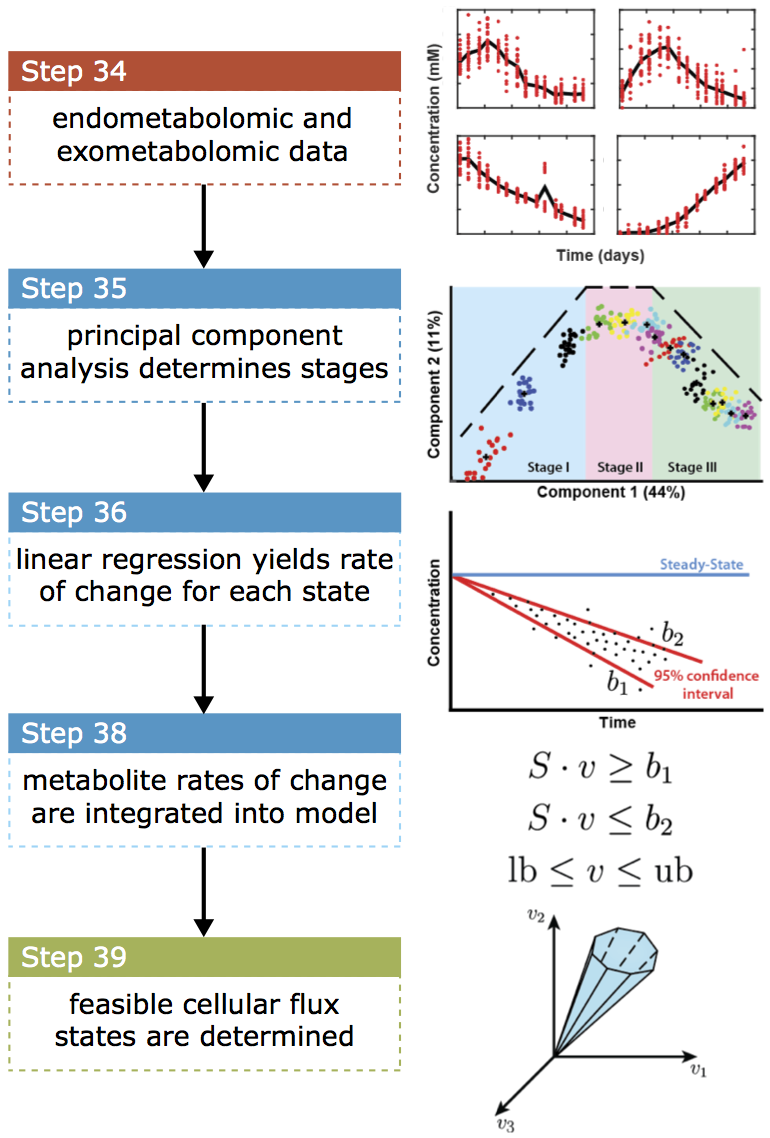}\caption{\label{fig:uFBAsteps}Conceptual overview of the main steps involved
in the unsteady-state flux balance analysis (uFBA\cite{bordbar_elucidating_2017})
method. }
}}
\end{figure}

The first step is to experimentally quantify the absolute concentrations
of a set of extracellular and intracellular metabolites at regular
time intervals\cite{bordbar_elucidating_2017}.

\pStep Plot the time-course metabolomic data. If the data is non-linear,
use principal component analysis to define a sequence of temporal
stages during which the time-course metabolomic data can be considered
piecewise linear. 

\pStep Use linear regression to estimate the rate of change of concentration
with respect to time for each measured metabolite and for each temporal
stage. 

\pStep Load a standard COBRA \mcode{model} structure containing
the fields \mcode{.S}, \mcode{.b}, \mcode{.lb}, \mcode{.ub}, \mcode{.mets},
and \mcode{.rxns}.

\pStep\label{step:intraMetabolomic-end} Integrate the rate of change
in concentration for each measured metabolite with a COBRA model with:

\begin{lstlisting}   
>> uFBAoutput = buildUFBAmodel(model, uFBAvariables);
\end{lstlisting}

The \mcode{uFBAvariables} structure must contain the following fields:
\mcode{.metNames} is a list of measured metabolites, \mcode{.changeSlopes}
provides the rate of change of concentration with respect to time
for each measured metabolite, \mcode{.changeIntervals} yields the
difference between the mean rate of change of concentration with respect
to time and the lower bound of 95\% percent confidence interval. The
list \mcode{ignoreSlopes} contains metabolites whose measurements
should be ignored due to insignificant rate of change.  

The output is a \mcode{uFBAoutput} structure that contains the following
fields: \mcode{.model}, a COBRA model structure with constraints
on the rate of change of metabolite concentrations, \mcode{.metsToUse}
with a list of metabolites with metabolomic data integrated into the
model, and \mcode{.relaxedNodes} with a list of metabolites that
deviate from steady-state along with the direction (i.e., accumulation
or depletion) and magnitude (i.e., reaction bound) of deviation. The
uFBA algorithm automatically determines sink or demand reactions needed
to return a model with at least one feasible flux balance solution,
by automatically reconciling potentially incomplete or inaccurate
metabolomic data with the model structure. The added sink or demand
reactions allow the corresponding metabolites, defined by \mcode{.relaxedNodes},
to deviate from a steady state to ensure model feasibility. The default
approach is to minimise the number of metabolites that deviate from
steady state. 

The \mcode{buildUFBAmodel} function integrates quantitative time
course metabolomic data with a model by setting rates of change with
respect to time for a set of measured intracellular and extracellular
metabolites. A set of sink reactions, demand reactions, or both, may
have been added to certain nodes in the network to ensure that the
model admits at least one feasible mass balanced flux. 

\pStep\label{step:optimizeuFBA} The obtained model can then be minimized
using \mcode{optimizeCbModel}:

\begin{lstlisting}   
>> model_ufba = optimizeCbModel(uFBAoutput.model);
\end{lstlisting}

\subsubsection[Integration of transcriptomic and proteomic data]{Integration of transcriptomic and proteomic data\timing $10^{2}-10^{4}$
s}

\pStep \label{step:transcriptomic}Given a generic reconstruction
of a biochemical network for a particular organism, some reactions
may only be active in a specific tissue, cell-type, or under specific
environmental conditions. It is necessary to extract a context-specific
model from a generic model in order to create a model that is representative
of the part of the biochemical network that is active within a particular
context. Each context-specific model is therefore a subset of a generic
model. A variety of experimental data can be used to determine the
set of reactions that must be part of a context-specific model, including
transcriptomic, proteomic, and metabolomic data, as well as complementary
experimental data from the literature. 

Several model extraction methods have been developed, with different
underlying assumptions, and each has been the subject of multiple
comparative evaluations\cite{blazier_integration_2012,opdam_systematic_2017,estevez_generalized_2014}.
The selection of a model extraction method and its parametrisation,
as well as the methods chosen to preprocess and integrate the aforementioned
omics data, significantly influences the size, functionality, and
accuracy of the resulting context-specific model. Currently, there
is insufficient evidence to assert that one model extraction method
universally gives the most physiologically accurate models. Therefore,
a pragmatic approach is to test the biochemical fidelity of context-specific
models generated using a variety of model extraction methods. 

The COBRA Toolbox offers six different model extraction methods, accessible
via a common interface:

\begin{lstlisting}
>> tissueModel = createTissueSpecificModel(model, options);
\end{lstlisting}

The different methods and associated parameters are selected via the
\mcode{options}\textbf{ }structure. The \mcode{.solver} field indicates
which method shall be used. The other fields of the \mcode{options}
structure vary depending on the method and often depend on bioinformatic
preprocessing of input omics data. There are additional optional parameters
for all algorithms, with the default being the values indicated in
the respective papers. Please refer to the original papers reporting
each algorithm for details on the requirements for preprocessing of
input data. Each of the six different model extraction methods can
be invoked using:
\begin{enumerate}
\item The \mcodeENUM{FASTCORE}\cite{vlassis_fast_2014} algorithm
\begin{enumerate}
\item One set of core reactions that is guaranteed to be active in the extracted
model is identified by \mcodeENUM{FASTCORE}. Then, the algorithm
finds the minimum number of reactions possible to support the core;
\mcode{.core} field provides the core reactions which have to be
able to carry flux in the resulting model.
\end{enumerate}
\item The \mcodeENUM{GIMME}\cite{becker_context-specific_2008} algorithm
\begin{enumerate}
\item With this algorithm, the usage of low-expression reactions is minimised
while keeping the objective (e.g., biomass) above a certain value;
\mcode{.expressionRxns} field provides the reaction expression, with
$-1$ for unknown reactions or reactions not linked to genes; \mcode{.threshold}
field sets the threshold above which a reaction is assumed to be active.
\end{enumerate}
\item The \mcodeENUM{iMAT}\cite{zur_imat:_2010} algorithm
\begin{enumerate}
\item \mcodeENUM{iMAT} finds the optimal trade-off between including high-expression
reactions and removing low-expression reactions; \mcode{.expressionRxns}
field is defined as above; \mcode{.threshold_lb} field is the threshold
below which reactions are assumed to be inactive; \mcode{.threshold_ub}
field is the threshold above which reactions are assumed to be active.
\end{enumerate}
\item The \mcodeENUM{INIT}\cite{agren_reconstruction_2012} algorithm
\begin{enumerate}
\item The optimal trade-off between including and removing reactions based
on their given weights is determined by this algorithm; \mcodeENUM{.weights}
field provides the weights $w$ for each reaction in the objective
of \mcodeENUM{INIT} ($max(\sum_{i\in R}w_{i}y_{i}+\sum_{j\in M}x_{j})$
). Commonly a high expression leads to higher positive values and
low or no detection leads to negative values.
\end{enumerate}
\item The \mcodeENUM{MBA}\cite{jerby_computational_2010} algorithm
\begin{enumerate}
\item \mcodeENUM{MBA} defines high-confidence reactions to ensure activity
in the extracted model. Medium confidence reactions are only kept
when a certain parsimony trade-off is met. \mcode{.medium_set} field
provides the set of reactions that have a medium incidence, while
\mcode{.high_set} field provides the set of reactions that have to
be in the final model. Any reaction not in the medium or high set
is assumed to be inactive and preferably not present in the final
model.
\end{enumerate}
\item The \mcodeENUM{mCADRE}\cite{wang_reconstruction_2012} algorithm
\begin{enumerate}
\item A set of core reactions is first found and all other reactions are
then pruned based on their expression, connectivity to core, and confidence
score. Reactions that are not necessary to support the core or defined
functionalities are thus removed. Core reactions are removed if they
are supported by a certain number of zero-expression reactions. \mcode{.confidenceScores}
field provides reliability for each reaction, generally based on literature,
while \mcode{.ubiquityScore} field provides the ubiquity score of
each reaction in multiple replicates, i.e., the number of times the
reaction was detected as active in experimental data under the investigated
condition.
\end{enumerate}
\end{enumerate}
\criticalStep When integrating omics data, parameter selection is
critical, especially the threshold for binary classification, e.g.,
the threshold for genes into active or inactive sets. Algorithmic
performance often strongly depends on parameter choices and on the
choice of data preprocessing method\cite{opdam_systematic_2017}.
However,  \mcode{createTissueSpecificModel} does not offer data
preprocessing tools, because the selection of the discretisation method
and parameters depend on the origin of the data. However, the COBRA
Toolbox offers functionality to map preprocessed expression data to
reactions via the function \newline\mcode{mapExpressionToReactions(model, expression)}.

\subsubsection[Adding biological constraints to a flux balance model]{Adding biological constraints to a flux balance model \timing $\sim10^{2}$
s}

\pStep \label{step:addingConstraints}A cell-type or organ-specific
model can be converted into a condition-specific model, based on the
imposition of experimentally derived constraints. There are several
types of constraints that can be imposed on a metabolic network, such
as biomass maintenance requirements, environmental constraints, or
maximal enzyme activities. In general, biomass constraints\cite{feist_biomass_2010}
are added as part of a biomass reaction. In some instances, however,
a cell-type (e.g., neuron) does not divide, but is only required to
turn over its biomass components. Turnover rates are commonly expressed
as half-lives and represent the time required for half of the biomass
precursor to be replaced\cite{kuhar_use_2008}. A model can be constrained
with inequality constraints so as to require a minimal rate of turnover
for a metabolite. If that metabolite possesses only one degradation
pathway, then it is sufficient to adjust the bounds on a reaction
in that pathway. However, if there are multiple possible degradation
pathways, then it is necessary to impose a lower bound on the total
rates of a set of irreversible degradation reactions, one for each
possible degradation pathway of the metabolite in question. 

The implementation of such a constraint is illustrated in the following
example. In the brain, phosphatidylcholine (PC) can be degraded by
three different metabolic pathways\cite{lajtha_handbook_2008}: 
\begin{itemize}
\item Phospholipase D acts on the choline/phosphate bond of PC to form choline
and phosphatidic acid (PCHOLP\_hs, \url{https://vmh.uni.lu/#reaction/PCHOLP_hs}). 
\item Phospholipase A2 acts on the bond between the fatty acid and the hydroxyl
group of PC to form a fatty acid (e.g., arachidonic acid or docosahexaenoic
acid) and lysophosphatidylcholine (PLA2\_2, \url{https://vmh.uni.lu/#reaction/PLA2_2}). 
\item Ceramide and PC can also be converted to sphingomyelin by sphingomyelin
synthetase (SMS, \url{https://vmh.uni.lu/#reaction/SMS}). 
\end{itemize}
Load a COBRA model and define the set of reactions that will represent
degradation of the metabolite in question:

\begin{lstlisting}
>> multipleRxnList = {'PCHOLP_hs', 'PLA2_2', 'SMS'};
\end{lstlisting}

\criticalStep Correctly converting the literature data into bound
constraints with the same units used for the model fluxes may be a
challenge. Indeed, the curation of biochemical literature to abstract
the information required to quantitatively bound turnover rates can
take between 4-8 weeks, when the target is to retrieve the biomass
composition and the turnover rates of each of the different biomass
precursors. Once all the constraints are available, imposing the corresponding
reaction bounds takes less than 5 minutes.

\pStep Verify that all the reactions are irreversible (the lower
and upper bounds should be greater or equal to 0).\begin{lstlisting}
>> rxnInd = findRxnIDs(model, multipleRxnList);
>> model.lb(rxnInd);
>> model.ub(rxnInd); 
\end{lstlisting}

\pStep Generate and add the constraint: \begin{lstlisting}
>> c = [1, 1, 1]; 
>> d = 2.674; 
>> ineqSense = 'G'; 
>> modelConstrained = constrainRxnListAboveBound(model, multipleRxnList, c, d, ineqSense); 
\end{lstlisting}where \mcode{c} is a vector forming the inequality constraint $c^{T}v\ge d$,
and \mcode{d} is a scalar. \mcode{ineqSense} encodes the sense of
these inequality (\mcode{'L'} for a lower inequality, or \mcode{'G'}
for an upper inequality). In this example, all entries of $c$ are
positive as we seek for the sum of the rates of the three reactions
(irreversible in the forward direction) to be greater than $d$. This
extra constraint is encoded in the \mcode{model.C} field.

\pStep Check that the constraints are correctly added to the model:\begin{lstlisting}
>> [nMet, nRxn] = size(modelConstrained.S);
\end{lstlisting}\pStep Solve the FBA problem with the extra constraint $c^{T}v\ge d$:\begin{lstlisting}
>> solution = optimizeCbModel(modelConstrained, 'max', 1e-6); 
\end{lstlisting}

\pStep \label{step:addingConstraints-end} Check the values of the
added fluxes. The sum of fluxes should be greater than or equal to
the value of $d$:\begin{lstlisting}
>> solution.v(rxnInd);
>> sum(c*FBAsolution.v(rxnInd)); 
\end{lstlisting}

\subsubsection[Qualitative chemical and biochemical fidelity testing]{Qualitative chemical and biochemical fidelity testing\timing $10^{2}-10^{3}$
s}

\pStep\label{Step:subsec:-It-is} Once a context-specific model is
generated, it is highly advisable to frequently compare preliminary
model predictions with published experimental data\cite{thiele_protocol_2010}.
Such predictions must be compared directly with an unbiased selection
of appropriate independent biological literature in order to identify
possible sources of misconception or computational misspecification.
It is challenging to compare genome-scale predictions with experimental
data that may only be available for a subset of a biochemical network.
In this context, it is important to first turn to literature relevant
to the aspect of the biological network being represented by a model
and then check if the literature result is correctly predicted by
the model. Inevitably, this is an iterative approach with multiple
rounds of iterative refinement of the reconstruction and the model
derived from it, before finalising a model version and comparison
of final predictions with independent experimental data. 

A draft model should be subjected to a range of quantitative and qualitative
chemical and biochemical fidelity tests. As described in Step \ref{step:stoichConsistency},
chemical fidelity testing includes testing for stoichiometric consistency.
This should not be necessary if one starts with a stoichiometrically
consistent generic model and extracts a context-specific model from
it. However, it is possible that misspecified reactions might have
been inadvertently added during refinement of a reconstruction, therefore
retest for stoichiometric consistency. Beyond chemical fidelity, it
is advised to test again for biochemical fidelity. Such tests are
very specific to the particular biological domain that is being modelled.
Here we focus on human metabolism and use \mcode{modelClosed}, the
\mcode{Recon3Dmodel}\cite{brunk_recon_2017} with all external reactions
closed, from Steps \ref{step:FBA}-\ref{step:optimizeCbModel}.

It is important to encode and conduct qualitative fidelity tests for
anticipated true negatives. The following test is for the production
of ATP from water alone in a closed model: 

\begin{lstlisting}
>> modelClosedATP = changeObjective(modelClosed, 'DM_atp[c]'); 
>> modelClosedATP = changeRxnBounds(modelClosedATP, 'DM[atp_c]', 0, 'l'); 
>> modelClosedATP = changeRxnBounds(modelClosedATP, 'EX_h2o[e]', -1, 'l'); 
>> FBAsol = optimizeCbModel(modelClosedATP);
\end{lstlisting}

If \mcode{FBAsol.stat == 0}, then the model is incapable of producing
ATP from water, as expected. If \mcode{FBAsol.stat == 1}, then the
supposedly closed model can produce ATP from water. This indicates
that there are stoichiometrically inconsistent reactions in the network,
which need to be identified. See Step \ref{step:stoichConsistency}
for instructions how to approach this analysis.

\pStep \label{step:qualitativeFidelity-end}It is also important
to encode and conduct qualitative fidelity tests for anticipated true
positives. The following metabolic function test is for the production
of mitochondrial succinate from \mbox{4-Aminobutanoate} in a model
that is closed to exchange of mass across the boundary of the system,
except for the metabolites \mcode{'gly[c]', 'co2[c]'}, and \mcode{'nh4[c]'}.

\begin{lstlisting}
>> modelClosed = addSinkReactions(modelClosed, {'gly[c]', 'co2[c]', 'nh4[c]'}, ...
                                  [-100, -1; 0.1, 100; 0.1, 100]);
>> modelClosed = changeObjective(modelClosed, 'sink_nh4[c]');
>> sol = optimizeCbModel(modelClosed, 'max', 'zero');
\end{lstlisting}

If \mcode{FBAsol.stat == 1}, then it is feasible for the model to
produce mitochondrial succinate from \mbox{4-Aminobutanoate}. If
\mcode{FBAsol.stat == 0}, then this metabolic function is infeasible.
This is not anticipated and indicates that further gap filling is
required (see Steps \ref{step:DeadEndMetabolites}-\ref{Step:-The-main}).

\subsubsection[Quantitative biochemical fidelity testing]{Quantitative biochemical fidelity testing\timing $10^{2}-10^{3}$
s}

\pStep \label{step:quantitativeFidelity}It is important to check
if a model can reproduce or closely approximate known quantitative
features of the biochemical network being represented. Here we illustrate
how to predict the ATP yield from different carbon sources under aerobic
or anaerobic conditions for Recon3D\cite{brunk_recon_2017}. These
are compared with the values for the corresponding ATP yields obtained
from the biochemical literature. This approach can be adapted for
condition- and cell-type specific models derived from Recon3D in order
to test whether these models are still able to produce physiologically
relevant ATP yields. Add and define the ATP hydrolysis reaction \mcode{DM_atp[c]}
to be the objective reaction in the model with:\begin{lstlisting}
>> modelClosed = addReaction(modelClosed, ...
                             'DM_atp[c]', 'h2o[c] + atp[c] -> adp[c] + h[c] + pi[c]');
>> modelClosed = changeObjective(modelClosed, 'DM_atp[c]');
\end{lstlisting}

\pStep Allow the model to uptake oxygen and water, then provide 1
mol/gdw/hr of a carbon source, e.g., glucose (VMH ID: \mcode{glc_D[e]}):

\begin{lstlisting}
>> modelClosed.lb(find(ismember(modelClosed.rxns, 'EX_o2[e]'))) = -1000; 
>> modelClosed.lb(find(ismember(modelClosed.rxns, 'EX_h2o[e]'))) = -1000; 
>> modelClosed.ub(find(ismember(modelClosed.rxns, 'EX_h2o[e]'))) = 1000; 
>> modelClosed.ub(find(ismember(modelClosed.rxns, 'EX_co2[e]'))) = 1000; 
>> modelClosed.lb(find(ismember(modelClosed.rxns, 'EX_glc_D[e]'))) = -1; 
>> modelClosed.ub(find(ismember(modelClosed.rxns, 'EX_glc_D[e]'))) = -1;
\end{lstlisting}

\pStep \label{step:quantitativeFidelity-end}Compute a flux balance
analysis solution with maximum flux through the \mcode{DM_atp[c]}
reaction:

\begin{lstlisting}
>> FBAsolution = optimizeCbModel(modelClosed, 'max', 'zero');
\end{lstlisting}

\subsubsection[MinSpan Pathways: a sparse basis of the nullspace of a stoichiometric
matrix]{\label{subsec:MinSpan-Pathways}MinSpan Pathways: a sparse basis
of the nullspace of a stoichiometric matrix\timing $10^{2}-10^{4}$
s}

\pStep \label{step:minSpanPathways}COBRA models are often mathematically
deconstructed into feasible steady state flux vectors in biochemical
networks that can be biologically conceptualized as pathways. Much
development and analysis has been done for such \emph{pathway vectors}
in terms of elementary flux modes\cite{schuster_elementary_1994},
extreme pathways\cite{schilling_theory_2000}, and elementary flux
vectors\cite{klamt_elementary_2017}. As the number of elementary
or extreme vectors scales exponentially with the size of a typical
metabolic network, increasingly efficient algorithms become essential
for enumerating elementary or extreme vectors at genome-scale. An
alternate approach\cite{bordbar_minimal_2014} is to approximately
compute a set of $n-\textrm{rank}(S)$ sparse linearly independent
flux vectors that together form a basis of the right nullspace of
a stoichiometric matrix and also satisfy specified constraints on
reaction directionality. This approach requires the solution of a
greedy sequence of mixed-integer linear optimisation problems, each
of which computes a sparse flux mode that is linearly independent
from the rest of the vectors within a nullspace basis. The end result
is a sparse set of linearly independent flux modes denoted \emph{MinSpan}
pathways. 

Given a model with a stoichiometric matrix, reaction bounds and reaction
identifiers, the \emph{MinSpan} algorithm may be invoked with:\begin{lstlisting}
>> Z = detMinSpan(model, params);
\end{lstlisting}The \mcode{params} structure provides the user ability to change
key parameters. Among others, the main parameters include the amount
of time \mcode{.timeLimit} for each iterative solve in seconds and
the number of threads for the MILP solver to use. The output $Z\in\mathbb{R}^{n\times(n-\textrm{rank}(S))}$
is a sparse set of $n-\textrm{rank}(S)$ linearly independent flux
modes, each corresponding to a \emph{MinSpan} pathway.

\subsubsection[Low dimensional flux variability analysis]{Low dimensional flux variability analysis\timing $1-10^{3}$ s}

\pStep \label{step:low-FVA}Flux balance analysis does not, in general,
return a unique optimal flux vector. That is, Problem$~$(\ref{eq:FBA})
returns an optimal flux vector, $v^{\star}\in\mathbb{R}^{n}$ with
one flux value for each reaction, but typically an infinite set of
steady state flux vectors exist that can satisfy the same requirement
for an optimal objective, $c^{T}v^{\star}=c^{T}v$, as well as the
other equalities and inequalities in Problem (\ref{eq:FBA}). Flux
variability analysis is a widely used method for evaluating the minimum
and maximum range of each reaction flux that can still satisfy the
aforementioned constraints using two optimisation problems for each
reaction of interest 

\begin{equation}
\begin{array}{ll}
\underset{v}{\textrm{max}\backslash\textrm{min}} & v_{j}\\
\text{s.t.} & Sv=0,\\
 & l\leq v\leq u,\\
 & c^{T}v=c^{T}v^{\star}.
\end{array}\label{eq:FVA}
\end{equation}

Just as there are many possible variations on flux balance analysis,
there are many possible variations on flux variability analysis. The
COBRA Toolbox offers a straightforward interface to implement standard
flux variability analysis and a wide variety of options to implement
variations on flux balance analysis.
\begin{enumerate}
\item Standard flux variability analysis
\begin{enumerate}
\item The following command can be invoked to compute standard flux variability
analysis: \begin{lstlisting}
>> [minFlux, maxFlux] = fluxVariability(model);
\end{lstlisting} The result is a pair of $n$ dimensional column vectors, \mcode{minFlux}
and \mcode{maxFlux}, with the minimum and maximum flux values satisfying
Problem (\ref{eq:FVA}). 
\end{enumerate}
\end{enumerate}
\begin{enumerate}[resume]
\item Advanced flux variability analysis
\begin{enumerate}[resume]
\item The full spectrum of flux variability analysis options can be accessed
using the command:\begin{lstlisting}
>> [minFlux, maxFlux, Vmin, Vmax] = fluxVariability(model, optPercentage, osenseStr, rxnNameList, verbFlag, allowLoops, method);
\end{lstlisting}
\end{enumerate}
\end{enumerate}
The \mcode{optPercentage} parameter allows one to choose whether
to consider solutions that give at least a certain percentage of the
optimal solution. For instance \mcode{optPercentage = 0} would just
find the flux range of each reaction, without of any requirement to
satisfy any optimality with respect to flux balance analysis. Setting
the parameters \mcode{osenseStr = 'min'} or \mcode{osenseStr = 'max'}
determines whether the flux balance analysis problem is first solved
as a minimisation or maximisation. The \mcode{rxnNameList} accepts
a cell array list of reactions to selectively perform flux variability
upon. This is useful for high dimensional models, for which the computation
of a flux variability for all reactions is more time consuming. The
additional $n\times k$ output matrices \mcode{Vmin} and \mcode{Vmax}
return the flux vector for each of the $k\le n$ fluxes selected for
flux variability. The \mcode{verbFlag} input determines how much
output shall be printed. The parameter \mcode{allowLoops == 0} invokes
a mixed integer linear programming implementation of thermodynamically
constrained flux variability analysis for each minimisation or maximisation
of a reaction rate. The \mcode{method} input argument determines
whether the output flux vectors also minimise the \mcode{0-norm},
\mcode{1-norm} or \mcode{2-norm} while maximising or minimising
the flux through one reaction. 

The default result is a pair of maximum and minimum flux values for
every reaction. Optional parameters may be set. For instance, parameters
can be set to control which subset of $k\le n$ reactions of interest
that shall be be obtained, or to determine the characteristics of
each of the $2\times k$ flux vectors.

\subsubsection[High dimensional flux variability analysis]{High dimensional flux variability analysis\timing $1-10^{5}$ s}

\pStep\label{step:FVA} Besides flux balance analysis, flux variability
analysis is the most widely used constraint-based modelling method
for high-dimensional models. However, its use in this setting requires
a more sophisticated computational approach, with a multi-core processor\cite{gudmundsson_computationally_2010},
or computational cluster\cite{heirendt_distributedfba.jl:_2017},
and a commercial-grade linear optimisation solver. In this setting,
advanced users have two options: 
\begin{enumerate}
\item Use \mcode{fastFVA} with MATLAB
\begin{enumerate}
\item Solve the $2\times k$ linear optimisation problems using multiple
threads running on parallel processors with \mcode{fastFVA}, which
depends on the CPLEX solver (IBM Inc.), using the command :\begin{lstlisting}
>> [minFlux, maxFlux, optsol] = fastFVA(model, optPercentage, osenseStr);
\end{lstlisting}The output argument \mcode{optsol} returns the optimal solution of
the initial FBA.
\end{enumerate}
\item Use\emph{ distributedFBA.jl} with Julia
\begin{enumerate}
\item An alternative is to solve the $2\times k$ linear optimisation problems
using multiple threads running on parallel processors or a cluster
using \emph{distributedFBA.jl}, an openCOBRA extension that permits
the solution of flux balance analysis, a distributed set of flux balance
problems, or a flux variability analysis using a common of solver
(GLPK, CPLEX, Clp, Gurobi, Mosek). Assuming that \emph{distributedFBA.jl}
has been correctly installed and configured, the commands to go back
and forth between a model or results in MATLAB and the computations
in Julia are: \begin{lstlisting}
>> save('high_dimensional_model.mat', unimodel);
\end{lstlisting}
\begin{lstlisting}[style=juliaStyle]
# -- switch to Julia -- 
julia> model = loadModel("high_dimensional_model.mat");
julia> workersPool, nWorkers = createPool(128);
julia> minFlux, maxFlux, optSol, fbaSol, fvamin, fvamax = distributedFBA(model, solver, nWorkers=nWorkers, optPercentage=optPercentage, preFBA=true);
julia> saveDistributedFBA("high_dimensional_FVA_results.mat");
# -- switch to MATLAB --
\end{lstlisting}
\begin{lstlisting}
>> load('high_dimensional_FVA_results.mat');
\end{lstlisting}Here, \mcode{nWorkers = 128} will distribute the flux variability
analysis problem amongst 128 Julia processes on one or more computing
nodes in a computational cluster.
\end{enumerate}
\end{enumerate}

\subsubsection[Uniform sampling of steady-state fluxes]{Uniform sampling of steady-state fluxes \timing $1-10^{3}$ s}

\begin{figure}
\begin{centering}
\includegraphics[width=0.75\textwidth]{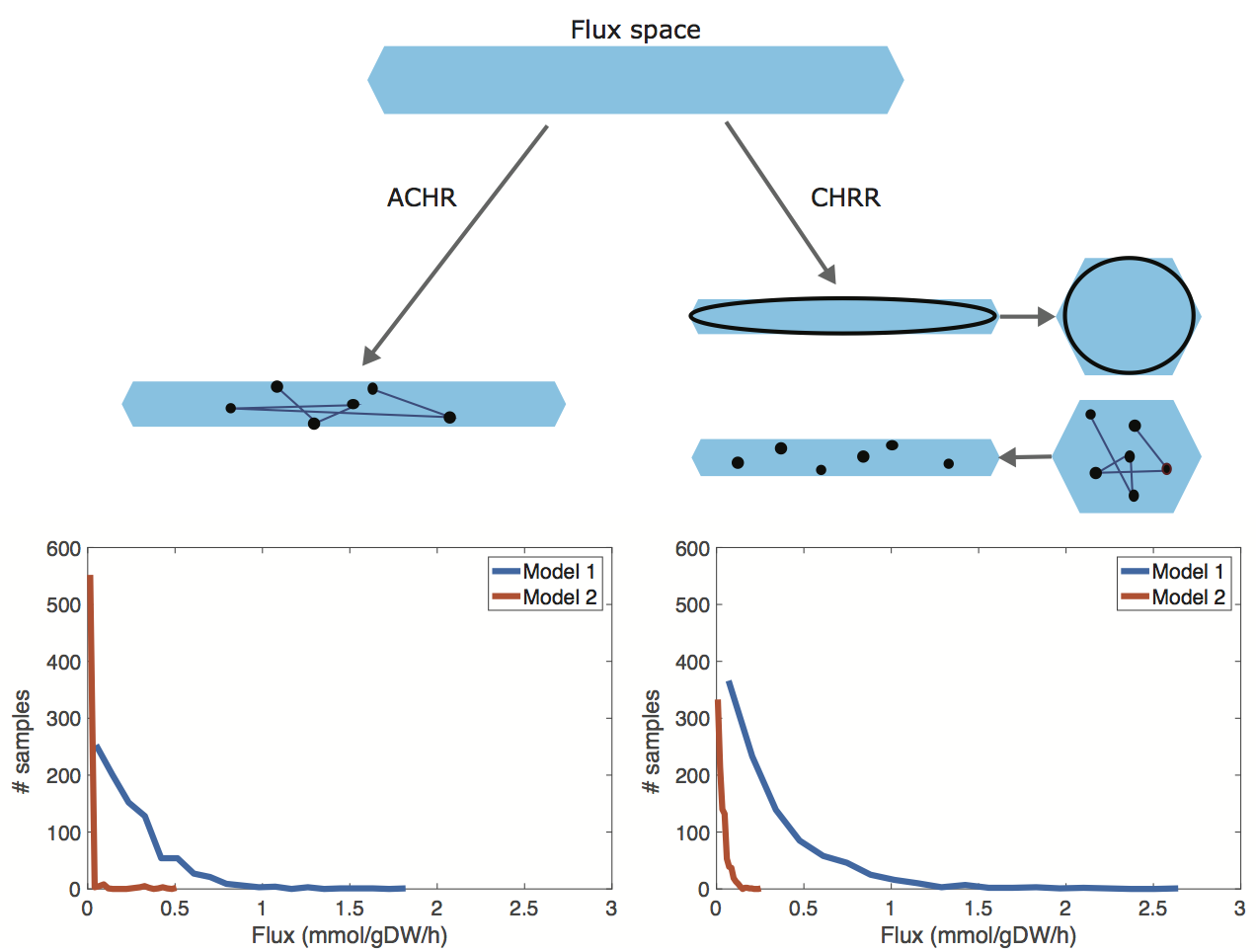}
\par\end{centering}
\caption{\label{fig:CHRR}Solution spaces from steady state fluxes are anisotropic,
that is, long in some directions and short in others. This impedes
the ability of any sampling algorithm taking a random direction to
evenly explore the full feasible set (\emph{artificial centering hit-and-run}
(ACHR) algorithm). The CHRR (\emph{coordinate hit-and-run with rounding})
algorithm first rounds the solution space based on the maximum volume
ellipsoid. Then, the rounded solution space is uniformly sampled using
a provably efficient coordinate hit-and-run random walk. Finally,
the samples are projected back onto the anisotropic feasible set.
This leads to a more distributed uniform sampling, so that the converged
sampling distributions for the selected reactions become smoother.
As an example, for both sampling distributions, the parameters were
defined as: $nSkip=8\times(dim(fluxSpace))^{2}$, $nSamples=1000$. }

\label{sampling}
\end{figure}

\pStep \label{step:uniformSampling}An unbiased characterisation
of the set of flux vectors consistent with steady state, mass balance,
and reaction bound constraints can be obtained by uniformly sampling
the \emph{feasible set} \\$\Omega\coloneqq\left\{ v\mid Sv=0;~l\leq v\leq u\right\} $.
The feasible set for sampling should be defined based on biochemically
justifiable constraints. These are the same conditions that apply
when formulating the flux balance analysis Problem (\ref{eq:FBA}),
except that there is no need to formulate a linear objective. To ensure
the sample is statistically representative of the entire feasible
set, a sufficiently large number of flux vectors and the flux vectors
must be collected randomly within the feasible set. Recently, we distributed
new software to uniformly sample feasible sets of steady state fluxes\cite{haraldsdottir_chrr:_2017}
based on a \emph{coordinate hit-and-run with rounding} (CHRR) algorithm\cite{cousins_bypassing_2014,cousins_practical_2015}
that is guaranteed to return a statistically uniform distribution
when appropriately utilised. The CHRR sampling algorithm is therefore
used by default. Figure \ref{fig:CHRR} illustrates the basics of
this algorithm.

Sampling of a model is invoked either by using the default setting
or by tailoring the parameters with more arguments to the interface.
A pragmatic approach is to first try Option (A) with the default parameters,
then check the quality of the marginal flux distribution for a subset
of reactions (see Figure \ref{fig:CHRR}). Especially for higher dimensional
models, it may be necessary to tune the parameters with Option (B).
\begin{enumerate}
\item Sampling of a mono-scale model with $\lesssim~$2000 variables
\begin{enumerate}
\item A Model that contains less than 2000 variables can usually be sampled
using the default settings: \begin{lstlisting}
>> [modelSampling, samples] = sampleCbModel(model);
\end{lstlisting}The \mcode{samples} output is an $n\times p$ matrix of sampled flux
vectors, where $p$ is the number of samples. In order to accelerate
any future rounds of sampling, use the \mcode{modelSampling} output.
This is a model storing extra variables acquired from preprocessing
the model for sampling (see Figure \ref{fig:CHRR}).
\end{enumerate}
\item Sampling of a model with $\lesssim~$10000 variables
\begin{enumerate}
\item Larger Models containing less than 10000 variables may be sampled
by tuning the optional input parameters: \begin{lstlisting}
>> [modelSampling, samples] = sampleCbModel(model, sampleFile, samplerName, options, ...
                                            modelSampling);
\end{lstlisting}The variable \mcode{sampleFile} contains the name of a \emph{.mat}
file used to save the sample vectors to disk. A string passed to \mcode{samplerName}
can be used to sample with non-default solvers. The \mcode{options}
structure contains fields that control the sampling density (\mcode{.nSkip})
and the number of samples (\mcode{.nSamples}). The total number of
\mcode{samples} returned is $p=nSkip\;\times\;nSamples$. The output
\mcode{modelSampling} may be used in subsequent rounds of sampling.
Although rounding large models is computationally demanding, the results
can be reused when sampling the same model more than once. The CHRR
algorithm provably converges to a uniform stationary sampling distribution
if enough samples are obtained and has been tested with mono-scale
metabolic models with up to $10000$ reactions. The default parameters
are set using heuristic rules to estimate a sufficiently large number
of samples, which balances this requirement against the desire to
complete the sampling procedure in a practically useful period of
time. \troubleshooting 
\end{enumerate}
\end{enumerate}

\subsubsection[Identify all genetic manipulations leading to targeted overproductions]{Identify all genetic manipulations leading to targeted overproductions
\timing $10-10^{5}~$s }

\begin{figure}
\begin{centering}
\includegraphics[width=0.7\textwidth]{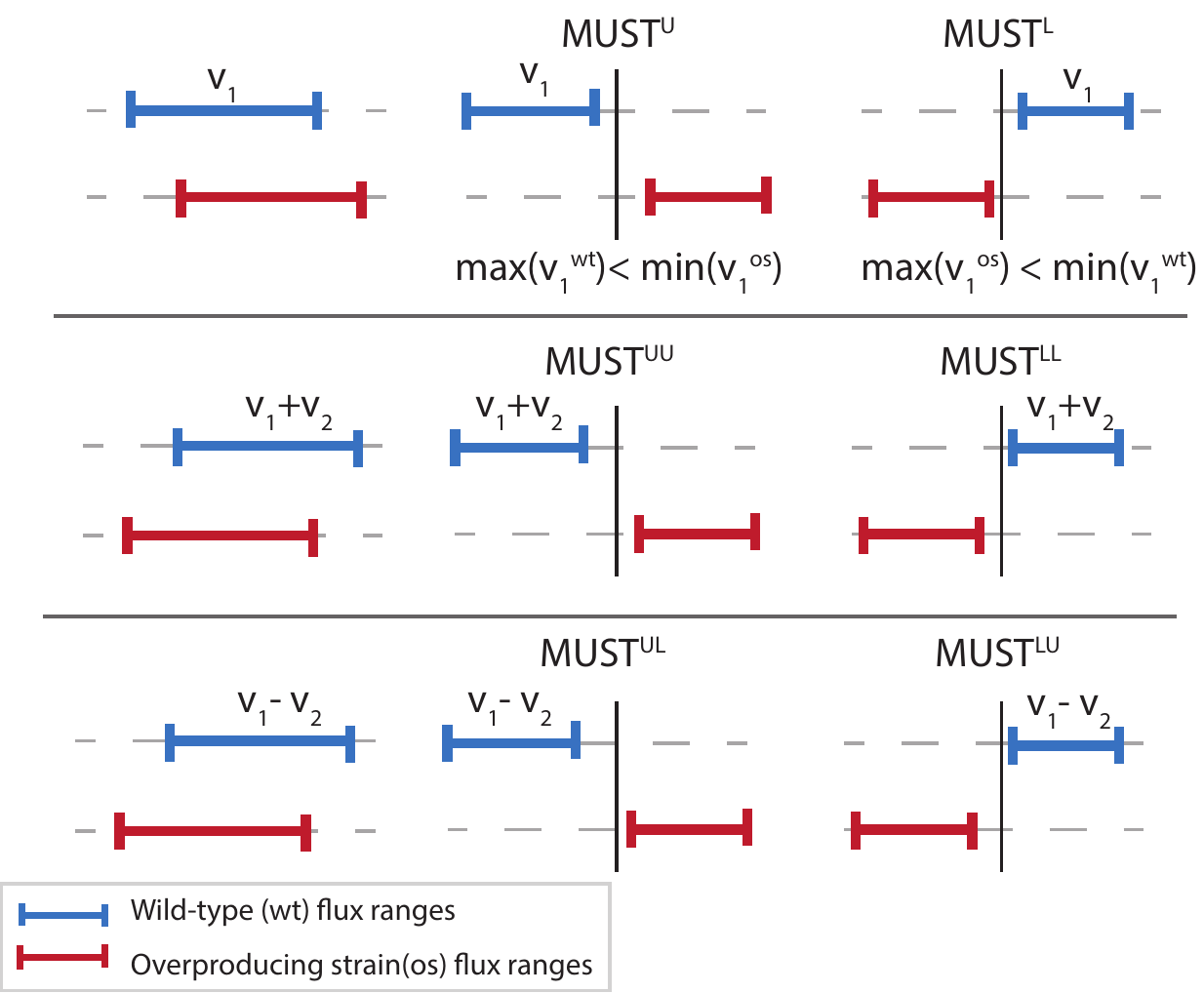}
\par\end{centering}
\caption{\label{fig:In-OptForce}In the OptForce procedure, the MUST sets are
determined by contrasting the flux ranges obtained using flux variability
analysis (FVA) of a wild-type (blue bars) and an overproducing strain
(red bars). The first order MUST sets (top panel) are denoted MUST$^{\textrm{L}}$
and MUST$^{\textrm{U}}$. For instance, a reaction belongs to the
MUST$^{\textrm{U}}$ set if the upper bound of the flux range in the
wild-type is less than the lower bound of the flux range of the overproducing
strain. The center and bottom panels show all possible second order
MUST sets. }
\end{figure}

\pStep\label{step:Select-solver} A variety of strain design algorithms\cite{maranas_optimization_2016}
are implemented within the COBRA Toolbox, including OptKnock\cite{burgard_optknock:_2003},
OptGene\cite{patil_evolutionary_2005}, Genetic Design Local Search
(GDLS\cite{lun_largescale_2009}), and OptForce\cite{ranganathan_optforce:_2010}.
While OptKnock, OptGene, and GDLS could identify gene deletion strategies,
the OptForce method can identify not only gene deletion but also up-
and down-regulation strategies. As the OptForce method is new to this
version of the COBRA Toolbox, we provide an illustrative example of
strain design using this method.

Consider the problem of finding a set of interventions of size \mcode{K}
such that when these interventions are applied to a wild-type strain,
the mutant created will produce a particular target of interest at
a higher yield than the wild-type strain. The interventions could
be knock-outs (zero out the flux for a particular reaction), up-regulations
(increase the flux for a particular reaction), or down-regulations
(decrease the flux for a particular reaction). As an example, we will
use the OptForce method to identify all genetic manipulations leading
to the overproduction of succinate in \emph{E. coli}\cite{ranganathan_optforce:_2010}.
The OptForce method consists of the following set of steps: define
the constraints for both wild-type and mutant strains, perform flux
variability analyses for both wild-type and mutant strains, find the
sets of reactions that must alter their flux in order to achieve the
desired phenotype in the mutant strain, and, finally, find the interventions
needed to ensure an increased production of the target of interest
(Steps \ref{step:-Define-theinter}-\ref{step:displayThe-reactions}).

First, select a commercial-grade solver and select the local directory
to save the generated results with:

\begin{lstlisting}
>> changeCobraSolver('gurobi', 'ALL'); 
\end{lstlisting}

\pStep Load an illustrative model that comprises only $90$ reactions,
describing the central metabolism in \emph{E. coli}\cite{antoniewicz_metabolic_2007}. 

\begin{lstlisting}
>> readCbModel('AntCore.mat'); 
\end{lstlisting}

\pStep Set the objective function to maximise the biomass reaction
(\mcode{R75}). Change the lower bounds such that \emph{E. coli} model
will be able to consume glucose, oxygen, sulfate, ammonium, citrate,
and glycerol. 

\begin{lstlisting}
>> model = changeObjective(model, 'R75', 1); 
>> for rxn = {'EX_gluc', 'EX_o2', 'EX_so4', 'EX_nh3', 'EX_cit', 'EX_glyc'}
      model = changeRxnBounds(model, rxn, -100, 'l');
   end
\end{lstlisting}

\pStep \label{step:Define-constraints}Define the constraints for
both wild-type and mutant strains:

\begin{lstlisting}
>> constrWT = struct('rxnList', {{'R75'}}, 'rxnValues', 14, 'rxnBoundType', 'b'); 
>> constrMT = struct('rxnList', {{'R75', 'EX_suc'}}, 'rxnValues', [0, 155.55], ...
                     'rxnBoundType', 'bb'); 
\end{lstlisting}

\criticalStep In this example, we provide the constraints for both
wild-type and mutant strains, but in a typical scenario the definition
of differential constraints on wild-type and mutant strains requires
additional research. This step could take a few days or weeks, depending
on the information available for the species of interest. Flux bounds
(i.e., uptake rate and minimum biomass yield target) are required
inputs. New experiments might be required to be performed in addition
to the literature curation task in order to obtain such data. Assumptions
may also be made when describing the phenotypes of both strains, which
will reduce the dependency on literature curation. It is important
that the two strains are sufficiently different in order to be able
to anticipate differences in reaction ranges.

\pStep \label{sec: FVA-Run-flux}Performing flux variability analysis
for both wild-type and mutant strains with:

\begin{lstlisting}
>> [minFluxesW, maxFluxesW, minFluxesM, maxFluxesM] = FVAoptForce(model, constrWT, constrMT); 
>> disp([minFluxesW, maxFluxesW, minFluxesM, maxFluxesM]); 
\end{lstlisting}

\pStep The MUST sets are the sets of reactions that must increase
or decrease their flux in order to achieve the desired phenotype in
the mutant strain. As shown in Figure \ref{fig:In-OptForce}, the
first order MUST sets are \mcode{MustU} and \mcode{MustL} while
second order MUST sets are denoted as \mcode{MustUU}, \mcode{MustLL},
and \mcode{MustUL}. After parameters and constraints are defined,
the functions \mcode{findMustL} and \mcode{findMustU} are run to
determine the \mcode{mustU} and \mcode{mustL} sets, respectively.
Define an ID of the run with:

\begin{lstlisting}
>> runID = 'TestoptForceM'; 
\end{lstlisting}

Each time the MUST sets are determined, folders are generated to read
inputs and store outputs, i.e., reports. These folders are located
in the directory defined by the uniquely defined \mcode{runID}.

\pStep In order to find the first order MUST sets, constraints should
be defined: 

\begin{lstlisting}
>> constrOpt = struct('rxnList', {{'EX_gluc', 'R75', 'EX_suc'}}, 'values', [-100; 0; 155.5]); 
\end{lstlisting}

\pStep \label{step:must-set}The first order MUST set \mcode{MustL}
is determined by running:

\begin{lstlisting}
>> [mustLSet, pos_mustL] = findMustL(model, minFluxesW, maxFluxesW, ...
                                     'constrOpt', constrOpt, 'runID', runID); 
\end{lstlisting}

If \mcode{runID} is set to \mcode{'TestoptForceL'}, a folder \emph{TestoptForceL}
is created, in which two additional folders \emph{InputsMustL} and
\emph{OutputsMustL} are created. The \emph{InputsMustL} folder contains
all the inputs required to run the function \mcode{findMustL}, while
the \emph{OutputsMustL} folder contains the \mcode{mustL} set found
and a report that summarises all the inputs and outputs. In order
to maintain a chronological order of computational experiments, the
report is timestamped.

\pStep Display the reactions that belong to the \mcode{mustL} set
using:

\begin{lstlisting}
>> disp(mustLSet)
\end{lstlisting}

\pStep The first order MUST set \mcode{MustU} is determined by running:

\begin{lstlisting}
>> [mustUSet, pos_mustU] = findMustU(model, minFluxesW, maxFluxesW, ...
                                     'constrOpt', constrOpt, 'runID', runID); 
\end{lstlisting}The results are stored and available in a format analogous to the
\mcode{mustL} set. The reactions that belong to the \mcode{mustU}
may be displayed in the same way as \mcode{mustL}.

\pStep Define the reactions that will be excluded from the analysis.
The reactions found in the previous step as well as exchange reactions
shall be included.

\begin{lstlisting}
>> constrOpt = struct('rxnList', {{'EX_gluc', 'R75', 'EX_suc'}}, 'values', [-100, 0, 155.5]'); 
>> exchangeRxns = model.rxns(cellfun(@isempty, strfind(model.rxns, 'EX_')) == 0); 
>> excludedRxns = unique([mustUSet; mustLSet; exchangeRxns]); 
\end{lstlisting}

\pStep The second order MUST set \mcode{MustUU} can be determined
by running:

\begin{lstlisting}
>> [mustUU, pos_mustUU, mustUU_linear, pos_mustUU_linear] = findMustUU(model, minFluxesW, ...
                      maxFluxesW, 'constrOpt', constrOpt, ...
                      'excludedRxns', excludedRxns,'runID', runID); 
\end{lstlisting}The results are stored and available in a format analogous to the
\mcode{mustL} set. The reactions of the \mcode{mustUU} can be displayed
using the \mcode{disp} function.

\pStep \label{step:-Repeat-the}Repeat the above steps to determine
the second order MUST sets \mcode{MustLL} and \mcode{MustUL} by
using the functions \mcode{findMustLL} and \mcode{findMustUL} respectively.
The results are stored and available in a format analogous to the
\mcode{mustL} set. In the present example, \mcode{mustLL} and \mcode{mustUL}
are empty sets.\troubleshooting

\pStep \label{step:-Define-theinter}In order to find the interventions
needed to ensure an increased production of the target of interest,
define the \mcode{mustU} set as the union of the reactions that must
be up-regulated in the first and second order MUST sets. Similarly,
\mcode{mustL} may be defined.

\begin{lstlisting}
>> mustU = unique(union(mustUSet, mustUU)); 
>> mustL = unique(union(mustLSet, mustLL)); 
\end{lstlisting}

\pStep Define the number of interventions \mcode{k} allowed, the
maximum number of sets to find \mcode{nSets}, the reaction producing
the metabolite of interest \mcode{targetRxn} (in this case, succinate),
and the constraints on the mutant strain \mcode{constrOpt}.

\begin{lstlisting}
>> k = 1; nSets = 1;  targetRxn = 'EX_suc';
>> constrOpt = struct('rxnList', {{'EX_gluc','R75'}}, 'values', [-100, 0]); 
\end{lstlisting}

\pStep \label{step:optForce}Run the OptForce algorithm and display
the reactions identified by \mcode{optForce} with:\begin{lstlisting}
>> [optForceSets, posoptForceSets, typeRegoptForceSets, flux_optForceSets] = optForce(model, targetRxn, mustU, mustL, minFluxesW, maxFluxesW, minFluxesM, maxFluxesM, 'k', k, 'nSets', nSets, 'constrOpt', constrOpt, 'runID', runID); 
>> disp(optForceSets)
\end{lstlisting}

\pStep In order to find non-intuitive solutions, increase the number
of interventions \mcode{k} and exclude the \mcode{SUCt} reaction
from up-regulations. Increase \mcode{nSets} to find the $20$ best
sets. Change the \mcode{runID} to save this second result in a separate
folder from the previous result, then run \mcode{optForce} again
as in Step \ref{step:optForce}.

\begin{lstlisting}
>> k = 2; nSets = 20; runID = 'TestoptForceM2'; 
>> excludedRxns = struct('rxnList', {{'SUCt'}}, 'typeReg','U'); 
\end{lstlisting}

\pStep \label{step:displayThe-reactions}The reactions determined
by \mcode{optForce} can be displayed using \mcode{disp(optForceSets)}.
The complete set of predicted interventions can be found in the folders
created inside the \mcode{runID} folder in which inputs and outputs
of \mcode{optForce} and associated \mcode{findMust*} functions are
stored.  The input folders \emph{InputsFindMust{*}} contain \emph{.mat}
files for running the functions to solve each one of the bilevel optimisation
problems. The output folders \emph{OutputsFindMust{*}} contain results
of the algorithms (saved as \emph{.xls} and \emph{.txt} files) as
well as a report (\emph{a .txt} file) that summarises the outcome
of the steps performed during the execution of each function. The
\mcode{optForce} algorithm will find sets of reactions that should
increase the production of a specified target. The first sets found
should be the best ones because the production rate will be the highest.
The last ones will be the worst, as the production rate is the lowest.
\caution Be aware that some sets may not guarantee a minimum production
rate for a target, so check the minimum production rate, e.g., using
the function \mcode{testoptForceSol}.

\subsubsection[Atomically resolve a metabolic reconstruction]{Atomically resolve a metabolic reconstruction\timing $10-10^{5}$
s}

\pStep \label{step:atomicallyResolve}In most genome-scale metabolic
models, it is not explicit that the stoichiometric matrix represents
a network of biochemical reactions. It is implicit that each row of
the stoichiometric matrix corresponds to some molecular species, but
when computing properties of the model, the atomic structure of each
molecular species is not represented. It is also implicit that each
column of the stoichiometric matrix corresponds to some biochemical
reaction. However, when computing properties of the model, the mechanisms
of the underlying biochemical reaction, in terms of the structures
of the metabolites and the atomically resolved chemical transformations
that take place, are not represented. Recent developments in genome-scale
metabolic modelling have generated genome-scale metabolic reconstructions
where the molecular structures are specified\cite{haraldsdottir_comparative_2014}
and the reaction mechanisms are represented by atom mappings between
substrate and product atoms\cite{preciat_gonzalez_comparative_2017,brunk_recon_2017}. 

An atom mapping is a one-to-one association between a substrate atom
and a product atom. An instance of a chemical reaction may be represented
by a set of atom mappings, with one atom mapping between each substrate
and product atom. A single chemical reaction can admit multiple chemically
equivalent atom mappings when chemically equivalent atoms are present
in a substrate, a product, or both. Therefore, each chemical reaction
may be represented by one set, or multiple chemically equivalent sets,
of atom mappings. Together, a set of atom mappings for a chemical
reaction specify key aspects of the reaction mechanism, e.g., chemical
bond change, breakage, and formation. The Virtual Metabolic Human
database (VMH, \url{http://vmh.life}) provides metabolites chemical
structures and atom mapped reactions for $9,610$ reactions in Recon3D\cite{brunk_recon_2017}
and $4,831$ metabolites from Recon3D\cite{brunk_recon_2017} and
the human gut microbiota\cite{magnusdottir_generation_2017}. Metabolite
structures are provided in canonically ordered MOL and SMILES formats.
Atom mapping data are provided in both RXN and SMILES formats. This
explicit representation of metabolite and reaction structure offers
the possibility of a broader range of biological, biomedical and biotechnological
applications than with stoichiometry alone.

In order to obtain chemical structures for each metabolite, there
are three main ways:
\begin{enumerate}
\item Use chemoinformatics software tools
\begin{enumerate}
\item Suitable cheminformatics software tools\cite{haraldsdottir_comparative_2014}
may be used to automatically obtain metabolite identifiers in metabolic
network reconstructions and download the corresponding structure from
a database.
\end{enumerate}
\item Manually interrogate metabolic databases
\begin{enumerate}
\item Databases such as VMH (\url{http://vmh.life}), PubChem\cite{kim_pubchem_2016},
KEGG\cite{kanehisa_kegg:_2000}, ChEBI\cite{hastings_chebi_2013},
LMSD\cite{sud_lmsd:_2007}, BioPath database\cite{forster_system_2002},
ChemSpider database\cite{williams_chemspider_2010}, HMDB\cite{wishart_hmdb:_2007},
etc provide chemical structures for metabolites in a network.
\end{enumerate}
\item Manually draw structures of metabolites
\begin{enumerate}
\item Based on chemical knowledge, one could manually draw structures of
metabolites using tools such as ChemDraw (PerkinElmer, \url{https://perkinelmer.com/ChemDraw}).
\end{enumerate}
\end{enumerate}
\pStep In order to obtain an atom mapping for a metabolic reaction,
the reaction stoichiometry and the chemical structures of the corresponding
metabolites must be available. To obtain atom mappings, there are
three main options:
\begin{enumerate}
\item Use software tools for prediction of atom mappings
\begin{enumerate}
\item A comparative study has been performed using Recon3D as a test case\cite{preciat_gonzalez_comparative_2017}\textcolor{black}{.}
 Due to its accuracy and availability, Reaction Decoder Tool (RDT\cite{rahman_reaction_2016})
is considered being the most suitable algorithm to atom map the reactions
from a genome-scale metabolic network. Nevertheless, note that the
Canonical Labelling for Clique Approximation (CLCA\cite{kumar_clca:_2014})
algorithm can map reactions with explicit hydrogen atoms for fully
protonated reactions, while RDT can only atom map reactions with implicit
hydrogen atoms. 
\end{enumerate}
\item Manually interrogate metabolic databases
\begin{enumerate}
\item Databases such as BioPath\cite{forster_system_2002} and KEGG RPAIR\cite{shimizu_generalized_2008}
disseminate manually curated atom mappings. VMH (\url{http://vmh.life})
also contains manually curated atom mappings for a subset of human
metabolic reactions.
\end{enumerate}
\item Manually draw atom mappings
\begin{enumerate}
\item Based on chemical knowledge, one could draw atom mappings using tools
such as ChemDraw (PerkinElmer, \url{https://perkinelmer.com/ChemDraw}).
\end{enumerate}
\end{enumerate}
\pStep\textcolor{black}{\label{Step:-Given-a} Given a }\mcode{model}\textcolor{black}{{}
structure and the directory containing the chemical structure files
(}\mcode{molFileDir}\textcolor{black}{) in MDL MOL file format},
RDT can be invoked to atom map a metabolic model using:\begin{lstlisting}  
>> balancedRxns = obtainAtomMappingsRDT(model, molFileDir, outputDir, ...
                                        maxTime, standardiseRxn);
\end{lstlisting}This\textcolor{black}{{} function computes atom mapping data for the
balanced and unbalanced reactions in the metabolic network and saves
it in the }\mcode{outputDir}\textcolor{black}{{} directory}. The optional
\mcode{maxTime} parameter sets a runtime limit for atom mapping of
a reaction. If \mcode{standardiseRxn == 1}, then atom mappings are
also canonicalised, which is necessary in order to obtain a consistent
interoperable set of atom mappings for certain applications, e.g.,
computation of conserved moieties in Step \ref{Step:First-compute}.
The output \mcode{balancedRxns} contains the balanced atom mapped
metabolic reactions.\troubleshooting 

\pStep\label{Step:First-compute} \label{subsec:Identify-conserved-moieties}With
a set of canonicalised atom mappings for a metabolic network, the
set of linearly independent conserved moieties for a metabolic network
can be identified\textcolor{black}{\cite{haraldsdottir_identification_2016}}.
Each of these conserved moieties corresponds to a molecular substructure
(set of atoms in a subset of a molecule)\textcolor{black}{{} }whose
structure remains invariant despite all the chemical transformations
in a given network. A conserved moiety is a group of atoms that follow
identical paths through metabolites in a metabolic network. Similarly
to a vector in the (right) nullspace of a stoichiometric matrix that
corresponds to a pathway (see Step \ref{step:minSpanPathways}), a
conserved moiety corresponds to a vector in the left nullspace of
a stoichiometric matrix.  Metabolic networks are hypergraphs\cite{klamt_hypergraphs_2009},
while most moiety subnetwork are graphs. Therefore\textcolor{black}{{}
}conserved moieties have both biochemical and mathematical significance
and once computed, can be used for a wide variety of applications.
Given a metabolic network of exclusively mass balanced reactions,
one can identify conserved moieties by a graph theory analysis of
its atom transition network\textcolor{black}{\cite{haraldsdottir_identification_2016}.}

First compute an atom transition network for a metabolic network using:

\begin{lstlisting}  
>> ATN = buildAtomTransitionNetwork(model, rxnfileDir);
\end{lstlisting}\textcolor{black}{where }\mcode{rxnfileDir}\textcolor{black}{{} is
a directory containing only atom mapped files from balanced reactions,
which can be obtained as explained in Step }\ref{Step:-Given-a}\textcolor{black}{.
}The output \mcode{ATN} is a structure with several fields: \mcode{.A}\textcolor{black}{{}
}is a $p\times q$ sparse incidence matrix for the atom transition
network, where $p$ is the number of atoms and $q$ is the number
of atom transitions, \mcode{.mets} is a $p\times1$ cell array of
metabolite identifiers to link each atoms to its corresponding metabolites,
\mcode{.rxns} is a $q\times1$ cell array of reaction identifiers
to link atom transitions to their corresponding reactions, and \mcode{.elements}
is a $p\times1$ cell array of element symbols for atoms in \mcode{.A}.

\criticalStep All the RXN files needed to compute the atom transition
network must be in a canonical format. This can be achieved by setting
\mcode{standardiseRxn = 1}.

\pStep \label{step:identifyConservedMoieties-end}In order to identify
the conserved moieties in the metabolic network, invoke:

\begin{lstlisting}
>> [L, M, moietyFormulas] = identifyConservedMoieties(model, ATN); 
\end{lstlisting}

where \mcode{L} represents the conserved moieties in the metabolic
network. That is, \mcode{L} is an $m\times r$ matrix of $r$ moiety
vectors in the left null space of the stoichiometric matrix, \mcode{M}
is the $u\times v$ incidence matrix of the moiety supergraph in which
each connected component is a moiety graph, and \mcode{moietyFormulas}
is an $m\times r$ cell array with chemical formulas of the computed
moieties.

\subsubsection[Thermodynamically constrain a metabolic model]{Thermodynamically constrain a metabolic model\timing$1-10^{3}$
s}

\pStep \label{step:componentContribution}In flux balance analysis
of genome-scale stoichiometric models of metabolism, the principal
constraints are uptake or secretion rates, the steady state mass conservation
assumption, and reaction directionality. The COBRA Toolbox extension
\emph{vonBertalanffy}\cite{fleming_von_2011} is a set of methods
for integration of thermochemical data with constraint-based models\cite{fleming_quantitative_2009,haraldsdottir_quantitative_2012,noor_consistent_2013}
as well as application of thermodynamic laws to increase the physicochemical
fidelity of constraint-based modelling predictions\cite{fleming_variational_2012}.
A full exposition of the method to thermodynamically constrain a genome-scale
metabolic model is beyond the scope of this protocol. Therefore, only
several key steps are highlighted. 

Given a set of experimentally derived \mcode{training_data} on standard
transformed Gibbs energies of formation, a state-of-the art quantitative
estimation of standard Gibbs energy of formation for metabolites with
similar chemical substructures can be obtained using an implementation
of the component contribution method\cite{noor_consistent_2013}.
We assume that the input \mcode{model} has been anatomically resolved
as described in Steps \ref{step:atomicallyResolve}-\ref{step:identifyConservedMoieties-end}.
Access to a compendium of stoichiometrically consistent metabolite
structures\cite{haraldsdottir_comparative_2014,haraldsdottir_identification_2016}
is a prerequisite. The component contribution method is then invoked
as follows:

\begin{lstlisting}   
>> model = componentContribution(model, training_data); 
\end{lstlisting}The \mcode{model.DfG0} field gives the estimated standard Gibbs energy
of formation for each metabolite in the model with \mcode{model.DfG0_Uncertainty}
field expressing the uncertainty in these estimates, which is smaller
for metabolites structurally related to metabolites in the training
set. All thermodynamic estimates are given in units of kJ/mol.

\pStep The standard Gibbs energy of formation for each metabolite
must be transformed according to the environment of each compartment
of the model\cite{haraldsdottir_quantitative_2012}, i.e., the temperature,
pH, ionic strength and electrical potential specific to each compartment.
Then the thermodynamic properties of reactions are estimated, given
\mcode{model.concMin} and \mcode{model.concMax} where one can supply
lower and upper bounds on compartment-specific metabolite concentrations
(mol/L), which may be achieved with:

\begin{lstlisting}   
>> model = setupThermoModel(model, confidenceLevel); 
\end{lstlisting}In the output, field \mcode{.DfGt0} of \mcode{model} gives the estimated
standard transformed Gibbs energy of formation for each metabolite
and \mcode{.DrGt0} gives the estimated standard transformed Gibbs
energy for each reaction. Subject to a \mcode{confidenceLevel} specified
as an input, the upper and lower bounds on standard transformed Gibbs
energy for each reaction are provided in \mcode{.DrGtMin} and \mcode{.DrGtMax}
respectively.

\criticalStep In a multi-compartmental model, this step must be done
for an entire network at once in order to ensure that thermodynamic
potential differences, arising from differences in the environment
between compartments, are properly taken into account. See\cite{haraldsdottir_quantitative_2012}
for a theoretical justification for this assertion.

\pStep Reaction directionality may be quantitatively assigned based
on the aforementioned thermodynamic estimates with:\begin{lstlisting}   
>> [modelThermo, directions]= thermoConstrainFluxBounds(model, confidenceLevel, DrGt0_Uncertainty_Cutoff);
\end{lstlisting}If \mcode{model.DrGtMax(j) < 0}, then the $j^{th}$ reaction is assigned
to be forward, and if \mcode{model.DrGtMin(j) > 0} then the $j^{th}$
reaction is assigned to be reverse, unless the uncertainty in estimation
of standard transformed reaction Gibbs energy exceeds a specified
cutoff (\mcode{DrGt0_Uncertainty_Cutoff}). In this case, the qualitatively
assigned reaction directionality, specified together by \mcode{model.lb(j)}
and \mcode{model.ub(j)}, takes precedence. The \mcode{directions}
output provides a set of boolean vector fields that can be used to
analyse the effect of qualitatively versus quantitatively assigning
reaction directionality using thermochemical parameters.

\pStep \label{step:thermodynamicallyConstraint}Thermodynamically
constrained flux balance analysis may then be invoked by disallowing
flux around stoichiometrically balanced cycles, also known as loops,
using the \mcode{allowLoops} parameter to \mcode{optimizeCbModel}
with:

\begin{lstlisting}   
>> allowLoops = 0;
>> solution = optimizeCbModel(model, [], [], allowLoops);
\end{lstlisting}The \mcode{solution} structure is the same as for flux balance analysis
(see Problem (\ref{eq:FBA})), except that this \mcode{solution}
satisfies additional constraints that ensure the predicted steady
state flux vector is thermodynamically feasible\cite{beard_energy_2002}.
The solution satisfies energy conservation and the second law of thermodynamics\cite{qian_thermodynamics_2005}.

\subsubsection[Convert a flux balance model into a kinetic model]{Convert a flux balance model into a kinetic model \timing$1-10^{3}$
s}

\pStep \label{step:convertFBAtoKinetic}In order to analyse biochemical
networks at genome scale, systems biologists often use a linear optimisation
technique called flux balance analysis (FBA). Linear approximation
to known nonlinear biochemical reaction network function is sufficient
to get biologically meaningful predictions in some situations. However,
there are many biochemical processes where a linear approximation
is insufficient, which motivates the quest for developing variational
kinetic modelling\cite{fleming_integrated_2010,schellenberger_elimination_2011,soh_network_2010}.
Certain conditions are required to be met in order to generate a
kinetic model that is internally consistent. First we describe those
conditions, then we demonstrate how to ensure that they are met. Consider
a biochemical network with $m$ molecular species and $n$ reversible
reactions. We define forward and reverse \emph{stoichiometric matrices},
$F,R\in\mathbb{\mathbb{Z}}_{+}^{m\times n}$, respectively, where
$F_{ij}$ denotes the \emph{stoichiometry} of the $i^{th}$ molecular
species in the $j^{th}$ forward reaction and $R_{ij}$ denotes the
stoichiometry of the $i^{th}$ molecular species in the $j^{th}$
reverse reaction. We assume that the network of reactions is stoichiometrically
consistent\cite{gevorgyan_detection_2008}, that is, there exists
at least one positive vector $l\in\mathbb{R}_{++}^{m}$ satisfying
$(R-F)^{T}l=0$. Equivalently, we require that \emph{every reaction
conserves mass}. The matrix $N:=R-F$ represents net reaction stoichiometry
and may be viewed as the incidence matrix of a directed hypergraph\cite{klamt_hypergraphs_2009}.
We assume that there are less molecular species than there are net
reactions, that is $m<n$. We assume the cardinality of each row of
$F$ and $R$ is at least one, and the cardinality of each column
of $R-F$ is at least two. The matrices $F$ and $R$ are sparse and
the particular sparsity pattern depends on the particular biochemical
network being modelled. Moreover, we assume that $\text{rank}([F,R])=m$,
which is a requirement for kinetic consistency\cite{fleming_conditions_2016}.

\subsubsection[Compute a non-equilibrium kinetic steady state]{Compute a non-equilibrium kinetic steady state \timing$1-10^{3}$
s}

\pStep \label{step:nonEquilibriumSteadyState}Let $c\in\mathbb{R}_{++}^{m}$
denote a variable vector of molecular species concentrations. Assuming
constant nonnegative elementary kinetic parameters $k_{f},k_{r}\in\mathbb{R}_{+}^{n}$,
we assume \emph{elementary reaction kinetics} for forward and reverse
elementary reaction rates as $s(k_{f},c):=\exp(\ln(k_{f})+F^{T}\ln(c))$
and $r(k_{r},c):=\exp(\ln(k_{r})+R^{T}\ln(c))$, respectively, where
$\exp(\cdot)$ and $\ln(\cdot)$ denote the respective component-wise
functions\cite{artacho_accelerating_2015,fleming_conditions_2016}.
Then, the deterministic dynamical equation for time evolution of molecular
species concentration is given by 
\begin{eqnarray}
\frac{dc}{dt} & \equiv & N(s(k_{f},c)-r(k_{r},c))\label{eq:dcdt2}\\
 & = & N\left(\exp(\ln(k_{f})+F^{T}\ln(c)\right)-\exp\left(\ln(k_{r})+R^{T}\ln(c))\right)=:-f(c).\nonumber 
\end{eqnarray}
A vector $c^{*}$ is a \emph{steady state} if and only if it satisfies
$f(c^{*})=0,$ leading to the nonlinear system 

\[
f(x)=0.
\]

There are many algorithms that can handle this nonlinear system by
minimising a nonlinear least-squares problem; however, particular
features of this mapping, such as sparsity of stoichiometric matrices
$F$ and $R$ and non-unique local zeros of mapping $f$, motivates
the quest for developing several algorithms for efficient dealing
with this nonlinear system. A particular class of such mappings, called
duplomonotone mapping, was studied for biochemical networks\cite{artacho_globally_2014}
and three derivative-free algorithms for finding zeros of strongly
duplomonotone mappings were introduced. Further, it is shown that
the function $\|f\left(x\right)\|^{2}$ can be rewritten as a difference
of two convex functions that is suitable to be minimised with DC programming
methods\cite{artacho_accelerating_2015}. Therefore, a DC algorithm
and its acceleration with adding a line search technique were proposed
for finding a stationary point of $\|f\left(x\right)\|^{2}$. Since
the mapping $f$ has locally non-unique solutions, it does not satisfy
classical assumptions (e.g., nonsingularity of the Jacobian) for convergence
theory. As a result, it was proved that the mapping satisfies the
so-called H\"older metric subregularity assumption\cite{ahookhosh_local_2017}
and an adaptive Levenberg-Marquardt method was proposed to find a
solution of this nonlinear system if the starting point is close enough
to a solution. In order to guarantee the convergence of the Levenberg-Marquardt
method with arbitrary starting point, it is combined with globalisation
techniques such as line search or trust-region, which leads to computationally
efficient algorithms. 

Compute a non-equilibirum kinetic steady state by running the function
\mcode{optimizeVKmodels}. The mandatory inputs for computing steady
states are a model \mcode{vKModel} containing $F$ and $R$, the
name of a solver to solve the nonlinear system, an initial point \mcode{x0},
and parameters for the considered solvers. For example, for specifying
a solver, we write \mcode{solver = 'LMTR';}. Optional parameters
for the selected algorithm may be given to \mcode{optimizeVKmodels}
by the \mcode{params} struct as follows

\begin{lstlisting}
>> params.MaxNumIter = 1000; params.adaptive = 1; params.kin = kin;
\end{lstlisting}Otherwise, the selected algorithm will be run with the default parameters
assigned for each algorithm. Running the function \mcode{optimizeVKmodels}
is done by typing 

\begin{lstlisting}
>> output = optimizeVKmodels(vKModel, solver, x0, params);
\end{lstlisting}The \mcode{output} struct contains information related to the execution
of the solver.

\subsubsection[Compute a moiety conserved non-equilibrium kinetic steady state]{Compute a moiety conserved non-equilibrium kinetic steady state
\timing$1-10^{3}$ s}

\pStep \label{step:moietyNonEquilibriumSteadyState}Let us note that
a vector $c^{*}$ is a steady state of the biochemical system if and
only if 
\[
s(k_{f},c^{*})-r(k_{r},c^{*})\in\mathcal{N}(N),
\]
where $\mathcal{N}(N)$ denotes the null space of $N$. Therefore,
the set of steady states $\Omega=\left\{ c\in\mathbb{R}_{++}^{m}|\,f(c)=0\right\} $
is unchanged if the matrix $N$ is replaced by a matrix $\bar{N}$
with the same kernel. Suppose that $\bar{N}\in\mathbb{Z}^{r\times n}$
is the submatrix of $N$ whose rows are linearly independent; then
$\mathrm{rank}\left(\bar{N}\right)=\mathrm{rank}(N)\eqqcolon r.$
If one replaces $N$ by $\bar{N}$ and transforms~(\ref{eq:dcdt2})
to logarithmic scale, and by letting $x\coloneqq\ln(c)\in\mathbb{R}^{m}$,
$k\coloneqq[\ln(k_{f})^{T},\,\ln(k_{r})^{T}]^{T}\in\mathbb{R}^{2n}$,
then the right-hand side of~(\ref{eq:dcdt2}) is equal to the function
\begin{equation}
\bar{f}(x):=\left[\bar{N},-\bar{N}\right]\exp\left(k+[F,\,R]^{T}x\right),\label{eq:f(x)}
\end{equation}
where $\left[\,\cdot\thinspace,\cdot\,\right]$ stands for the horizontal
concatenation operator. Let $L\in\mathbb{R}^{m-r,m}$ denote a basis
for the left nullspace of $N$, which implies $LN=0$. We have $\mathrm{rank}(L)=m-r$.
We say that the system satisfies \emph{moiety conservation} if for
any initial concentration $c_{0}\in\mathbb{R}_{++}^{m}$,
\[
L\,c=L\,\mathrm{exp}(x)=l_{0},
\]
where $l_{0}\in\mathbb{R}_{++}^{m}.$ It is possible to compute $L$
such that each corresponds to a structurally identifiable conserved
moiety in a biochemical network\textcolor{black}{\cite{haraldsdottir_identification_2016}}.
The problem of finding the \emph{moiety conserved steady state} of
a biochemical reaction network is equivalent to solving the nonlinear
system of equations 
\begin{equation}
h(x):=\left(\begin{array}{c}
\bar{f}(x)\\
L\,\mbox{exp}(x)-l_{0}
\end{array}\right)=0.\label{eq:steadyStateEquation}
\end{equation}
Among algorithms mentioned in the previous section, the local and
global Levenberg-Marquardt methods\cite{ahookhosh_local_2017} are
designed to compute either a solution of the nonlinear system (\ref{eq:steadyStateEquation})
or a stationary point of the merit function $\frac{1}{2}\|h\left(x\right)\|^{2}$.
The computation of a moiety conserved non-equilibrium kinetic steady
state is made by running the \mcode{optimizeVKmodels} function in
the same way as in previous section. A model \mcode{vKModel} containing
$F$ and $R$, $L$ and $l_{0}$ is then passed to \mcode{optimizeVKmodels}
together with the name of one of the Levenberg-Marquardt solvers.

\subsubsection[Human metabolic network visualisation with ReconMap]{Human metabolic network visualisation with ReconMap\timing $1-10^{2}$
s}

\pStep \label{step: ReconMap}The visualisation of biochemical pathways
is an important tool for biologically interpreting predictions generated
by constraint-based models. It can be an invaluable aid for developing
an understanding of the biological meaning implied by a prediction.
Biochemical network maps permit the visual integration of model predictions
with the underlying biochemical context. Patterns that are very difficult
to appreciate in a vector can often be much better appreciated by
studying a generic map contextualised with model predictions. Genome-scale
biochemical network visualisation is particularly demanding. No currently
available software satisfies all of the requirements that might be
desired for visualisation of predictions from genome-scale models.
Automatic layout of genome-scale biochemical networks is insufficiently
developed to generate an aesthetically pleasing map, yet manual layout
of such maps is very labour intensive and there is no global reference
coordinate system for such maps, so each human might layout a global
map differently. Software applications for graph visualisation are
often not suited to displaying metabolic hypergraphs\cite{shannon_cytoscape:_2003}.
Client-server software models have to trade off between highly interactive
display of subsystem maps\cite{king_escher:_2015} and less interactive
display of genome-scale maps\cite{gawron_minervaplatform_2016}. An
additional challenge with genome-scale models is that there is too
much detail to visually appreciate if an entire genome-scale map is
visualised at once, necessitating the application of techniques to
dimensionally reduce the presentation, e.g., semantic zooming\cite{kuperstein_navicell:_2013}.
With these caveats in mind, we present a method for genome-scale visualisation
of human metabolic network predictions using ReconMap 2.01\cite{noronha_reconmap:_2017},
a manual layout of the reactions in the human metabolic reconstruction
Recon 2.04\cite{thiele_community-driven_2013}, visualised with the
Molecular Interaction NEtwoRk visualisation (MINERVA\cite{gawron_minervaplatform_2016}),
a stand-alone web service built on the Google Maps (Google Inc.) application
programming interface, that enables low latency web display and navigation
of genome-scale molecular interaction networks. 

Visualisation of context-specific predictions in ReconMap via a web
browser depends on access to a server running MINERVA, which requests
user credentials for remote access. Public access to this server is
provided free of charge. To request user credentials, navigate with
a web browser to \url{http://vmh.life/#reconmap}, select ADMIN (bottom
left), and click on \textit{Request an account} to send an email to
the MINERVA team and subsequently receive your user credentials.

\pStep In order to prepare for remote access from within MATLAB,
load the details of the MINERVA instance on the remote server, which
are provided within the COBRA Toolbox during installation, then add
your user credentials to it: \begin{lstlisting}
>> load('minerva.mat');
>> minerva.login = 'username'; 
>> minerva.password = 'password'; minerva.map = 'ReconMap-2.01';
\end{lstlisting}

\pStep Load a human metabolic model into MATLAB with:\begin{lstlisting}
>> model = readCbModel('Recon2.v04.mat');
\end{lstlisting}

\pStep Change the objective function to maximise ATP production through
complex V (ATP synthase, \mcode{'ATPS4m'}) in the electron transport
chain with:

\begin{lstlisting}
>> modelATP = changeObjective(model, 'ATPS4m'); 
\end{lstlisting}

\pStep Although the optimal objective value of the flux balance analysis
Problem (\ref{eq:FBA}) is unique, the optimal flux vector itself
is most likely not. When visualising a flux vector, it is important
that a unique solution to some optimisation problem is displayed.
For example, we can predict a unique network flux by regularising
the flux balance analysis Problem (\ref{eq:FBA}) by redefining $\rho(v)\coloneqq c^{T}v-\frac{\sigma}{2}v^{T}v$
and $\sigma=10^{-6}$ (see Step \ref{step:FBA}). In order to obtain
a unique optimal flux vector, run:\begin{lstlisting}
>> solution = optimizeCbModel(modelATP, 'max', 1e-6);
\end{lstlisting}

\pStep\label{step:reconMap-layout} Build the context-specific overlay
of a flux vector on ReconMap by instructing the COBRA Toolbox to communicate
with the remote MINERVA server using: \begin{lstlisting} 
>> identifier = 'your_overlay_title';
>> response = buildFluxDistLayout(minerva, model, solution, identifier);
\end{lstlisting}The only new input variable is the text string in the \mcode{identifier}
that enables you to name each overlay according to a unique title.
The \mcode{response} status will be set to $1$ if the overlay was
successfully received by the MINERVA server.\troubleshooting

\pStep Visualise context-specific ReconMaps using a web browser.
Navigate to \url{http://vmh.life/#reconmap}, login with your credentials
above then select 'OVERLAYS' and the list of USER-PROVIDED OVERLAYS
appears. In order to see the map from Step \ref{step:reconMap-layout},
check the box adjacent to the unique text string provided by \mcode{identifier}. 

\pStep \label{step:exportReconMap}In order to export context-specific
ReconMaps as publishable graphics, two options are possible: portable
document format (\textit{.pdf}) or portable network format (\textit{.png}).
The former is useful for external editing whereas the latter essentially
produces a snapshot of the visual part of the map.
\begin{enumerate}
\item PDF export
\begin{enumerate}
\item Zoom out until the entire map is visible. Right click on the map \textendash >
Export as image \textendash > PDF. A file named \textit{model.pdf}
will be downloaded to the default directory of the browser. This PDF
is a scalable network graphic, so optionally one can use a PDF editor
to zoom in or crop the PDF as desired.
\end{enumerate}
\item PNG export
\begin{enumerate}
\item Navigate and zoom until the desired region of the map is visible.
Right click on the map \textendash > Export as image \textendash >
PNG and a file named \textit{model.png} will be downloaded to the
default directory of the browser. 
\end{enumerate}
\end{enumerate}

\subsubsection[Variable scope visualisation of a network with Paint4Net]{Variable scope visualisation of a network with Paint4Net\timing
$1-10^{3}$ s}

\pStep \label{step:paint4Net}During model validation or optimisation,
visualisation of a small-scale fragment of the network area of interest
is often sufficient and is especially convenient during network reconstruction
when a manual layout may not yet be available. Automatic generation
of a hypergraph layout for a chosen subset of network can be achieved
with the COBRA Toolbox extension Paint4Net{\small{}\cite{kostromins_paint4net:_2012}}.
A subset of a network may be visualised, and the directionality and
the fluxes for selected reactions may be shown. Details on each reaction
(ID, name and synonyms, and formula) and metabolite (ID, name and
synonyms, and charged formula) pop-up when a cursor is placed over
the corresponding item. 

First compute a flux vector, e.g., with flux balance analysis, using:

\begin{lstlisting}   
>> FBAsolution = optimizeCbModel(model); 
\end{lstlisting}

\pStep Visualise a selected network fragment around a list of reactions
in a \mcode{model}, contextualised using a flux vector \mcode{flux},
by running:\begin{lstlisting}  
>> flux = FBAsolution.v;
>> involvedMets = draw_by_rxn(model, rxns, drawMap, direction, initialMet, excludeMets, flux);
\end{lstlisting} The \mcode{rxns} input provides a selection of reactions of interest.
The remaining inputs are optional and control the appearance of the
automatic layout. For example, \mcode{excludeMets} provides a list
of metabolites that may be excluded from the network visualisation,
e.g., cofactors such as NAD and NADP. 

\pStep \label{step:paint4NetRadius}In order to visualise a model
fragment with a specified \mcode{radius} around a specified metabolite
of interest, such as \mcode{'etoh[c]'}, run:\begin{lstlisting}   
>> metAbbr = {'etoh[c]'};
>> [involvedRxns, involvedMets] = draw_by_met(model, metAbbr, 'true', 1, 'struc', {''}, flux); 
\end{lstlisting}

\subsubsection[Contributing to the COBRA Toolbox with MATLAB.devTools]{Contributing to the COBRA Toolbox with MATLAB.devTools \timing
$1-30$ s}

\pStep \label{subsec:Installation-of-the} A comprehensive code base
such as the COBRA Toolbox evolves constantly. The open-source community
is very active, and collaborators submit their contributions frequently.
The more a new feature or bug fix is interlinked with existing functions,
the higher the risk of a new addition breaking instantly code that
is heavily used on a daily basis. In order to decrease this risk,
a continuous integration setup interlinked with the version control
system \emph{git} has been set up. A \emph{git}-tracked repository
is essentially a folder with code or other documents of which all
incremental changes are tracked by date and user.

Any incremental changes to the code are called commits. The main advantage
of \emph{git} over other version control systems is the availability
of branches. In simple terms, a branch contains a sequence of incremental
changes to the code. A branch is also commonly referred to as a feature.
Consequently, a contribution generally consists of several commits
on a branch.

\begin{figure}[H]
\begin{centering}
\includegraphics[width=1\textwidth]{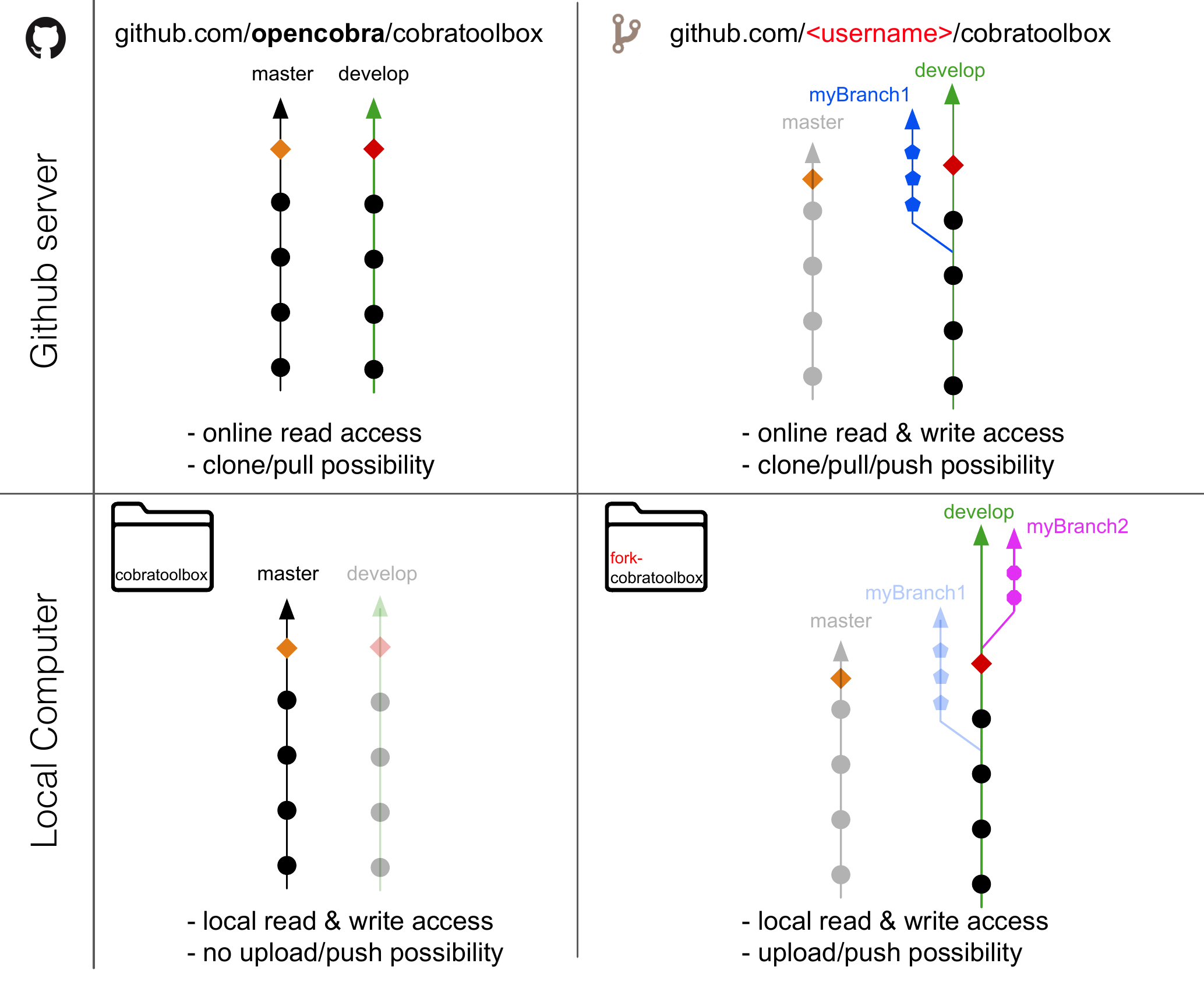}
\par\end{centering}
\caption{The openCOBRA repository and the fork of a contributor located on
the Github server can be cloned to the local computer as \emph{cobratoolbox}
and \emph{fork-cobratoolbox} folders, respectively. Each repository
might contain different branches, but each repository contains the
\emph{master} and \emph{develop} branches. Note that contributors
only have read on the openCOBRA repository. The stable branch is the
\emph{master} branch (black branch), while the development of code
is made on the \emph{develop }branch ({\color{greenContrib}green}
branch). The \emph{master} branch shall be checked out when using
the \emph{cobratoolbox} repository, whereas contributors shall create
new branches originating from the \emph{develop} branch (local \emph{fork-cobratoolbox}
directory and online \emph{<username>/cobratoolbox} repository). In
the present example, \emph{myBranch1} ({\color{blueContrib}blue}
branch) has already been pushed to the forked repository on the Github
server, while \emph{myBranch2} ({\color{pinkContrib}pink} branch)
is only present locally. The branch \emph{myBranch1 }may be merged
into the \emph{develop} branch of the openCOBRA repository through
opening a pull request. In order to submit the contributions (commits)
on \emph{myBranch2}, the contributor must first push the commits to
the forked repository (\protect\url{https://github.com/<username>/cobratoolbox})
before opening a pull request. Any commit made on the \emph{develop}
branch ({\color{redContrib}red} square) will be merged to the \emph{master}
branch if the \emph{develop} branch is stable overall ({\color{orangeContrib}orange}
square). \label{fig:The-setup-of}}
\end{figure}

Contributing to the COBRA Toolbox is straightforward. As a contributor
to the COBRA Toolbox is likely more familiar with MATLAB than with
the internal mechanics of \emph{git}, the MATLAB.devTools (\url{https://github.com/opencobra/MATLAB.devTools})
have been developed specifically to contribute to a \emph{git}-tracked
repository located on the Github server. In Figure \ref{fig:The-setup-of},
an overview of the two online repositories as well as their local
copies is given.

There are two ways of using the COBRA Toolbox, which depends on the
type of user. 

\begin{enumerate}
\item A user of the COBRA Toolbox
\begin{enumerate}
\item The openCOBRA repository (\url{https://github.com/opencobra/cobratoolbox})
is a public repository that is read-only. Once the openCOBRA repository
has been installed (as explained in Steps \ref{step:Init1}-\ref{subsec:Verify-and-test})
in the folder \emph{cobratoolbox}, all branches (including \emph{master}
and \emph{develop}) are available locally. In the local folder \emph{cobratoolbox},
the user has read and write access, but \emph{cannot} push eventual
changes back to the openCOBRA repository. It is the default and stable
\emph{master} branch only that should be used. The local copy located
in the \emph{cobratoolbox} directory can be updated (both branches). 
\end{enumerate}
\item A contributor or a developer of the COBRA Toolbox
\begin{enumerate}
\item In order to make changes to the openCOBRA repository, or, in other
words, \mcode{contribute}, you must obtain your own personal copy
first. You must register on the Github website (\url{https://github.com})
in order to obtain a \emph{username}. First, click on the button \textbf{FORK}\emph{
}at the top right corner of the official openCOBRA repository website
(\url{https://github.com/opencobra/cobratoolbox}) in order to create
a personal copy (or \emph{fork}) with write and read access of the
openCOBRA\emph{ }repository. This copy is accessible under \url{https://github.com/<username>/cobratoolbox}.
These branches can be accessed by following the procedure {[}2{]}
(see Step \ref{subsec:Continue-with-an}).
\end{enumerate}
\end{enumerate}
After initialisation of the MATLAB.devTools, the user and developer
may have two folders: a \emph{cobratoolbox} folder with the stable
\emph{master} branch checked out, and a \emph{fork-cobratoolbox} folder
with the \emph{develop} branch checked out. Detailed instructions
for troubleshooting and/or contributing to the COBRA Toolbox using
the terminal (or shell) are provided in Supplementary Manual 3.

After the official openCOBRA version of the COBRA Toolbox has been
installed, it is possible to install the MATLAB.devTools from within
MATLAB:

\begin{lstlisting}
>> installDevTools
\end{lstlisting}

With this command, the directory \emph{MATLAB.devTools} is created
next to the \emph{cobratoolbox} installation directory. The MATLAB.devTools
can also be installed from the terminal (or shell):

\begin{lstlisting}[style=bashStyle]
$ git clone git@github.com:opencobra/MATLAB.devTools
\end{lstlisting}\criticalStep A working internet connection is required and \emph{git}
and \emph{curl} must be installed. Installation instructions are provided
on the main repository page of the MATLAB.devTools. A valid passphrase-less
SSH key must be set in the Github account settings in order to contribute
without entering a password while securely communicating with the
Github server. \troubleshooting 

\pStep \label{subsec:Initialise-a-new}The MATLAB.devTools are configured
on the fly or whenever the configuration details are not present.
The first time a user runs \mcode{contribute}, the personal repository
(fork) is downloaded (cloned) into a new folder named \emph{fork-cobratoolbox}
at the location specified by the user. In this local folder, both
\emph{master} and \emph{develop} branches exist, but it is the \emph{develop}
branch that is automatically selected (\emph{checked out}). Any new
contributions are derived from the \emph{develop} branch. 

Initialising a contribution using the MATLAB.devTools is straightforward.
In MATLAB, type:

\begin{lstlisting}    
>> contribute % then select procedure [1]
\end{lstlisting}

If the MATLAB.devTools are already configured, procedure $[1]$ updates
the fork (if necessary) and initialises a new branch with a name requested
during the process. Once the contribution is initialised, files can
be added, modified or deleted in the folder \emph{fork-cobratoolbox.}
A contribution is successfully initialised when the user is presented
with a brief summary of configuration details. Instructions on how
to proceed are also provided.

\criticalStep The location of the fork must be specified as the root
directory. There will be a warning issued if the path already contains
another git-tracked repository.\troubleshooting 

\pStep \label{subsec:Continue-with-an}An existing contribution can
be continued after a while. This step is particularly important in
order to retrieve all changes that have been made to the openCOBRA
repository in the meantime.

\begin{lstlisting}   
>> contribute % then select procedure [2]
\end{lstlisting}

Procedure {[}2{]} pulls all changes from the openCOBRA repository,
and rebases the existing contribution. In other words, existing commits
are shifted forward and placed after all commits made on the \emph{develop}
branch of the openCOBRA repository.

\criticalStep Before attempting to continue working on an existing
feature, make sure that you published your commits as explained in
Step \ref{subsec:Publish-an-existing}.\troubleshooting 

\pStep \label{subsec:Publish-an-existing}Publishing a contribution
means uploading the incremental code changes to the fork. The changes
are available in public, but not yet available in the openCOBRA repository.
A contribution may only be accepted in the official repository once
a pull request has been submitted. It is not necessary to open a pull
request if you want to simply upload your contribution to your fork.

\begin{lstlisting}   
>> contribute % then select procedure [3]
\end{lstlisting}

When running procedure $[3]$, you have two options:
\begin{enumerate}
\item Simple contribution without opening a pull request
\begin{enumerate}
\item All changes to the code are individually listed and the user is asked
explicitly which changes shall be added to the commit. Once all changes
have been added, a commit message must be entered. Upon confirmation,
the changes are pushed to the online fork automatically. 
\end{enumerate}
\item Publishing and opening a pull request
\begin{enumerate}
\item The procedure for submitting a pull request is the same as Option
(A) with the difference that when selecting to open a pull request,
a link is provided that leads to a pre-configured website according
to the contributing guidelines. The pull request is then one click
away.
\end{enumerate}
\end{enumerate}
\criticalStep After following procedures $[1]$ and $[2]$, all changes
should be published using procedure $[3]$ before stopping to work
on that contribution. When following procedure $[3]$, the incremental
changes are uploaded to the remote server. It is advised to publish
often, and make small incremental changes to the code. There is no
need for opening a pull request immediately (Option A) if there are
more changes to be made. A pull request may be opened at any time,
even manually and directly from the Github website.\troubleshooting 

\pStep \label{subsec:Delete-an-existing}If a contribution has been
merged into the \emph{develop} branch of the openCOBRA repository
(accepted pull request), the contribution (feature or branch) can
be safely deleted both locally and remotely on the fork by running
\mcode{contribute} and selecting procedure $[4]$.

Note that deleting a contribution deletes all the changes that have
been made on that feature (branch). It is not possible to selectively
delete a commit using the MATLAB.devTools. Instead, create a new branch
by following procedure $[1]$ (see Step \ref{subsec:Initialise-a-new}),
and follow the instructions to \emph{cherry-pick} in the Supplementary
Manual 3.

\criticalStep Make sure that your changes are either merged or saved
locally if you need them. Once procedure {[}4{]} is concluded, all
changes on the deleted branch are removed, both locally and remotely.
No commits can be recovered.\troubleshooting 

\pStep \label{subsec:Update-the-fork} It is sometimes useful to
simply update the fork without starting a new contribution. The local
fork can be updated using procedure {[}5{]} of the \mcode{contribute}
menu.

\begin{lstlisting}   
>> contribute  % then select procedure [5]
\end{lstlisting}

\criticalStep Before updating your fork, make sure that no changes
are present in the local directory \emph{fork-cobratoolbox}. You can
do so by typing:

\begin{lstlisting}    
>> checkStatus
\end{lstlisting}

If there are changes listed, publish them by selecting procedure $[3]$
of the \mcode{contribute} menu as explained in Step \ref{subsec:Publish-an-existing}.
\troubleshooting

\subsubsection[Engaging with the COBRA Toolbox forum]{Engaging with the COBRA Toolbox forum\timing $1-10^{2}$ s}

\pStep\label{subsec:Search-for-existing} The Frequently Asked Questions
(FAQ) section of the documentation (\url{https://opencobra.github.io/cobratoolbox/docs/FAQ.html})
is a good starting point to find answers to questions or issues one
may face.

The public forum associated with the COBRA Toolbox, available at \url{https://groups.google.com/forum/#!forum/cobra-toolbox},
is a great way to search for solutions to previously recognised problems
that are similar to problems novel to the user. This is especially
so with respect to recent installation and configuration issues that
have arisen due to asynchronous development of the many software packages
integrated with the COBRA Toolbox. 

\pStep \label{subsec:Suggest-new-solutions}Suggest new solutions
to problems
\begin{enumerate}
\item Post your question to the online COBRA Toolbox forum

Questions posted in the forum are welcome provided that some simple
guidelines are followed:
\begin{enumerate}
\item Before a question can be posted, an application for membership at
\url{https://groups.google.com/forum/#!forum/cobra-toolbox} is required
to eliminate spam. 
\item Make the question as detailed as possible to increase the probability
of a rapid and helpful reply. 
\item Append your message with the result of running \mcode{generateSystemConfigReport}
so that repository maintainers are aware of the system configuration.
That is often the first question that comes to mind when considering
to respond.
\end{enumerate}
\item Reply to a question online COBRA Toolbox forum
\begin{enumerate}
\item Community contributions are welcomed to help users overcome any issues
they face and are noticed by existing COBRA community members.
\end{enumerate}
\end{enumerate}
Generally, responses to questions can be expected within 1-2 days
of posting, provided that posting guidelines are followed.

\pagebreak{}

\section[TROUBLESHOOTING]{{\normalsize{}\troubleshooting} \label{sec:troubleshooting}}

\begin{longtable}{>{\raggedright}p{0.9cm}|>{\raggedright}p{3.5cm}|>{\raggedright}p{3cm}|>{\raggedright}p{7cm}}
\hline 
\textbf{Step} & \textbf{Problem} & \textbf{Possible reason} & \textbf{Solution}\tabularnewline
\hline
\endhead
\hline 
\pStepRef{step:Init1} & The \mcodeENUM{initCobraToolbox} function displays warnings or error
messages. & Incompatible third-party software or improperly configured system. & First, read the output of the initialisation script in the command
window. Any warning or error messages, though often brief, may point
toward the source of the problem during initialisation if read literally.
Second, verify that all software versions are supported and have been
correctly installed, as described in the MATERIALS section. Third,
ensure that you are using the latest version of the COBRA Toolbox,
cf. Steps \ref{subsec:Installation-of-the}-\ref{subsec:Update-the-fork}.
Fourth, verify and test the COBRA Toolbox, as described in Step \ref{subsec:Verify-and-test}.
Finally, if nothing else works, consult the COBRA Toolbox forum, as
described in Steps \ref{subsec:Search-for-existing}- \ref{subsec:Suggest-new-solutions}.\tabularnewline
\hline 
\pStepRef{subsec:Verify-and-test} & Some tests are listed as failed when running \mcodeENUM{testAll}. & Some third party dependencies are not properly installed or the system
is improperly configured. & Verify that all required software has been correctly installed as
described in the MATERIALS section. The specific test can then be
run individually to determine the exact cause of the error. If the
error can be fixed, try to use the MATLAB.devTools and contribute
a fix. Further details on how to approach submitting a contribution
are given in Steps \ref{subsec:Installation-of-the}-\ref{subsec:Update-the-fork}.
If the error cannot be determined, reach out to the community as explained
in Steps \ref{subsec:Search-for-existing}- \ref{subsec:Suggest-new-solutions}.\tabularnewline
\hline 
\pStepRef{step:ImportAReconstruction} & The \mcodeENUM{readCbModel}function fails to import a model. & The input file is not correctly formatted or the SBML file format
is not supported. & Specifications for Excel sheets accepted by the COBRA Toolbox can
be found on Github (\url{opencobra.github.io/cobratoolbox/docs/COBRAModelFields.html}).
Files with legacy SBML formats can be imported, but some information
from the SBML file might be lost. In addition to constraint-based
information encoded by fields of the \emph{fbc} package, the COBRA-style
annotations introduced in the COBRA Toolbox 2.0\cite{schellenberger_quantitative_2011}
are supported for backward compatibility. Some information is still
stored in this type of annotations. The data specified with the latest
version of the \emph{fbc} package is used in preference to other fields,
e.g., legacy COBRA-style notes which may contain similar data. \tabularnewline
\hline 
\pStepRef{step:ExportingAReconstruction}  & The \mcodeENUM{writeCbModel} function fails to export a model. & Some of the required fields of the \mcodeENUM{model} structure are
missing or the model contains invalid data. & Before a reconstruction or model is exported, a summary of invalid
data in the model can be obtained by running \mcodeENUM{verifyModel(model)}.
A list of required fields for the \mcodeENUM{model} structure is
presented in Table \ref{tab:Fields,-dimensions,-data}.\tabularnewline
\hline 
\pStepRef{subsec:selectSolver}  & The \mcodeENUM{dqqMinos} or \mcodeENUM{quadMinos} interfaces are
not working as intended. & The binaries might not be compatible with your operating system. & Make sure that all relevant system requirements described in the MATERIALS
section are satisfied. If you are still unable to use the respective
interfaces, reach out to the community as explained in Steps \ref{subsec:Search-for-existing}-
\ref{subsec:Suggest-new-solutions}.\tabularnewline
\hline 
\pStepRef{step:stoichConsistency}

A) & The \mcodeENUM{findSExRxnInd} function fails to identify some exchange,
demand and sink reactions. & Some exchange, demand and sink reactions do not start with any of
anticipated prefixes. & Try an alternative approach.\tabularnewline
\hline 
\pStepRef{step:stoichConsistency}

B) & The function \mcodeENUM{checkMassChargeBalance} returns wrong results. & Some formulae are missing or a formula is incorrectly specified, leading
to one or more reactions to be incorrectly identified as being elementally
balanced. & Try an alternative approach.\tabularnewline
\hline 
\pStepRef{step:stoichConsistency}

C) & Erroneous predictions. & Inadvertent violation of the steady-state mass conservation constraint. & Manually inspect the reaction formulae for each reaction to identify
any obviously mass imbalanced reactions, omit them from the reconstruction
and run \mcodeENUM{findStoichConsistentSubset} again. \tabularnewline
\hline 
\pStepRef{step:optimizeCbModel} & The solution status given by \mcodeENUM{FBAsolution.stat}

is $-1$. & A too short runtime limit has been set or numerical issues happened
during the optimisation procedure. & Check the value of \mcodeENUM{FBAsolution.origStat} and compare it
with the documentation provided by the solver in use for further information.
If one is solving with a double precision solver a model that could
be multi-scale but is not yet recognised as such, then \mcodeENUM{FBAsolution.stat == -1}
can be symptomatic of this situation. In that case, refer to Steps
\ref{step:checkScaling}-\ref{subsec:selectSolver} to learn how to
numerically characterise a reconstruction model. \tabularnewline
\hline 
\pStepRef{step:uniformSampling} & The sampling distribution is not uniform (revealed by a non-uniform
marginal flux distribution). & The sampling parameters \mcodeENUM{options.nSkip} and \mcodeENUM{options.nSamples}
are set too low. & Increase the values of the \mcodeENUM{options.nSkip} and \mcodeENUM{options.nSamples}
parameters until smooth and unimodal marginal flux distributions are
obtained.\tabularnewline
\hline 
\pStepRef{step:-Repeat-the} & No reaction is found in the MUST sets. & The wild-type or mutant strain may not be enough constrained. & A solution is to add more constraints to the strains until differences
in the reaction ranges are shown. If no differences are found, another
algorithm might be better suited. If there is an error when running
the \mcodeENUM{findMust*} functions, a possible reason is that the
inputs are not well defined or a solver may not be set. Verify the
inputs, use \mcodeENUM{changeCobraSolver} to change to a commercial-grade
optimisation solver (see Table \ref{tab:Optimisation-Solvers.} for
a list of supported solvers).\tabularnewline
\hline 
\pStepRef{Step:-Given-a} & Some reactions could not be mapped. & Too short runtime limit or a reaction that the algorithm could not
atom map. & Increase the runtime limit of the algorithm.\tabularnewline
\hline 
\pStepRef{step:reconMap-layout} & The remote MINERVA server refuses to build a new overlay.  & The text string in the \mcodeENUM{identifier} input variable is not
uniquely defined in your account.  & Change the \mcodeENUM{identifier} text string of your overlay.\tabularnewline
\hline 
\pStepRef{subsec:Initialise-a-new} & An error occurs when run running \mcodeENUM{contribute} claiming
that the fork cannot be reached or that the local fork cannot be found. & The local forked folder cannot be found, has been moved, or the remote
fork cannot be reached. & It may occur that the configuration of the MATLAB.devTools is faulty
or has been mistyped. In that case, try to reset the configuration
by typing:

\begin{lstlisting}    
>> resetDevTools
\end{lstlisting}\tabularnewline
\hline 
\pStepRef{subsec:Installation-of-the} & An error might be thrown claiming \mcodeENUM{permission denied}. & The SSH key of the computer is not configured properly.  & The installation of the MATLAB.devTools is dependent on a correctly
configured Github account. The SSH key of the computer must be set
in the Github account settings or otherwise errors will be thrown.
If the \emph{git clone} command works, the SSH key is properly set.
In that case, delete the SSH key locally (generally located in the
folder \emph{.ssh }in the home directory) and remotely on Github,
and generate a new SSH key.\tabularnewline
\hline 
\pStepRef{subsec:Initialise-a-new} & Procedure $[1]$ when running \mcodeENUM{contribute} might not be
successful. & The local \emph{fork-cobratoolbox} folder is too old or has not been
updated for a while. & In that case and if no local changes are present, backup and remove
the local \emph{fork-cobratoolbox }folder\emph{ }and run the \mcodeENUM{contribute}
command again. Alternatively, try to delete the forked repository
online and re-fork the openCOBRA repository. When one is sure that
everything is fine, the backup can be safely deleted, but it is wise
to store it for some time, in case later one realises that some updates
to code have gone missing.\tabularnewline
\hline 
\pStepRef{subsec:Initialise-a-new} & Procedure $[1]$ when running \mcodeENUM{contribute} might not be
successful. & There are changes in the local \emph{fork-cobratoolbox} folder. & Contribute the changes manually as described in the Supplementary
Manual 3.\tabularnewline
\hline 
\pStepRef{subsec:Initialise-a-new} & Procedure $[1]$ when running \mcodeENUM{contribute} might not be
successful. & The forked repository cannot be reached online or the SSH key is not
configured properly. & Set the SSH key in your Github account and make sure that the forked
repository can be reached. This can easily be checked by re-cloning
the MATLAB.devTools in the terminal as explained in Step \ref{subsec:Installation-of-the}
and by browsing to the forked repository online.\tabularnewline
\hline 
\pStepRef{subsec:Continue-with-an} & Procedure $[2]$ when running \mcodeENUM{contribute} might fail. & Your contribution has been deleted online, or is no longer available
locally. & When the rebase process fails, the user is asked to reset the contribution,
which will reset the contribution to the online version of the branch
in the fork. In general, when the rebase fails there have been changes
made on the openCOBRA repository that are in conflict with the local
changes. You can check the status of the local repository by typing:

\begin{lstlisting}    
>> checkStatus
\end{lstlisting}

If there are conflicts that you do not know how to resolve, check
the official repository or ping the developers in \url{https://groups.google.com/forum/#!forum/cobra-toolbox}
as explained in Steps \ref{subsec:Search-for-existing}- \ref{subsec:Suggest-new-solutions}.
If you already have published changes, try to submit a pull request
as explained in Step \ref{subsec:Publish-an-existing} for developers
to understand the situation. Alternatively, you can try to resolve
the conflicts manually. More information on how to solve conflicts
is given as Supplementary Manual 3. \tabularnewline
\hline 
\pStepRef{subsec:Publish-an-existing} & Procedure $[3]$ when running \mcodeENUM{contribute} might not be
successful. & The forked repository cannot be reached online or if the SSH key is
not configured properly. & Check to set the SSH key in your Github account and make sure that
the forked repository can be reached. \tabularnewline
\hline 
\pStepRef{subsec:Publish-an-existing} & When opening a pull request, Github cannot automatically merge. & There have been changes made on the\emph{ }openCOBRA repository and
on your local fork.  & Submit however the pull request; another developer will help you rebase
your contribution manually.\tabularnewline
\hline 
\pStepRef{subsec:Delete-an-existing} & Procedure $[4]$ when running \mcodeENUM{contribute} might not be
successful. & Your local changes are not yet published (committed). & Follow procedure $[3]$ of the \mcodeENUM{contribute} menu in order
to publish your changes first as explained in Step \ref{subsec:Publish-an-existing}. \tabularnewline
\hline 
\pStepRef{subsec:Update-the-fork} & Procedure $[5]$ when running \mcodeENUM{contribute} might not be
successful. & There are some local changes that have not yet been published (committed). & Backup eventual modifications, remove the \emph{fork-cobratoolbox
folder},\emph{ }and run the \mcodeENUM{contribute} command again. \tabularnewline
\hline 
\pStepRef{subsec:Update-the-fork} & Procedure $[5]$ when running \mcodeENUM{contribute} might not be
successful. & Too many changes have been made in the openCOBRA repository. & Backup your modified files to a separate location, and reset your
branch manually by typing in the terminal (be careful - this will
delete all your changes locally, but not remotely): 

\begin{lstlisting}[style=bashStyle]
$ git reset --hard origin/<yourBranch>
\end{lstlisting}

Then, copy your file back into the \emph{fork-cobratoolbox} folder
and \mcodeENUM{contribute} normally.\tabularnewline
\hline 
\end{longtable}
\pagebreak{}

\section[TIMING]{{\normalsize{}\timing \label{sec:timing}}}

Steps \ref{step:Init1}-\ref{subsec:checkSolvers}, \nameref*{step:Init1}:
$5-30$ s

Step \ref{subsec:Verify-and-test}, \nameref*{subsec:Verify-and-test}:
$\sim10^{3}$ s

Step \ref{step:ImportAReconstruction}, \nameref*{step:ImportAReconstruction}:
$10-10^{2}$ s

Step \ref{step:ExportingAReconstruction}, \nameref*{step:ExportingAReconstruction}:
$10-10^{2}$ s

Step \ref{step:rBioNet}, \nameref*{step:rBioNet}: $1-10^{3}$ s

Steps \ref{subsec:Load-reactions}-\ref{step:Merge-recon}, \nameref*{subsec:Load-reactions}:
$1-10^{3}$s

Steps \ref{step:reconstrFunctions}-\ref{Step.-Remove-trivial}, \nameref*{step:reconstrFunctions}:
$1-10^{2}$s

Step \ref{step:checkScaling}, \nameref*{step:checkScaling}: $1-10^{2}$
s

Step \ref{subsec:selectSolver}, \nameref*{subsec:selectSolver}:
$1-5$ s

Step \ref{step:stoichConsistency}, \nameref*{step:stoichConsistency}:
$1-10^{5}$ s

Step \ref{step:ConsistentSpecies}, \nameref*{step:ConsistentSpecies}:
$1-10^{3}$ s

Step \ref{step:set-simulation}, \nameref*{step:set-simulation}:
$1-10^{3}$ s

Step \ref{step:identify-molecular-species}, \nameref*{step:identify-molecular-species}:
$1-10^{3}$ s

Step \ref{step:inconsflux}, \nameref*{step:inconsflux}: $1-10^{3}$
s

Steps \ref{step:FBA}-\ref{step:optimizeCbModel}, \nameref*{step:FBA}:
$1-10^{2}$ s

Step \ref{step:relaxedFBA}, \nameref*{step:relaxedFBA}: $1-10^{3}$
s

Step \ref{step:sparseFBA}, \nameref*{step:sparseFBA}: $1-10^{3}$
s

Steps \ref{Step:subsec:-Given-a}-\ref{step:blockedReaction}, \nameref*{step:blockedReaction}:
$\sim$$10^{2}$ s

Steps \ref{step:DeadEndMetabolites}-\ref{Step:-The-main}, \nameref*{Step:-The-main}:
$10^{2}-10^{5}$ s

Steps \ref{step:extraMetabolomic}-\ref{step:quantConstraint}, \nameref*{step:extraMetabolomic}:
$10^{3}-10^{5}$ s

Steps \ref{step:intraMetabolomic}-\ref{step:optimizeuFBA}, \nameref*{step:intraMetabolomic}:
$10^{2}-10^{4}$ s

Step \ref{step:transcriptomic}, \nameref*{step:transcriptomic}:
$10^{2}-10^{4}$ s

Steps \ref{step:addingConstraints}-\ref{step:addingConstraints-end},
\nameref*{step:addingConstraints}: $\sim10^{2}$ s

Steps \ref{Step:subsec:-It-is}-\ref{step:qualitativeFidelity-end},
\nameref*{Step:subsec:-It-is}: $10^{2}-10^{3}$ s

Steps \ref{step:quantitativeFidelity}-\ref{step:quantitativeFidelity-end},
\nameref*{step:quantitativeFidelity}: $10^{2}-10^{3}$ s

Step \ref{step:minSpanPathways}, \nameref*{step:minSpanPathways}:
$10^{2}-10^{4}$ s

Step \ref{step:low-FVA}, \nameref*{step:low-FVA}: $1-10^{3}$ s

Step \ref{step:FVA}, \nameref*{step:FVA}: $1-10^{5}$ s

Step \ref{step:uniformSampling}, \nameref*{step:uniformSampling}:
$1-10^{3}$ s

Steps \ref{step:Select-solver}-\ref{step:displayThe-reactions},
\nameref*{step:Select-solver}: $10-10^{5}$ s

Steps \ref{step:atomicallyResolve}-\ref{step:identifyConservedMoieties-end},
\nameref*{step:atomicallyResolve}: $10-10^{5}$ s

Steps \ref{step:componentContribution}-\ref{step:thermodynamicallyConstraint},
\nameref*{step:componentContribution}: $1-10^{3}$ s

Step \ref{step:convertFBAtoKinetic}, \nameref*{step:convertFBAtoKinetic}:
$1-10^{3}$ s

Step \ref{step:nonEquilibriumSteadyState}, \nameref*{step:nonEquilibriumSteadyState}:
$1-10^{3}$ s

Step \ref{step:moietyNonEquilibriumSteadyState}, \nameref*{step:moietyNonEquilibriumSteadyState}:
$1-10^{3}$ s

Steps \ref{step: ReconMap}-\ref{step:exportReconMap}, \nameref*{step:exportReconMap}:
$1-10^{2}$ s

Steps \ref{step:paint4Net}-\ref{step:paint4NetRadius}, \nameref*{step:paint4Net}:
$1-10^{3}$ s

Steps \ref{subsec:Installation-of-the}-\ref{subsec:Update-the-fork},
\nameref*{subsec:Installation-of-the}: $1-30$ s

\section{ANTICIPATED RESULTS}

\normalsize

\normalsize

\subsubsection{\textbf{\nameref*{step:Init1}}}

\begin{figure}[H]
\begin{centering}
\includegraphics[width=1\textwidth]{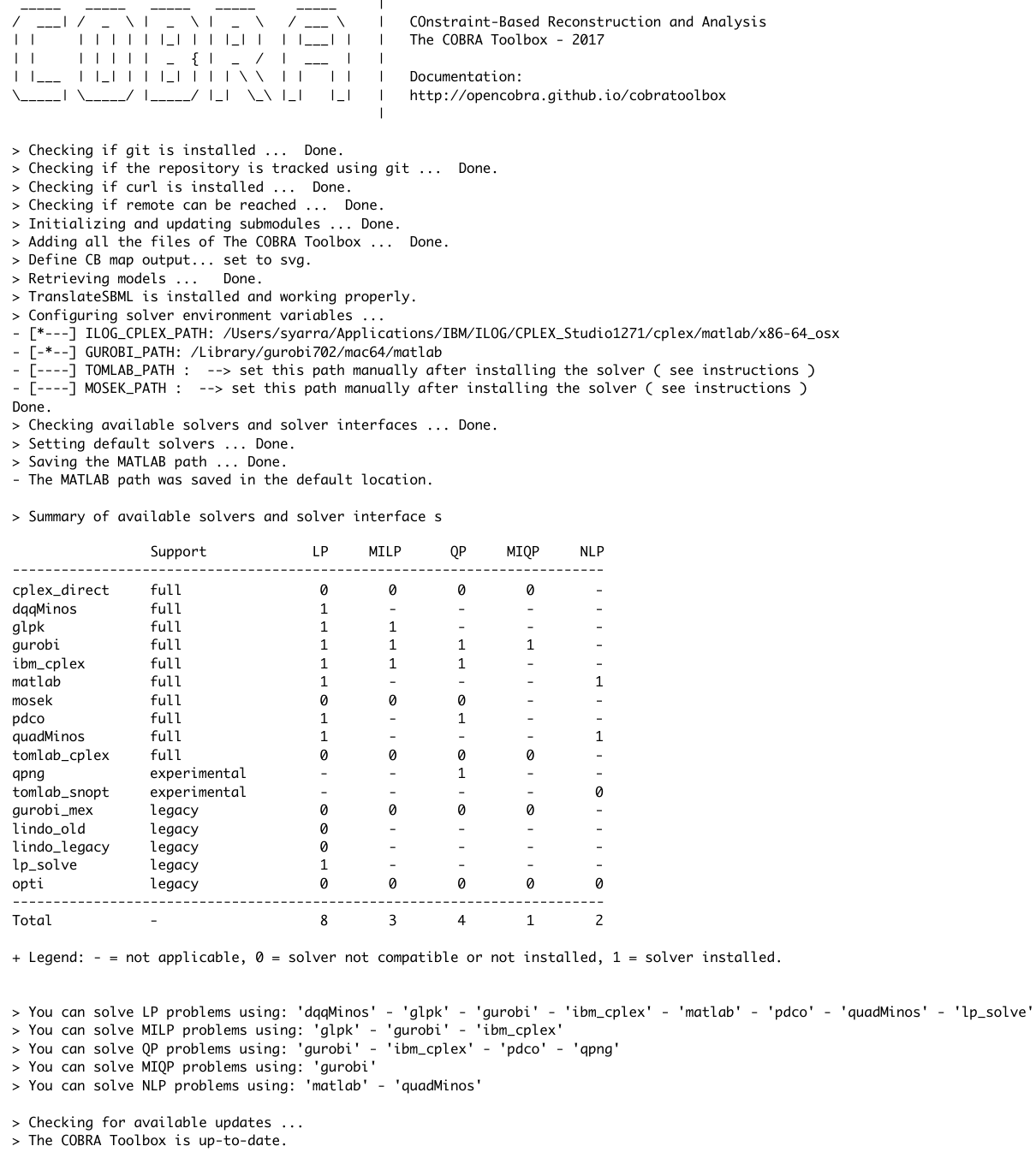}
\par\end{centering}
\caption{\label{fig:Output-of-initCobraToolbox}Output of initialisation of
the COBRA Toolbox with \mcode{initCobraToolbox}.}
\end{figure}

\pStepRef{step:Init1}The initialisation step automatically checks
the configuration of all of the required and some of the optional
software dependencies. During initialisation, all \emph{git} submodules
are updated. The solver paths are set when available and compatible.
A system-dependent table with the solver status is returned, together
with solver suggestions as shown in Figure \ref{fig:Output-of-initCobraToolbox}.
The user is also presented with options to update the COBRA Toolbox.

\pStepRef{subsec:checkSolvers}A list of solvers assigned to solve
each class of optimisation solvers is returned:

\begin{lstlisting}    
Defined solvers are:     
CBT_LP_SOLVER: gurobi     
CBT_MILP_SOLVER: gurobi     
CBT_QP_SOLVER: qpng     
CBT_MIQP_SOLVER: gurobi     
CBT_NLP_SOLVER: matlab
\end{lstlisting}\pStepRef{subsec:Verify-and-test}The test suite starts by initialising
the COBRA Toolbox and thereafter, all of the tests are run. At the
end of the test run, a comprehensive summary table is presented in
which the respective tests and their test outcome is shown. On a fully
configured system that is compatible with the most recent version
of the COBRA Toolbox, all tests should pass. It may not be necessary
to have a fully configured system to use one's particular subset of
methods.

\subsubsection{\textbf{\nameref*{step:ImportAReconstruction}}}

\pStepRef{step:ImportAReconstruction} The reconstruction or model
is loaded into the MATLAB workspace within a structure named \mcode{model},
irrespective of whether the \mcode{fileName} specified a reconstruction
or model. The model structure should contain all of the information
in different fields. Table \ref{tab:Fields,-dimensions,-data} provides
an overview of the individual model fields and their content. Very
large SBML models may take some time to load. 

\pStepRef{step:ExportingAReconstruction} An exported file containing
the information from the model in the location and format specified
by the \mcode{fileName}. 

\subsubsection{\textbf{\nameref*{step:checkScaling}}}

\pStepRef{step:checkScaling} The \mcode{checkScaling} function returns
a \mcode{precisionEstimate} string that is either \mcode{'double'}
or \mcode{'quad'}. The scaling estimate is based on the order of
magnitude of the ratio of the maximum and minimum row and column scaling
coefficients, which are determined such that the scaled stoichiometric
matrix has entries close to unity. In addition, a summary of scaling
properties included in \mcode{scalingProperties} may be returned. 

\subsubsection{\textbf{\nameref*{subsec:selectSolver}}}

\pStepRef{subsec:selectSolver} If the selected solver is installed,
\mcode{solverStatus == 1} will be returned, the solver interface
to MATLAB is configured correctly, and the solver is compatible with
the system environment. If the \mcode{dqqMinos} solver has been selected
and \mcode{solverStatus == 1}, then LP problem solutions are computed
somewhat slower than with a double precision solver, but with the
advantage that solutions are computed with a feasibility and optimality
tolerance of $10^{-15}$, which becomes an advantage for a multi-scale
model, where the typical tolerance of $10^{-6}$ for a double precision
solver may be insufficient. 

\subsubsection{\textbf{\nameref*{step:ConsistentSpecies}}}

\pStepRef{step:ConsistentSpecies} Any molecular species corresponding
to a non-zero entry within \mcode{SConsistentMetBool} is always involved
in mass imbalanced reactions indicating that the stoichiometry is
misspecified. Double check the chemical formulae involved in the corresponding
reactions to ensure that, e.g., the stoichiometry for protons, cofactors,
etc., leads to balanced reactions.

\pStepRef{step:inconsflux}Any non-zero entry in \mcode{fluxInConsistentRxnBool}
indicates a flux inconsistent reaction, i.e., a reaction that does
not admit a non-zero flux. Blocked reactions can be resolved by manual
reconstruction\cite{thiele_protocol_2010}, algorithmic reconstruction\cite{thiele_fastgapfill:_2014},
or a combination of both.

\subsubsection{\textbf{\nameref*{step:sparseFBA}}}

\begin{figure}[H]
\begin{centering}
\includegraphics[width=1\textwidth]{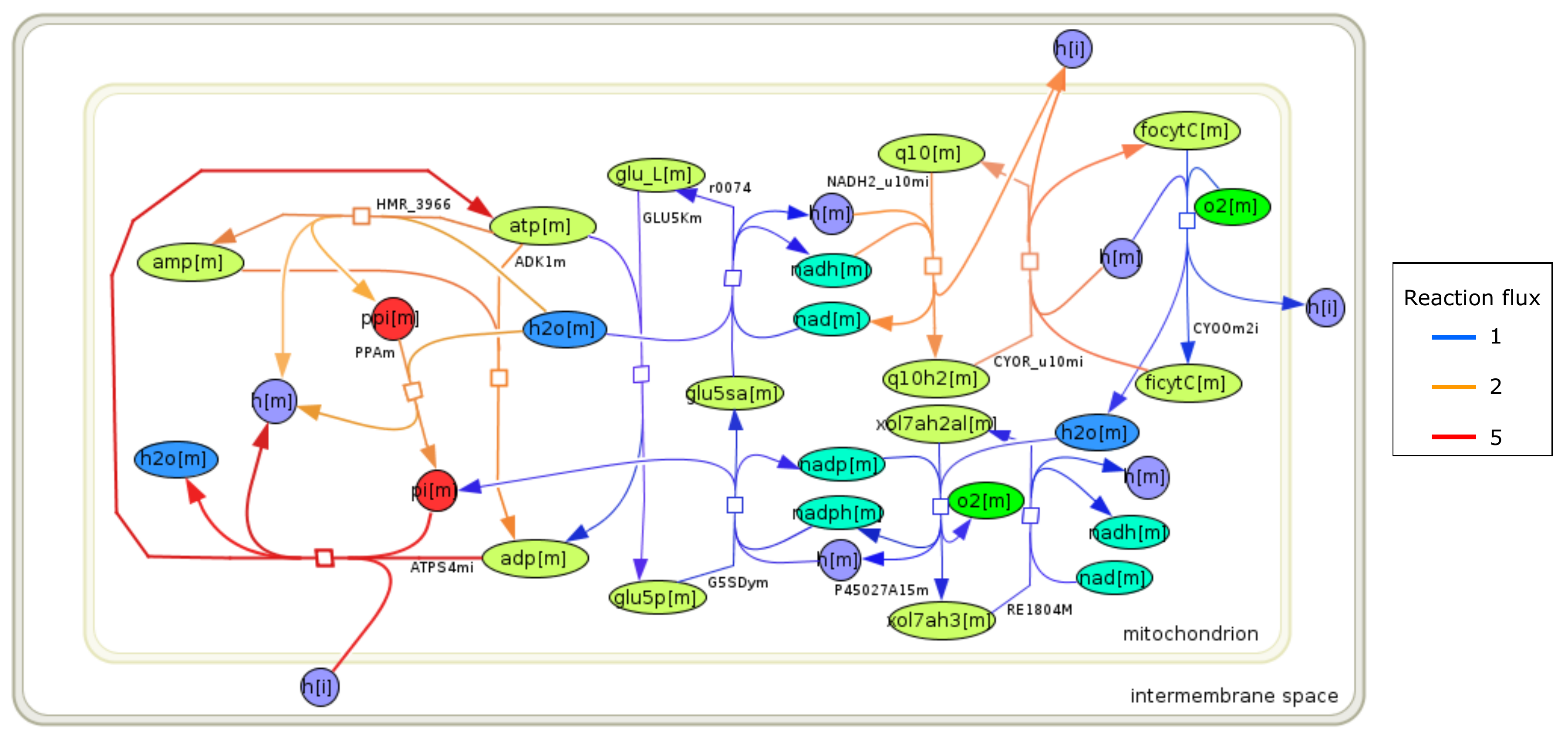}
\par\end{centering}
\caption{\label{fig:minATPSyn}\textbf{An energy generating stoichiometrically
balanced cycle.} The smallest stoichiometrically balanced cycle that
produces ATP at a maximal rate using the ATP synthase reaction, in
Recon3D, with all internal reactions. All metabolite and reaction
abbreviations are primary keys in the Virtual Metabolic Human database
(\protect\url{https://vmh.uni.lu}): reaction abbreviation, reaction
name: ADK1m, adenylate kinase, mitochondrial; G5SDym, glutamate-5-semialdehyde
dehydrogenase, mitochondrial; GLU5Km, glutamate 5-kinase, mitochondrial;
P45027A15m, 5-beta-cytochrome P450, family 27, subfamily A, polypeptide
1; PPAm, inorganic diphosphatase; r0074, L-glutamate 5-semialdehyde:NAD+
oxidoreductase; HMR\_3966, nucleoside-triphosphate giphosphatase;
ATPS4mi, ATP synthase (four protons for one ATP); CYOR\_u10mi, ubiquinol-6
cytochrome c reductase, Complex III; NADH2\_u10mi, NADH dehydrogenase,
mitochondrial; CYOOm2i, cytochrome c oxidase, mitochondrial complex
IV.}
\end{figure}

\pStepRef{step:sparseFBA} There should be no such cycle in a network
with bounds that are sufficiently constrained. Figure$~$\ref{fig:minATPSyn}
illustrates the cycle obtained from Recon3D with all internal reaction
bounds set to zero.

\subsubsection{\textbf{\nameref*{step:transcriptomic}}}

\pStepRef{step:transcriptomic} \mcode{createTissueSpecificModel}
returns a COBRA model which is constrained with the context of the
data provided to it. Usually, this means enrichment of reactions with
high expression and omission of reactions with low expression profiles.
Each method returns a flux consistent model, hence it is likely that
certain reactions, without experimental evidence, are added to the
context-specific model in order to enable non-zero net flux through
reactions for which supporting experimental evidence for activity
exists.

\subsubsection{\textbf{\nameref*{step:quantitativeFidelity-end}}}

\pStepRef{step:quantitativeFidelity-end} For the \mcode{Recon3Dmodel},
the anticipated yield is $32$ ATP per unit of glucose, which compares
favourably to the ATP yield of $31$ ATP obtained from the biochemical
literature.

\subsubsection{\textbf{\nameref*{step:uniformSampling}}}

\pStepRef{step:uniformSampling} The marginal flux distributions for
each reaction should be smooth and uni-modal for any biochemical network
with feasible set $\Omega\coloneqq\left\{ v\mid Sv=0;~l\leq v\leq u\right\} $. 

\subsubsection{\textbf{\nameref*{step:optForce}}}

\pStepRef{step:optForce} The identified reaction is \mcode{SUCt},
i.e., a transporter for succinate (an intuitive solution). However,
changing the parameters will enable \mcode{optForce} to find non-intuitive
solutions.

\subsubsection{\textbf{\nameref*{step:displayThe-reactions}}}

\begin{figure}[H]
\begin{centering}
\includegraphics[width=0.5\textwidth]{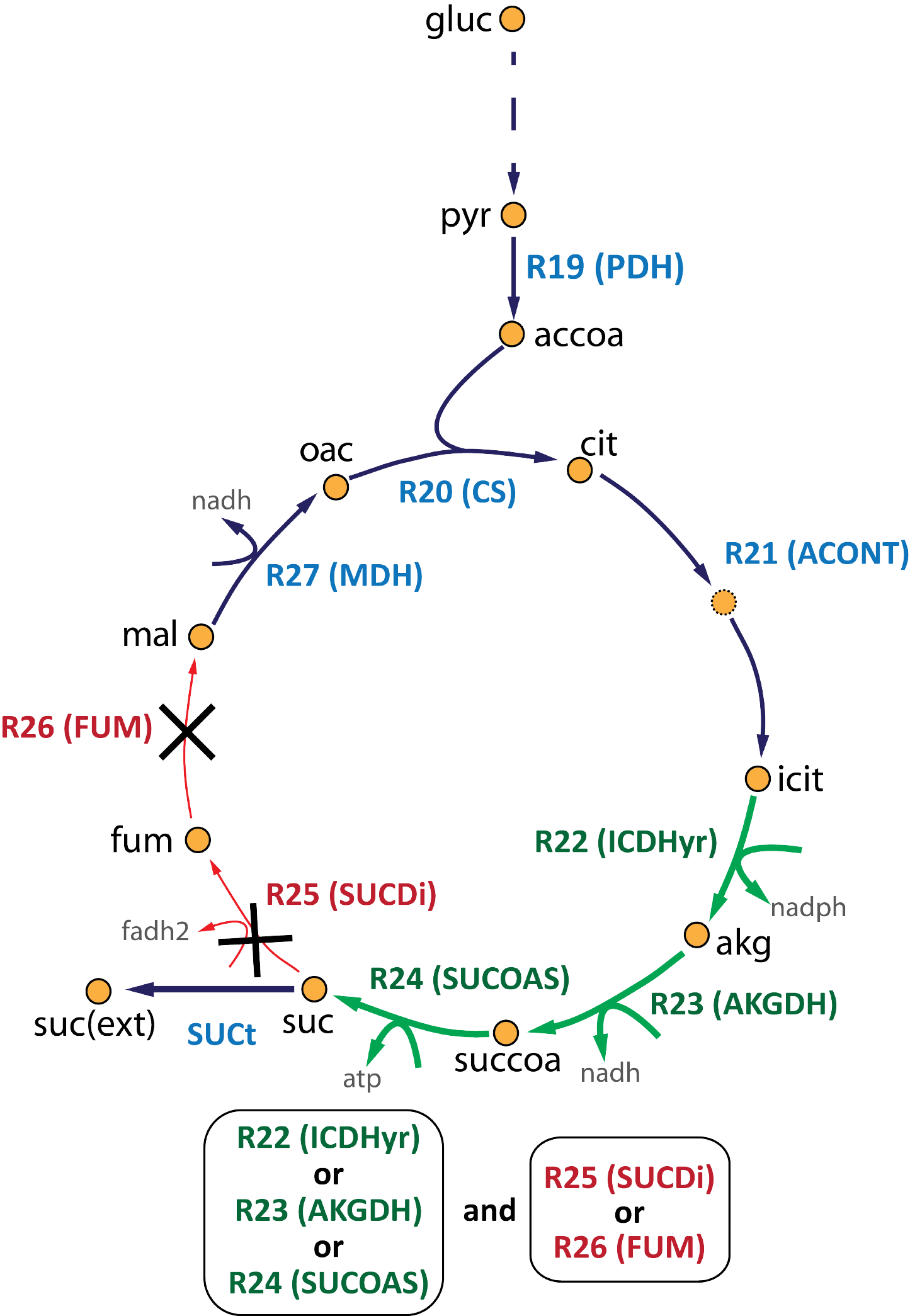}
\par\end{centering}
\caption{\label{fig:optForceResult}The interventions predicted by the OptForce
method for succinate overproduction in \emph{E. coli} (AntCore model)
under aerobic conditions. Reactions that need to be up-regulated (\textcolor{lime}{\textcolor{green_optForce}{green}}
arrows and labels) and knocked out (\textcolor{red}{red} arrows and
labels) are shown in this simplified metabolic map. The strategies
include up-regulation of reactions generating succinate such as isocitrate
dehydrogenase (R2), $\alpha$-ketoglutarate dehydrogenase or succinyl-CoA
synthetase, along with knockout of reactions draining succinate such
as succinate dehydrogenase or fumarate hydratase. Note that each of
these reactions may associate with one or more genes in \emph{E. coli}.}
\end{figure}

\pStepRef{step:displayThe-reactions} Figure \ref{fig:optForceResult}
illustrates the interventions predicted by the OptForce method for
succinate overproduction in the AntCore \emph{E. coli} model under
aerobic conditions.

\subsubsection{\textbf{\nameref*{step:thermodynamicallyConstraint}}}

\begin{figure}[H]
\begin{centering}
\includegraphics[width=1\textwidth]{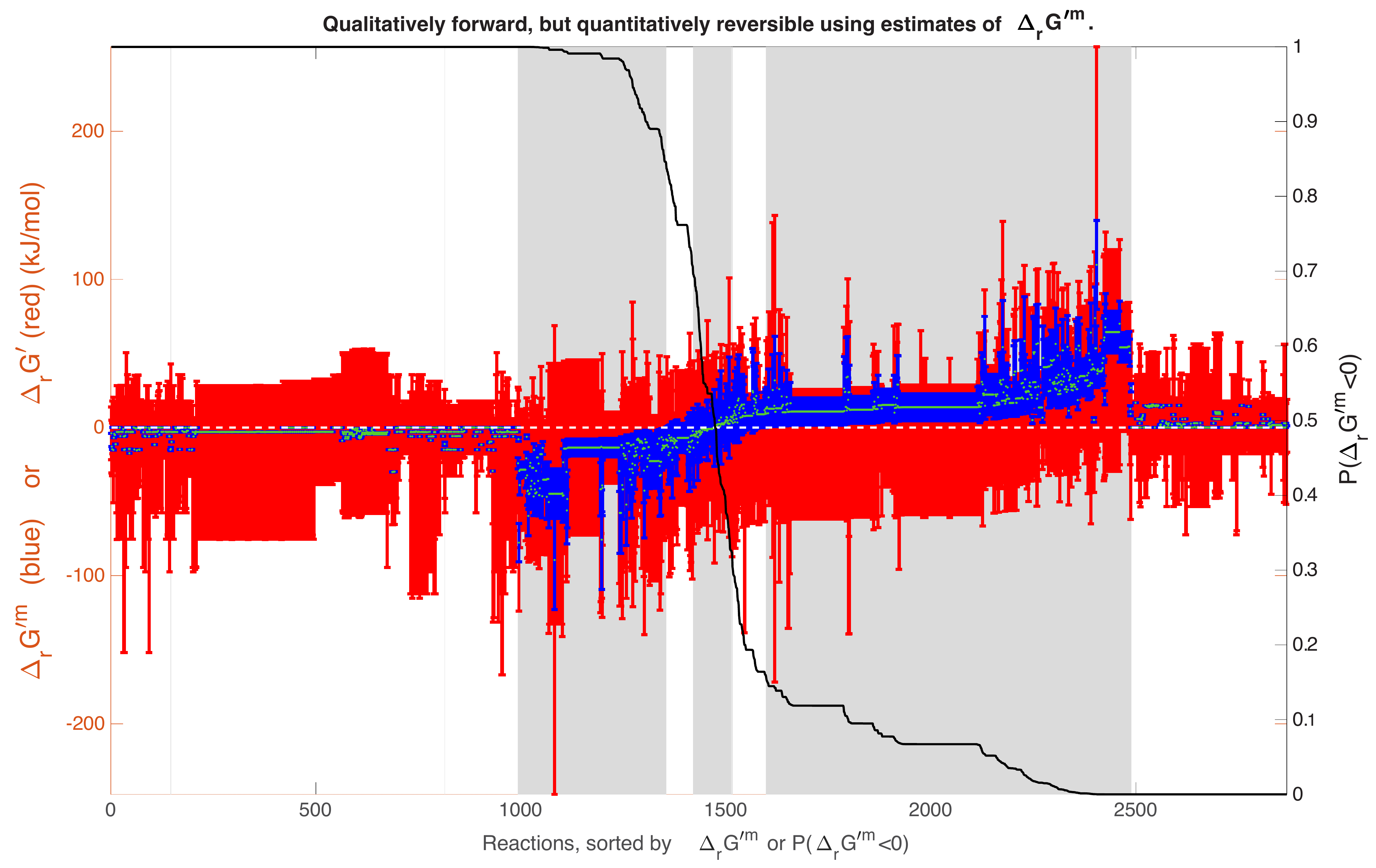}
\par\end{centering}
\caption{\label{fig:fwdReverse}\textbf{Qualitatively forward, quantitatively
reverse reactions in a multi-compartmental, genome-scale model.} In
Recon3D, the transformed reaction Gibbs energy could be estimated
for $7,215$ reactions. Of these reactions, $2,868$ reactions were
qualitatively assigned to be forward in the reconstruction, but were
quantitatively assigned to be reversible using subcellular compartment
specific thermodynamic parameters, the component contribution method,
and broad bounds on metabolite concentrations ($10^{-5}-0.02$ mol/L),
except for certain cofactors. The geometric mean (\textcolor{green}{green})
and feasible range (between maximum and minimum) of estimated millimolar
standard transformed reaction Gibbs energy ($\Delta_{r}G^{\prime m}$,
\textcolor{blue}{blue}) and transformed reaction Gibbs energy ($\Delta_{r}G^{\prime}$,
\textcolor{red}{red}) are illustrated. The relative uncertainty in
metabolite concentrations versus uncertainty in thermochemical estimates
is reflected by the relative breadth of the red and blue bars for
each reaction, respectively. The reactions are rank ordered by the
cumulative probability that millimolar standard transformed reaction
Gibbs energy is less than zero, $P(\Delta_{r}G^{\prime m}<0)$, (black
descending line from left to right). This assumes that all metabolites
are at a millimolar concentration ($1mM$) and a Gaussian error is
assumed in component contribution estimates. In this ordering, forward
transport reactions have $P(\Delta_{r}G^{\prime m}<0)=1$ (far left)
and reverse transport reactions have $P(\Delta_{r}G^{\prime m}<0)=0$
(far right). In between, from left to right are biochemical reactions
with decreasing cumulative probability of being forward in direction,
subject to the stated assumptions. Alternative rankings are possible.
The key point is to observe that the COBRA Toolbox is primed for quantitative
integration of metabolomic data as the uncertainty in transformed
reaction Gibbs energy associated with thermochemical estimates using
the component contribution method is now significantly lower than
the uncertainty associated with the assumption of broad concentration
range.}
\end{figure}

\pStepRef{step:thermodynamicallyConstraint} Thermodynamically constrained
flux predictions can differ markedly from those obtained with flux
balance analysis. An open challenge is acquisition of sufficient thermochemical
training data as well as sufficient quantitative metabolomic data,
such that estimates of transformed reaction Gibbs energies can be
made with sufficiently low uncertainty to constrain reaction directionality
with high confidence. The degree of confidence typically differs markedly
between reactions. Therefore, a pragmatic approach to rank order reaction
directionality assignments by the probability that the thermodynamically
assigned reaction directionality is forward, reversible or reverse
(see Figure \ref{fig:fwdReverse} for an application to Recon3D). 

\subsubsection{\textbf{\nameref*{step:exportReconMap}}}

\begin{figure}[H]
\begin{centering}
\includegraphics[width=1\textwidth]{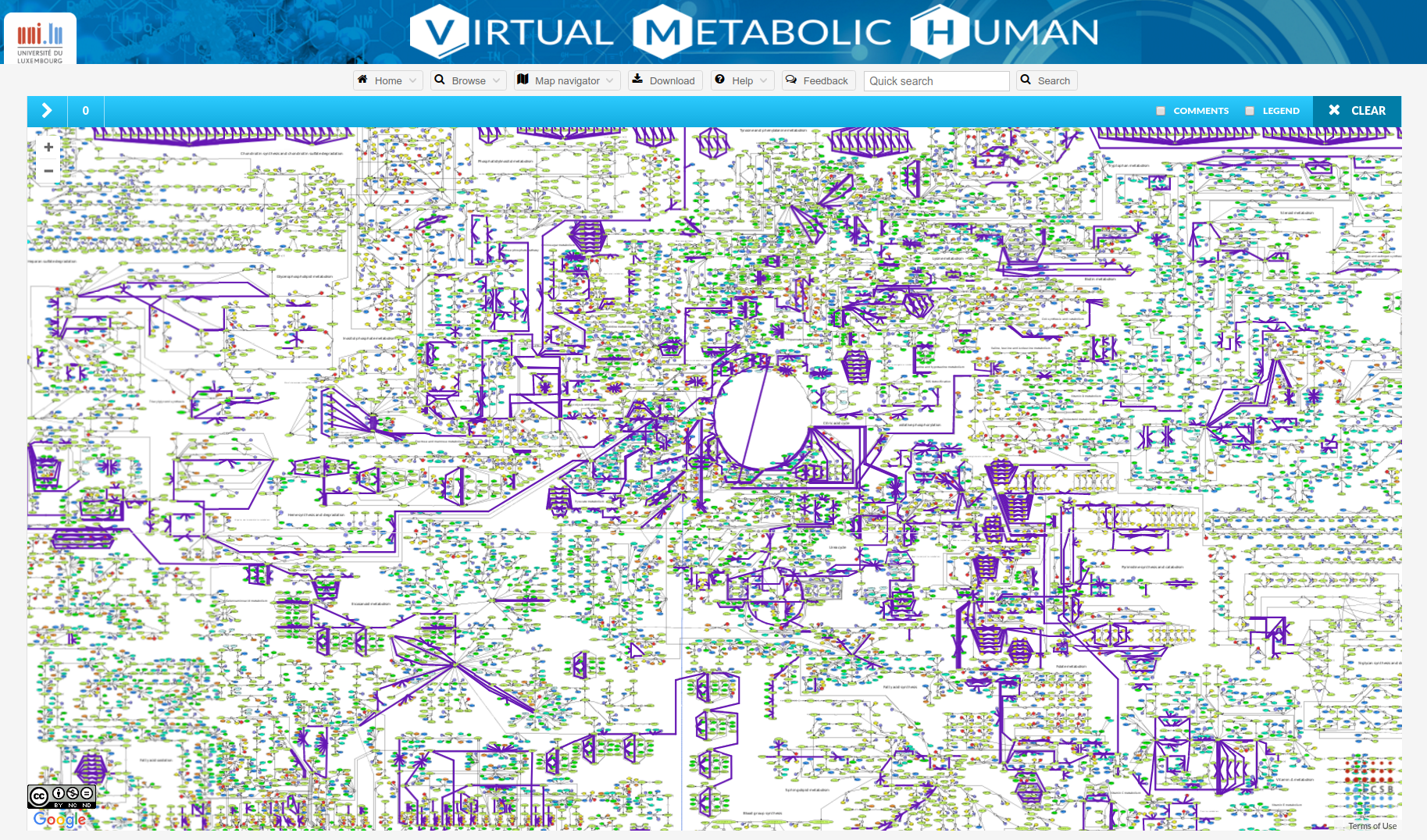}
\par\end{centering}
\caption{\label{fig:Overlay-of-the}Overlay of the flux vector for maximum
ATP synthase flux, using flux balance analysis with regularisation
of the flux vector. Active fluxes are highlighted (purple).}
\end{figure}

\pStepRef{step:exportReconMap} Figure \ref{fig:Overlay-of-the} illustrates
the overlay of an optimal regularised flux balance analysis solution
overlain within ReconMap within a web browser window.

\subsubsection{\textbf{\nameref*{step:paint4NetRadius}}}

\begin{figure}[H]
\begin{centering}
\includegraphics[width=1\textwidth]{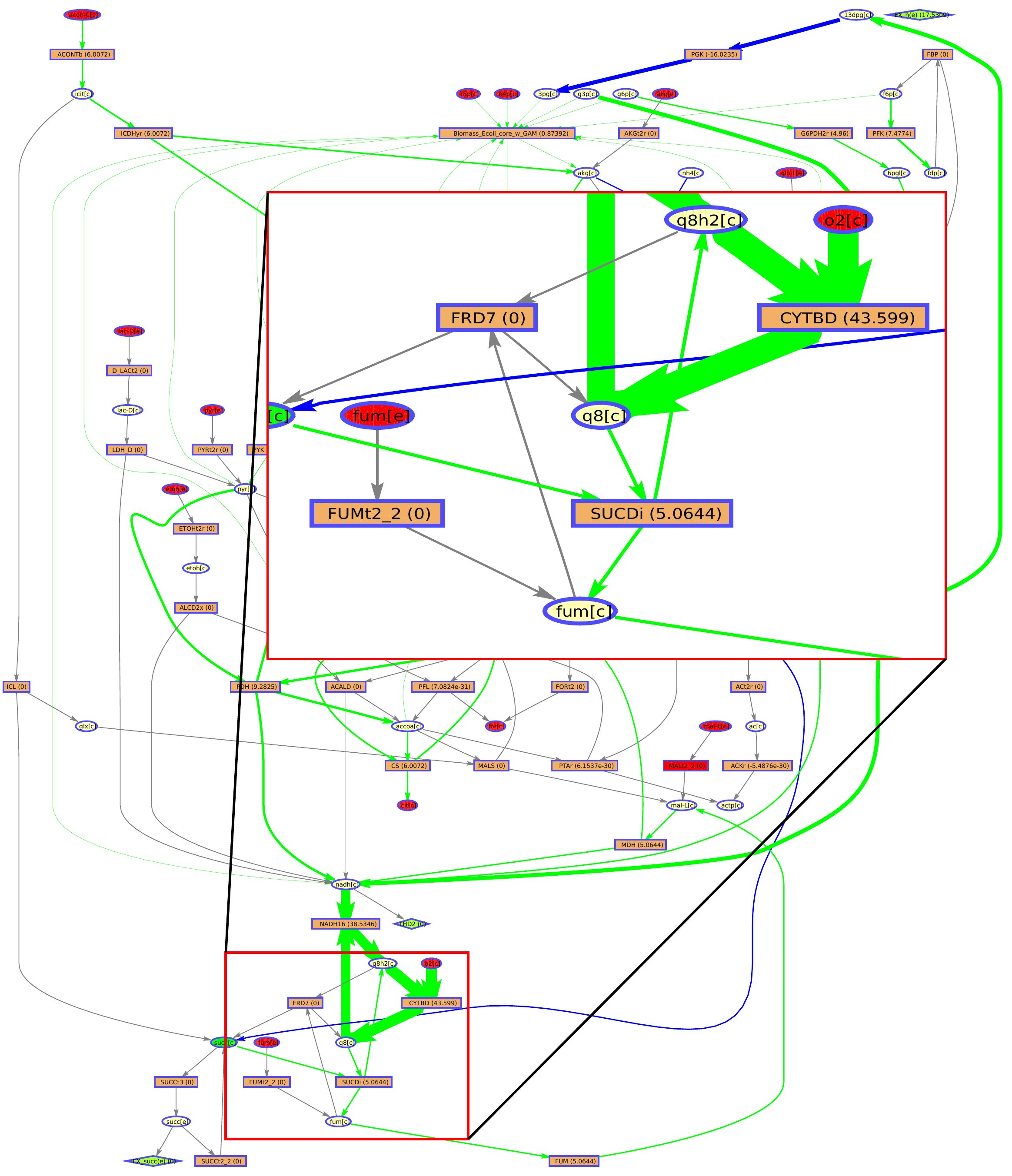}
\par\end{centering}
\caption{\label{fig:Selective-scope-visualisation}Selective scope visualisation
of the \emph{E. coli} core model model by Paint4Net. Rectangles represent
reactions with rates of fluxes in brackets; the red rectangles represent
reactions with only one metabolite; ellipses represent metabolites;
the \textcolor{red}{red} ellipses represent dead end metabolites;
\textcolor{darkgray}{grey} edges represent zero-rate fluxes; \textcolor{green}{green}
edges represent positive-rate (forward) fluxes; and \textcolor{blue}{blue}
edges represent negative-rate (backward) fluxes. Network visualisation
also enables zoom to specific regions.}
\end{figure}

\pStepRef{step:paint4NetRadius} Figure \ref{fig:Selective-scope-visualisation}
illustrates a fragment of a Paint4Net visualisation contextualised
using a flux vector to control the thickness and colour of edges representing
reactions. In the visualisation, one can discover the isolated parts
of network without any flux, as well as cycles running without any
substrate.

\vfill{}


\paragraph*{ACKNOWLEDGEMENTS}

The Reproducible Research Results (R3) team, in particular Christophe
Trefois and Yohan Jarosz, of the Luxembourg Centre for Systems Biomedicine,
is acknowledged for their help in setting up the virtual machine and
the Jenkins server. This study was funded by the National Centre of
Excellence in Research (NCER) on Parkinson\textquoteright s disease,
the U.S. Department of Energy, Offices of Advanced Scientific Computing
Research and the Biological and Environmental Research as part of
the Scientific Discovery Through Advanced Computing program, grant
no. DE-SC0010429. This project has also received funding from the
European Union\textquoteright s HORIZON 2020 research and innovation
programme under grant agreement No 668738, the Fonds National de la
Recherche grant O16/11402054, and the Luxembourg National Research
Fund (FNR) ATTRACT program (FNR/A12/01) and OPEN (FNR/O16/11402054)
grants. Nathan E. Lewis has been supported by NIGMS (R35 GM119850)
and the Novo Nordisk Foundation (NNF10CC1016517). Anne Richelle has
been supported by the Lilly Innovation Fellows Award. Francisco Planes
has been supported by the Minister of Economy and Competitiveness
of Spain (BIO2016-77998-R) and ELKARTEK Programme of the Basque Government
(KK-2016/00026). I\~{o}igo Apaolaza has been supported by the Basque Government
predoctoral grant (PRE\_2016\_2\_0044). Bernhard \O. Palsson has
been supported by the Novo Nordisk Foundation through the Center for
Biosustainability at the Technical University of Denmark (NNF10CC1016517).

\textbf{These authors contributed equally to this work}: Laurent Heirendt
\& Sylvain Arreckx

\pagebreak{}

\paragraph*{AUTHOR CONTRIBUTIONS}
\begin{center}
\begin{longtable}{>{\raggedright}p{4cm}|>{\raggedright}p{10.5cm}}
\hline 
\textbf{Author} & \textbf{Contributions}\tabularnewline
\hline 
Sylvain  Arreckx & Continuous integration, code review, opencobra.github.io/cobratoolbox,
Jenkins, Documenter.py, changeCobraSolver, pull request support, tutorials,
tests, coordination, manuscript, initCobraToolbox.\tabularnewline
\hline 
Laurent  Heirendt & Continuous integration, code review, fastFVA (new version, test \&
integration), MATLAB.devTools, opencobra.github.io, tutorials, tests,
pull request support, coordination, manuscript, initCobraToolbox,
forum support.\tabularnewline
\hline 
Thomas  Pfau & IO and transcriptomic integration, tutorials, tutorial reviews, IO
and transcriptomic integration sections of manuscript, forum support,
pull request support, code review.\tabularnewline
\hline 
Sebasti\'{a}n  N. Mendoza & Development and update of strain design algorithms, GAMS and MATLAB
integration, tutorials.\tabularnewline
\hline 
Anne  Richelle & Transcriptomic data integration methods, tutorials, transcriptomic
integration section of manuscript, RuMBA, pFBA, metabolic tasks, tutorial
review.\tabularnewline
\hline 
Almut  Heinken & Multispecies modelling code contribution, tutorial review, testing.\tabularnewline
\hline 
Hulda S. Haraldsd\'{o}ttir & Thermodynamics, conserved moiety and sampling methods.\tabularnewline
\hline 
Jacek  Wachowiak & Documentation.\tabularnewline
\hline 
Sarah M. Keating & SBML input-output support.\tabularnewline
\hline 
Vanja  Vlasov & Tutorials.\tabularnewline
\hline 
Stefania  Magnusd\'{o}ttir & Multispecies modelling, tutorial review, testing.\tabularnewline
\hline 
Chiam Yu  Ng & Strain design code review, tutorial review, manuscript (OptForce/biotech
introduction).\tabularnewline
\hline 
German  Preciat & Tutorials and chemoinformatics for metabolite structures and atom
mapping data.\tabularnewline
\hline 
Alise  \v Zagare & Metabolic cartography.\tabularnewline
\hline 
Siu H.J. Chan & Solution navigation, multispecies modelling code, tutorial review.\tabularnewline
\hline 
Maike K. Aurich & Metabolomic data integration.\tabularnewline
\hline 
Catherine M. Clancy & Tutorials, testing.\tabularnewline
\hline 
Jennifer  Modamio & Metabolic cartography and human metabolic network visualisation tutorials.\tabularnewline
\hline 
John T. Sauls & modelBorgifier code and tutorial.\tabularnewline
\hline 
Alberto  Noronha & Virtual metabolic human interoperability.\tabularnewline
\hline 
Aarash  Bordbar & MinSpan method and tutorial, supervision on uFBA method and tutorial.\tabularnewline
\hline 
Benjamin  Cousins & CHRR uniform sampling.\tabularnewline
\hline 
Diana  C. El Assal  & Tutorials.\tabularnewline
\hline 
Luis V. Valcarcel & Tutorials and genetic MCSs implementation.\tabularnewline
\hline 
I\~{n}igo  Apaolaza & Tutorials and genetic MCSs implementation.\tabularnewline
\hline 
Susan  Ghaderi & Interoperability with CellNetAnalyzer.\tabularnewline
\hline 
Masoud  Ahookhosh & Adaptive Levenberg-Marquardt solver.\tabularnewline
\hline 
Marouen  Ben Guebila & Tutorial reviews.\tabularnewline
\hline 
Andrejs  Kostromins & Paint4Net code and tutorial.\tabularnewline
\hline 
Nicolas Sompairac & Development of metabolomic cartography tool and tutorial.\tabularnewline
\hline 
Hoai M. Le & Cardinality optimisation solver.\tabularnewline
\hline 
Ding  Ma & Quad precision solvers.\tabularnewline
\hline 
Yuekai  Sun & Multiscale flux balance analysis reformulation.\tabularnewline
\hline 
Lin  Wang & Strain design code review, tutorial review, manuscript (OptForce).\tabularnewline
\hline 
James T. Yurkovich & uFBA method and tutorial.\tabularnewline
\hline 
Miguel A.P. Oliveira & Tutorial.\tabularnewline
\hline 
Phan T. Vuong & Adaptive Levenberg-Marquardt solvers, boosted difference of convex
optimisation solver.\tabularnewline
\hline 
Lemmer P. El Assal  & Chemoinformatic data integration, documentation.\tabularnewline
\hline 
Inna Kuperstein & Development of metabolomic cartography tool and tutorial.\tabularnewline
\hline 
Andrei Zinovyev & Development of metabolomic cartography tool and tutorial.\tabularnewline
\hline 
H. Scott  Hinton & Tutorials.\tabularnewline
\hline 
William A. Bryant & Code refinement.\tabularnewline
\hline 
Francisco J. Arag\'{o}n Artacho & Duplomonotone equation solver, boosted difference of convex optimisation
solver, adaptive Levenberg-Marquardt solvers.\tabularnewline
\hline 
Francisco  J. Planes & Academic Supervision, tutorials and genetic MCSs implementation.\tabularnewline
\hline 
Egils  Stalidzans & Academic supervision, Paint4Net, tutorial.\tabularnewline
\hline 
Alejandro  Maass & Academic supervision.\tabularnewline
\hline 
Santosh  Vempala & Academic supervision, CHRR uniform sampling algorithm.\tabularnewline
\hline 
Michael  Hucka & Academic supervision, SBML input-output support.\tabularnewline
\hline 
Michael A.  Saunders & Academic supervision, quad precision solvers, nullspace computation,
convex optimisation.\tabularnewline
\hline 
Costas D. Maranas & Academic supervision, strain design algorithms.\tabularnewline
\hline 
Nathan  E. Lewis & Academic supervision and coding, transcriptomic data integration.
RuMBA, pFBA, metabolic tasks, tutorial review.\tabularnewline
\hline 
Thomas  Sauter & Academic supervision, FASTCORE algorithm.\tabularnewline
\hline 
Bernhard \O. Palsson & Academic supervision, openCOBRA stewardship.\tabularnewline
\hline 
Ines  Thiele & Academic supervision, tutorials, code contribution, manuscript.\tabularnewline
\hline 
Ronan M.T. Fleming & Conceptualisation, lead developer, academic supervision, software
architecture, code review, sparse optimisation, nullspace computation,
thermodynamics, variational kinetics, fastGapFill, sampling, conserved
moieties, network visualisation, forum support, tutorials, manuscript.\tabularnewline
\hline 
\end{longtable}
\par\end{center}

\paragraph*{\emph{}COMPETING FINANCIAL INTERESTS }

The authors declare that they have no competing financial interests.

\pagebreak{}

\normalsize

\textsf{\textcolor{black}{\renewcommand{\refname}{REFERENCES}}}

\textsf{\textcolor{black}{\pagebreak{}}}

\normalsize

\section*{Supplementary information}

\normalsize

\subsection*{\label{sec:MATLAB-basics}Supplementary Manual 1 - MATLAB basics}

\quad{}\quad{}Brief tutorial and guidelines on how to use the MATLAB
software.

\subsection*{\label{sec:Shell-or-Terminal}Supplementary Manual 2 - Shell or Terminal
basics}

\quad{}\quad{}Brief tutorial and guidelines on how to use the Terminal
or shell.

\subsection*{\label{subsec:Contributing-to-The}Supplementary Manual 3 - Contributing
to the COBRA Toolbox using git}

\quad{}\quad{}Guidelines on how to use git and contribute to the
COBRA Toolbox.
\begin{center}
\par\end{center}

\section*{}

\pagebreak{}

\setcounter{page}{1}

\section{Supplementary Manual 1 - MATLAB basics}

The following introductory material and online tutorials available
at \url{https://mathworks.com/help/matlab/getting-started-with-matlab.html}
are recommended to learn some of the most useful features of MATLAB.
In order to execute MATLAB in a Unix environment, at the prompt, type

\begin{lstlisting}[style=bashStyle]
$ matlab
\end{lstlisting}

and return. Alternatively, double-click on the MATLAB icon in your
installation folder or the desktop (or dock). In MATLAB, each function
has the following general form

\begin{lstlisting}
>> [output1, ~, output3] = functionName(input1, input2)
\end{lstlisting}

where \mcode{input1} and \mcode{input2} denote mandatory inputs
supplied to a function named \mcode{functionName}. The tilde character
\mcode{\~} denotes a placeholder for an output that is not stored
in the workspace. The significant inequality sign \mcode{>>} denotes
the MATLAB command line; anything directly following \mcode{>>} and
before a semicolon is meant to be either entered on the MATLAB command
line or included in a script that implements a sequence of functions
in an analysis pipeline. Such a script may be standard \emph{.m} file,
but preferably a MATLAB live script \url{https://mathworks.com/help/matlab/matlab_prog/what-is-a-live-script.html},
which is an interactive document that combines MATLAB code with embedded
output, formatted text, equations, and images in a single environment.
In this manuscript, all MATLAB code and commands are typed in \mcode{monospaced}
characters. The COBRA Toolbox, as well as MATLAB, are documented extensively.
For instance, more information on \mcode{commandName} can be retrieved
by typing:

\begin{lstlisting}
>> help commandName
\end{lstlisting}

\subsubsection*{Basic Commands}

Load a file \emph{filenam}\textit{\emph{e}}\textit{.mat}:

\begin{lstlisting}
>> load(filename.mat)
\end{lstlisting}

Save the workspace in \emph{filename} (as \textit{.mat} file):

\begin{lstlisting}
>> save filename
\end{lstlisting}

Save only the variable in \emph{filename} (as \textit{.mat} file):

\begin{lstlisting}
>> save filename variable
\end{lstlisting}

Clear workspace:

\begin{lstlisting}
>> clear
\end{lstlisting}

Delete \emph{only one variable}: 

\begin{lstlisting}
>> clear variable
\end{lstlisting}

Call .m file which contains a MATLAB script:

\begin{lstlisting}
>> scriptname
\end{lstlisting}

\subsubsection*{Matrix-related commands}

Create matrix \mcode{X} with 3 columns and 2 rows. The elements of
this matrix are denoted \mcode{X(i,j)}, where \mcode{i} are the
rows and \mcode{j} are the columns:

\begin{lstlisting}
>> X = [1 1 1; 1 1 1]
\end{lstlisting}

Create vector with words \mcode{(name1, names2)} as entry:

\begin{lstlisting}
>> names = {'name1' 'name2'} 
\end{lstlisting}

Calculate the rank of matrix \mcode{X}:

\begin{lstlisting}
>> rank(X)
\end{lstlisting}

Calculate the null space of matrix \mcode{X}:

\begin{lstlisting}
>> null(X)
\end{lstlisting}

Calculate the singular value decomposition of matrix \mcode{X} where \mcode{U}
contains the left singular vectors, \mcode{S} contains singular values,
and \mcode{V} contains the right singular vectors:

\begin{lstlisting}
>> [U, S, V] = svd(X) 
\end{lstlisting}

\subsubsection*{Basic Graphic commands}

Open a graphic window:

\begin{lstlisting}
>> figure
\end{lstlisting}

Plot vector \mcode{x} versus \mcode{y} (type \mcode{help plot}
to get more option information):

\begin{lstlisting}
>> plot(x, y)
\end{lstlisting}

Create histograms of matrix \mcode{X} (\emph{optional}: \mcode{hist(X, bins)}
where bins is a number of how many points are averaged (default \mcode{100})):

\begin{lstlisting}
>> hist(X)
\end{lstlisting}

Label the x-axes of the plot with \emph{name}:

\begin{lstlisting}
>> xlabel('name')
\end{lstlisting}

Label the y-axes of the plot with \emph{name}:

\begin{lstlisting}
>> ylabel('name')
\end{lstlisting}

Add title \emph{name} on plot:

\begin{lstlisting}
>> title('name')
\end{lstlisting}

Plot multiple graphs in one figure:

\begin{lstlisting}
>> hold on
\end{lstlisting}

Disable multiple graphs:

\begin{lstlisting}
>> hold off
\end{lstlisting}

\pagebreak{}

\setcounter{page}{1}

\section{Supplementary Manual 2 - Shell or Terminal basics}

The shell command-line, sometimes referred to as the terminal, is
a tool into which you can type text commands to perform specific tasks\textemdash in
contrast to a graphical user interface. You can navigate between folders,
act on files inside those folders, or perform other actions. The symbol
\mcode{\$} denotes the shell command line.

In order to list files inside the active directory, use the \mcode{$ ls}
command. Directories can be changed using the \mcode{$ cd} command.
You can navigate to either full or relative paths. For example, the
following command navigates to a relative path\textemdash one above
the current directory:

\begin{lstlisting}[style=bashStyle]
$ cd ..
\end{lstlisting} 

If you want to navigate to a directory \textit{myDirectory} inside
of the current directory, type:

\begin{lstlisting}[style=bashStyle]
$ cd myDirectory
\end{lstlisting}

You can always use the \mcode{$ pwd} command to retrieve the current
absolute path. In order to create a new directory, use the \mcode{$ mkdir folderName}
command. You can remove any directory with the \mcode{$ rm -rf folderName}
command. Use the \mcode{$ rm} command to delete a file called \textit{fileName}:

\begin{lstlisting}[style=bashStyle]
$ rm fileName
\end{lstlisting}

\pagebreak{}

\setcounter{page}{1}

\section{Supplementary Manual 3 - Contributing to the COBRA Toolbox using
git}

\criticalStep This supplementary information is tailored to users
who feel comfortable using the terminal (or shell). It is recommended
for other users to use the MATLAB.devTools described in Steps \ref{subsec:Installation-of-the}-\ref{subsec:Update-the-fork}.
A Github account is required and \emph{git} must be installed. You
also must already have forked the \emph{opencobra/cobratoolbox} repository
by clicking on the fork button on the main \emph{opencobra/cobratoolbox}
repository page.

The repository of the COBRA Toolbox is version controlled with the
open-source standard \emph{git} on the public code development site
\url{https://github.com}. Any incremental change to the code is wrapped
in a commit, tagged with a specific tag (called SHA1), a commit message,
and author information, such as the email address and the user name.
Contributions to the COBRA Toolbox are consequently commits that are
made on branches.

The development scheme adopted in the repository of the COBRA Toolbox
has 2 branches: a \emph{master }and a \emph{develop} branch. The stable
branch is the \emph{master} branch, while it is the \emph{develop}
branch that includes all new features and to which new contributions
are merged. Contributions are submitted for review and testing through
pull requests, the \emph{git} standard. The \emph{develop} branch
is regularly merged into the \emph{master} branch once testing is
concluded. The development scheme has been adopted for obvious reasons:
the COBRA Toolbox is heavily used on a daily basis, while the development
community is active. The key advantage of this setup is that developers
can work on the next stable release, while users can enjoy a stable
version. Developers and users are consequently working on the same
code base without interfering. Understanding the concept of branches
is key to submitting hassle-free pull requests and starting to contribute
using \emph{git}.

\caution The following commands should only be run from the terminal
(or the shell). An SSH key must be set in your Github account settings.

\medskip{}

In order to get started, clone the forked repository:

\begin{lstlisting}[style=bashStyle]
$ git clone git@github.com:<username>/cobratoolbox.git fork-cobratoolbox
\end{lstlisting}

This will create a folder called \emph{fork-cobratoolbox}. Make sure
to replace \emph{<username>} with your Github username. Any of the
following commands are meant to be run from within the folder of the
fork called \emph{fork-cobratoolbox}.

\begin{lstlisting}[style=bashStyle] 
$ cd fork-cobratoolbox
\end{lstlisting}

In order to complete the cloned repository with external code, it
is recommended to clone all submodules:

\begin{lstlisting}[style=bashStyle]   
$ git submodule update --init
\end{lstlisting}

Note that your fork is a copy of the \emph{opencobra/cobratoolbox}
repository and is not automatically updated. As such, you have to
configure the address of the \emph{opencobra/cobratoolbox }repository:

\begin{lstlisting}[style=bashStyle]   
$ git remote add upstream git@github.com:opencobra/cobratoolbox.git
\end{lstlisting}

Now, there are two addresses (also called remotes) configured: \emph{origin}
and \emph{upstream}. You can verify this by typing:

\begin{lstlisting}[style=bashStyle]
$ git remote -v
\end{lstlisting}

In order to update your fork, run the following commands:

\begin{lstlisting}[style=bashStyle]
$ git fetch upstream
\end{lstlisting}

First, update the \emph{master} branch:

\begin{lstlisting}[style=bashStyle] 
$ git checkout master # checkout the <master> branch locally
$ git merge upstream/master # merge the changes from the upstream repository
$ git push origin master # push the changes to the <master> branch of the fork
\end{lstlisting}

Then, update the \emph{develop} branch

\begin{lstlisting}[style=bashStyle]
$ git checkout develop # checkout the <develop> branch
$ git merge upstream/develop # merge the changes from the upstream repository
$ git push origin develop # push the changes to the <develop> branch of the fork
\end{lstlisting}

\troubleshooting Should the step fail to checkout the develop branch,
you should create the develop branch first based on the \emph{develop}
branch of the upstream repository:

\begin{lstlisting}[style=bashStyle]
$ git checkout -b develop upstream/develop
\end{lstlisting}

\textbf{Create a contribution and submit a pull request}

Now, as the fork is up-to-date with the upstream repository, start
a new contribution. A new contribution must be made on a new branch,
that originates from the \emph{develop} branch\emph{.} Create the
new branch:

\begin{lstlisting}[style=bashStyle]
$ git checkout -b <myBranch> develop
\end{lstlisting}

Now, you can make changes in the folder \emph{fork-cobratoolbox}.
Once you are done making changes, you can contribute the files. An
important command that lists all changes is to retrieve the repository
status:

\begin{lstlisting}[style=bashStyle]
$ git status
\end{lstlisting}

A list is displayed with new, modified, and deleted files. You can
add the changes (even deletions) by adding the file:

\begin{lstlisting}[style=bashStyle]
$ git add <fileName>.<fileExtension>
\end{lstlisting}

\criticalStep Contrary to what is sometimes provided as a shortcut,
it is not advised to add all files all at once using as this command
will add \textbf{all} files, even hidden files and binaries.

\begin{lstlisting}[style=bashStyle]
$ git add . # bad practice
\end{lstlisting}

Then, commit the changes by setting a commit message \emph{<yourMessage>}:

\begin{lstlisting}[style=bashStyle]
$ git commit -m "<myMessage>"
\end{lstlisting}

Finally, push your commit to Github:

\begin{lstlisting}[style=bashStyle]
$ git push origin <myBranch>
\end{lstlisting}

You should then see your commit online, and if ready, you can open
a pull request. You can select your branch in the dropdown menu and
list all commits by clicking on COMMITS.

\paragraph{Continue working on your branch after a while (rebase)}

If there have been major changes or if you want to continue working
on a branch after a while, it is recommended to do a rebase. In simple
terms, rebasing your branch shifts your commits to the top of the
branch and includes all changes from the upstream repository. Before
doing so, make sure that you do not have any uncommitted or local
changes (\emph{git status}).

\begin{lstlisting}[style=bashStyle]
$ git checkout develop
$ git fetch upstream
$ git merge upstream/develop
$ git submodule update
$ git checkout <myBranch>
$ git rebase develop
\end{lstlisting}

If you do not have any conflicts, you should see messages showing
that your changes have been applied.

If however there are conflicts, it is advised to use a merge tool
such as \emph{kdiff3}. In order to install a merge tool or abort the
rebase process, type:

\begin{lstlisting}[style=bashStyle]
$ git rebase --abort
\end{lstlisting}

In order to have the changes on \emph{<myBranch>} reflected in the
online repository, push the changes with force. Pushing with force
is required as the history of the branch has been rewritten during
rebase.

\begin{lstlisting}[style=bashStyle]
$ git push <myBranch> --force
\end{lstlisting}

\paragraph{Selectively use a commit on your branch (cherry-pick)}

Imagine having two branches called \emph{<myBranch-1>} and \emph{<myBranch-2>}.
On branch \emph{<myBranch-1>} is a commit with a SHA1 that you need
on \emph{<myBranch-2>. }You can cherry-pick the commit from \emph{<myBranch-1>
}to \emph{<myBranch-2>} by typing:

\begin{lstlisting}[style=bashStyle]
$ git checkout myBranch-2
$ git cherry-pick SHA1
\end{lstlisting}

If there are no conflicts, the displayed message should contain the
commit message and author information. In order to have the commit
listed online, conclude the cherry-pick by pushing the commit to the
remote repository:

\begin{lstlisting}[style=bashStyle]
$ git push myBranch-2
\end{lstlisting}

\paragraph*{Displaying the history of a file}

Sometimes, the history of a file is not correctly displayed online.
You can however display the history by typing:

\begin{lstlisting}[style=bashStyle]
$ git log --follow --pretty=short <fileName>.<fileExtension>
\end{lstlisting}

You can exit the screen by typing the letter \emph{q}.

When the MATLAB.devTools are installed, you can also display the history
of a file from within MATLAB:

\begin{lstlisting}[style=bashStyle]
>> history('fileName.fileExtension')
\end{lstlisting}

\end{document}